\documentclass[reqno]{amsart}
\usepackage{geompsfi}
\usepackage{latexsym}
\usepackage{amstext}
\usepackage {amsmath}
\usepackage {amsfonts}
\usepackage {amssymb}
\usepackage {amsthm}
\usepackage {bbm}
\usepackage{enumerate}
\usepackage{array}
\usepackage{comment}
\usepackage{xspace}
\DeclareMathOperator{\dist}{dist}

\DeclareMathOperator{\spt}{supp}
\DeclareMathOperator\supp{supp}

\def\loc{{\mathrm{loc}}}

\DeclareMathOperator{\BV}{BV}
\DeclareMathOperator{\Lip}{Lip}

\DeclareMathOperator{\diam}{diam}
\DeclareMathOperator{\modulo}{\, mod}
\DeclareMathOperator{\Int}{Int}
\DeclareMathOperator{\length}{length}

\newcommand{\eps}{\varepsilon}
\newcommand{\Wasserstein}{Monge-Kantorovich\xspace}
\newcommand{\multiplicity}{multiplicity\xspace}
\newcommand{\Chi}{\mathcal{X}}
\newcommand{\R}{\ensuremath{\mathbb{R}}}
\newcommand{\Rn}{\ensuremath{\mathbb{R}^2}}
\newcommand{\N}{\ensuremath{\mathbb{N}}\ }
\newcommand{\LL}{\ensuremath{\mathcal{L}}}
\newcommand{\K}{\ensuremath{\mathcal K}}
\newcommand{\sphere}{\ensuremath{\mathcal S}}
\newcommand{\F}{\ensuremath{\mathcal F}}
\newcommand{\G}{\ensuremath{\mathcal G}}
\newcommand{\Fone}{\ensuremath{\mathcal F_1}}
\newcommand{\A}{\ensuremath{\mathcal{A}}}
\newcommand{\Ha}{\ensuremath{\mathcal{H}}}
\newcommand{\ma}{\ensuremath{\mathfrak{m}}}
\newcommand{\te}{\ensuremath{\mathfrak{t}}}
\newcommand{\alphaprime}{\ensuremath{\alpha^\prime}}

\newcommand{\schwto}{\ensuremath{\to}}

\let\ds\displaystyle
\def\Mod#1{\left\|#1\right\|}
\def\mod#1{\left|#1\right|}
\def\vc#1{\ensuremath{\vcenter{\hbox{#1}}}}
\newcounter{counter-liste}
\newcounter{counter-liste2}
\newenvironment{liste2}{\ 
\begin{list}{{\arabic{counter-liste2}.}\hfill}{\usecounter{counter-liste2}
\setlength{\topsep}{0bp}
\setlength{\labelwidth}{10bp}
\setlength{\leftmargin}{12bp}
\setlength{\labelsep}{2bp}
\setlength{\itemindent}{0bp}
\setlength{\parsep}{0bp}}
}{\end{list}}
\newenvironment{liste}%
  {\ \begin{list}{{(\arabic{counter-liste})}\hfill}%
  {\topsep2mm\itemindent1ex\leftmargin0cm\usecounter{counter-liste}}
  }%
  {\end{list}}
\newenvironment{liste(a)}%
  {\ \begin{list}{{(\alph{counter-liste})}\hfill}%
  {\topsep2mm\itemindent1ex\leftmargin0cm\usecounter{counter-liste}}
  }%
  {\end{list}}
\def\R{\mathbb R}
\def\H{\mathcal H}
\def\W{\mathcal W}
\let\e\varepsilon
\def\mod#1{\left|#1\right|}
\def\BV{\mathrm{BV}}
\let\ds\displaystyle
\def\pref#1{(\ref{#1})}
\theoremstyle{plain}
\numberwithin{equation}{section}
\newtheorem{lemma}{Lemma}[section]
\newtheorem{theorem}[lemma]{Theorem}
\newtheorem{proposition}[lemma]{Proposition}
\newtheorem{definition}[lemma]{Definition}

\newtheorem*{prob-mass-transport}{Optimal mass transport}
\newtheorem*{prob-Kantorovich}{Optimal Kantorovich potential}
\theoremstyle{definition}
\newtheorem{remark}[lemma]{Remark}
\def\nw{{N_w}}
\def\nl{{N_\ell}}
\def\X{{\mathcal{X}}}
\def\E{{\mathcal{E}}}
\def\Fid{F^{\textrm{id}}}
\def\Fnid{F^{\textrm{nid}}}
\def\Hid{H^{\textrm{id}}}

\begin{document}
\title{Partial Localization, Lipid Bilayers, and the Elastica Functional}
\author{Mark A. Peletier \and Matthias R\"oger}
\begin{abstract}
\emph{Partial localization} is the phenomenon of self-aggregation of
mass into high-density structures that are thin in one direction and
extended in the others.  
We give a detailed study of an energy functional that arises in a
simplified model for lipid bilayer membranes. We demonstrate that this
functional, defined on a class of two-dimensional spatial mass
densities, exhibits partial localization and displays the `solid-like'
behavior of cell membranes.

Specifically, we show that density fields of moderate energy are
partially localized, \emph{i.e.} resemble thin structures. Deviation
from a specific uniform thickness, creation of `ends', and the bending of such
structures all carry an energy penalty, of different orders in terms
of the thickness of the structure. 

These findings are made precise in a Gamma-convergence result. We prove
that a rescaled version of
the energy functional converges in the zero-thickness limit  to a
functional that is defined on 
a class of planar curves. Finiteness of the limit enforces both
optimal thickness and non-fracture; if these conditions are met, then
the limit value is given by the classical Elastica
(bending) energy of the curve. 
\end{abstract}
\maketitle

\section{Introduction}

In this paper we study the asymptotic expansion as $\eps\to 0$ of the
functional  
\begin{equation}
\label{def:Fe}
\F_\eps(u,v) := \begin{cases}
  \;\ds\e\! \int \mod{\nabla u} + \frac1\e d_1(u,v) \qquad\qquad &
  \text{if }(u,v)\in \K_\eps, \\ 
  \;\infty & \text{otherwise.}
\end{cases}
\end{equation}
Here $d_1(\cdot,\cdot)$ is the \Wasserstein distance 
(Definition~\ref{def:wasserstein}), and
\begin{multline}
\label{def:Ke}
\K_\eps := \biggl\{ (u,v)\in 
   \BV\bigl(\R^2;\{0,\e^{-1}\}\bigr)\times L^1\bigl(\R^2;\{0,\eps^{-1}\}\bigr): \\
  \left.\int u = \int v =M , \ uv = 0\text{ a.e.}\right\}.
\end{multline}
How this singular-perturbation problem is related to the terms in the title---partial
localization, lipid bilayers, and the Elastica functional---we explain in the rest
of this introductory section. 

\subsection{Lipid Bilayers}
\begin{figure}[t]
\centering
\includegraphics[width=\textwidth]{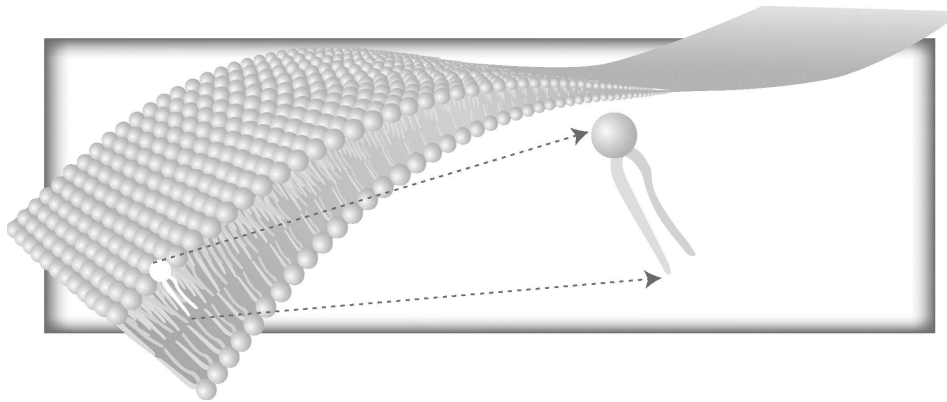}
\caption{Lipid molecules aggregate into macroscopically surface-like structures}
\label{fig:bl}
\end{figure}

Lipid bilayers, biological membranes,
are the living cell's main separating structure. 
They shield the interior of the cell from the outside, and their
mechanical properties determine a large part of the interior
organization of living cells. 
The main component is a lipid molecule (Fig.~\ref{fig:bl}) which consists
of a head and two tails. The head is usually charged, and therefore
hydrophilic, while the tails are hydrophobic. This difference in water
affinity causes lipids to aggregate, and in the biological setting
the lipids are typically found in a bilayer
structure as shown in Fig.~\ref{fig:bl}. In such a structure the
energetically unfavorable tail-water interactions are avoided by
grouping the tails together in a water-free zone, shielded from the water
by the heads. Note that, despite the structured appearance of the bilayer,
there is no covalent bonding between any two lipids; the bilayer structure is
entirely the result of the hydrophobic effect.
%
%

As planar structures, lipid bilayers have a remarkable combination of solid-like and
liquid-like properties. On one hand they resist various types
of deformation, such as extension, bending, and fracture, much in the way
a sheet of rubber or another elastic material does: 
for instance, in order to stretch the bilayer, a tensile
force has to be applied, and when this force is removed the structure will again relax to
its original length. 
With respect to in-plane rearrangements of the lipid molecules,
however, the material behaviour is viscous, and there is no penalty to large 
in-plane deformations.

While this interesting combination of properties is clearly related to the chemical
makeup of the lipids---most importantly, the hydrophobic character
of the tails---quantitative and detailed understanding of the phenomenon is
still lacking. Here we focus on a simple question that has already been alluded to above:
how can we understand the stability of these planar structures,
and their pseudo-solid behaviour, if they are constructed
from independent, non-bound molecules? This is the main question behind 
the analysis of this paper. 

\subsection{Partial localization}

The self-aggregation  of lipids into bilayers is fundamentally different
from the self-aggregation of `simple' hydrophobic compounds in
water. 
Lipid bilayers are \emph{thin structures}, in the sense that
there is a separation of length scales: the thickness of a lipid bilayer is fixed
to approximately two lipid lengths, while the in-plane spatial 
extent is only limited by the surroundings. By contrast, 
`simple' hydrophobic compounds aggregate in  water to form
drops that lack the small intrinsic length scale of lipid bilayers.

We use the term \emph{partial localization} for the self-aggregation
into structures that are thin in one or more directions and
`large' in others. The word \emph{localization} is taken from
the literature of reaction-diffusion equations, in which localized 
solutions are those that are concentrated, in a well-defined way, in a 
small neighbourhood of a point. By \emph{partial} localization
we refer to localization to the neighbourhood of a set that
has intermediate dimension, \emph{i.e.} dimension
larger than zero (a point) but
smaller than that of the ambient space.

\subsection{Energy on a mesoscale: The functional \boldmath $\F_\eps$}
The functional $\F_\eps$ that is defined above is
the result of an attempt to capture enough of the essence
of lipid bilayers to address the issues of stability
and pseudo-solid behaviour while
keeping the description as simple as possible. 
%
The derivation of $\F_\eps$, which is given in Appendix~\ref{app:derivation},
starts with a simple two-bead model of lipid molecules, inspired by
well-known models of block copolymers~\cite{Leibler80,BatesFredrickson90,FredricksonBates96}, and 
proceeds to reduce complexity through a number of sometimes
radical simplifications. Despite these simplifications the physical
origin of the various elements of $\F_\eps$ remains identifiable:
\begin{itemize}
\item The functions $u$ and $v$ represent densities of the (hydrophobic) tail and (hydrophilic) headb eads;
\item The term $\int \mod{\nabla u}$, coupled with the restriction to 
functions $u$ and $v$ that take only two values, and have disjoint
support, represents an interfacial energy that arises from the hydrophobic effect;
\item The \Wasserstein distance $d_1(u,v)$ between $u$ and 
$v$ is a weak remnant of the covalent bonding between head and tail
particles.
\end{itemize}
The parameter $\eps$ appears in \eqref{def:Ke} as a density scaling.
We prove that $\eps$ in fact is related
to the \emph{thickness} of structures with moderate energy.

We show in this paper that 
the functional $\F_\eps$ favours partial localization, \emph{i.e.}
that pairs $(u,v)$ for which $\F_\eps(u,v)$ is not too large
are necessarily partially localized. We will also show that $\F_\eps$
displays three additional
properties that were already mentioned: resistance to stretching,
to fracture, and to bending. To describe the last of these in more
detail we briefly return to biophysics. 

\subsection{Energy on the macroscale: The Helfrich Hamiltonian and the Elastica functional}

Intrigued by the shape of red blood cells
Canham, Helfrich, and Evans pioneered the modelling
of lipid bilayer vesicles by energy 
methods~\cite{Canham70,Helfrich73,Evans74}.
The name of Helfrich is now associated with 
a surface energy for closed vesicles, 
represented by a smooth boundaryless surface $S$, of the form
\begin{equation}
\label{def:EHelfrich}
E_{\mathrm{Helfrich}}(S) =  
\int_S \bigl[ k(H - H_0)^2 + \overline k K\bigr]\, d\H^2.
\end{equation}
Here $k,\ \overline k$, and $H_0$ are constants, $H$ and $K$
are the (scalar) total and Gaussian curvature, and $\H^2$ is the 
two-dimensional Hausdorff measure. This energy functional, 
and many generalizations in the same vein, have been 
remarkably successful in describing the 
wide variety of vesicle shapes that are observed in laboratory 
experiments~\cite{Seifert97}.
If we add an assumption of symmetry, in which
both `sides' of the surface have the same properties, the `spontaneous
curvature' $H_0$ vanishes.
For closed surfaces in the same homotopy class, by the Gau\ss-Bonnet
Theorem, $E_{\mathrm{Helfrich}}$
is essentially equal to the
\emph{Willmore functional}~\cite{Willmore65}
\begin{gather}
 W(S) \,=\, \frac{1}{2}\int_S H^2 \, d\H^2,\label{eq:willmore}
\end{gather}
which arises in a wide variety of
situations other than that mentioned here.


\medskip

In this paper we discuss the  above mentioned bilayer
models in two space dimensions, 
\emph{i.e.}\ we consider functions
whose support resembles fattened curves in $\R^2$;
in the limit $\e\to0$ the thickness of these curves tends to zero. 
If we consider the limit curves to be two-dimensional restrictions
of cylindrical surfaces, we observe that the
Gaussian curvature $K$ vanishes.
If we also assume that the bilayer is symmetric
then the Helfrich energy reduces to the Elastica functional $\W$, the
classical  bending energy of the curve, 
\begin{gather}
  \W(\gamma) \,=\, \frac{1}{2}\int_\gamma \kappa^2 \,
  d\H^1,\label{eq:def-elastica} 
\end{gather}
where $\kappa$ is the curvature of the curve. This functional has a
long history going back at least to Jakob Bernoulli; critical points of
this energy are known as \emph{Euler elastica}, see \cite{truesdell} for
a historical review. 

\subsection{The main result: the singular limit $\e\to0$} 

In the main result of this paper we study the limit $\e\to0$, and
connect all the aspects that were discussed above. The result revolves
around a rescaled functional 
\[
\G_\e = \frac{\F_\eps - 2M}{\e^2},
\]
where $M$ is the mass of $u$ and $v$ (see~\pref{def:Ke}).

Take any sequence \mbox{$(u_\e,v_\e)_{\eps>0}\subset \K_\eps$}. 
\emph{If the rescaled energy $\G_\e(u_\e,v_\e)$ is
  bounded}, then 
\begin{enumerate}
\item the sequence $u_\e$ converges as measures;
\item the limit measure can be represented as a collection of curves (which may overlap);
\item the support of $u_\e$ is of `thickness' approximately $2\e$;
\item in particular the \emph{total length} of the curves is $M/2$;
\item each curve in the collection is \emph{closed}.
\end{enumerate}
Boundedness of $\G_\e$ along the sequence therefore implies partial
localization: the support of $u_\e$ resembles a tubular
$\e$-neighbourhood of a curve.  
In addition, \emph{stretching} and \emph{fracture} are also represented
in this result: stretching corresponds to deviation in total length 
from the optimal value $M/2$, and fracture to creation of non-closed
curves. Neither is possible for a sequence of bounded $\G_\e$; both are
penalized in $\G_\e$ at an order larger than $O(1)$, or equivalently, at
an order larger than $O(\e^2)$ in $\F_\e$. 

The resistance to \emph{bending} arises as the Gamma-limit of the
functional $\G_\e$ itself: 
\begin{enumerate}
\item For any sequence \mbox{$(u_\e,v_\e)_{\eps>0}\subset \K_\eps$}, 
\[
\W(\Gamma) := \sum_{\gamma\in\Gamma} \W(\gamma) \leq
\liminf_{\e\to0} \G_\e(u_\e,v_\e),
\]
where $\Gamma $ is the collection of limit curves introduced above, and $\W(\gamma)$ is the curve bending energy defined in 
  \eqref{eq:def-elastica}; 
\item For any given collection of curves $\Gamma$, there exists a
  sequence \mbox{$(u_\e,v_\e)_{\eps>0}\subset \K_\eps$} such that 
$\W(\Gamma) = \lim_{\e\to0} \G_\e(u_\e,v_\e)$.
\end{enumerate}
The details of this result are given as Theorem~\ref{the:main}. 

Summarizing, only partially localized structures can have moderate
energy, and 
such thin structures also display resistance to bending, stretching, and
fracture. 

\medskip
So far we have described the result that we prove in this paper. The
proof, however, also suggests a more refined result, that we present
here as a conjecture: if a sequence \mbox{$(u_\e,v_\e)_{\eps>0}\subset
  \K_\eps$} converges 
to a collection of curves~$\Gamma$,
then
\begin{align}\label{eq:asym-Fe}
  \F_\eps(u_\eps,v_\eps)\,\geq\, 2M + \mathfrak L_M(\Gamma) 
  + \eps \mathfrak{B}(\Gamma)
  + \eps^2 \sum_{\gamma\in\Gamma}\W(\gamma) + O(\e^3).
\end{align}
In this development,
\begin{itemize}
\item the function  $\mathfrak L_M(\Gamma)$ is a measure of
the deviation of $\sum_{\gamma\in\Gamma}\length(\gamma)$ from length
$M/2$: $\mathfrak L_M(\Gamma)\geq0$, 
and $\mathfrak L_M(\Gamma)=0 $ only if this sum equals $M/2$;
\item $\mathfrak B(\Gamma)$ is zero if all curves $\gamma\in\Gamma$ are 
closed, and strictly positive (a positive constant times the
number of endpoints) if at least one $\gamma\in\Gamma$ is open.
\end{itemize}
As for the Gamma-convergence above,
the inequality in this statement should be interpreted as 
a `best-possible' behaviour; while for given limit curves $\gamma$ 
sequences exist for which this inequality is an equality, 
it is also possible to construct sequences along which the difference
between the left- and right-hand sides is unbounded. 
\subsection{Strategy of the analysis and overview of the paper}
At first glance it is not obvious that the functionals $\F_\eps$
encode bending stiffness effects. Curvature terms in fact appear at 
order $\eps^2$ of an asymptotic development, which asks for a
careful analysis. Most information is hidden in the
nonlocal \Wasserstein 
distance term. The special property that makes our analysis work is that
this distance decouples into one-dimensional problems. This reduction is
the major step in our analysis and the connection to the \emph{optimal mass
transport problem} is the most important tool here. Whereas the idea
becomes very clear in simple situations which resemble the lim-sup
construction in the Gamma-convergence proof, some effort is required to
prove the lim-inf estimate in the general case, where certain `defects'
such as high curvature and vanishing thickness of approximating structures
have to be controlled simultaneously.   

In the next section we motivate our choice of scaling in the functional
$\F_\eps$ and illustrate the behaviour of $\F_\eps$ and $\G_\e$ in two
particular examples. In Section \ref{sec:wass} we give definitions of
the 
\Wasserstein distance and `systems of curves' which describe the class of
limit structures.
A precise statement of our results is given in Section
\ref{sec:main}. We briefly review the optimal mass transport problem
in Section \ref{sec:mtp}. In Section~\ref{sec:overview} we provide a short guide to the structure of the proof of our 
main result, Theorem~\ref{the:main}, and the proof itself is given 
in Sections \ref{sec:liminf} and \ref{sec:limsup}.
We conclude in Section \ref{sec:discussion} with some discussions of generalisations and the wider implications of this result.

\bigskip
\textbf{Acknowledgement.} The authors are grateful for several inspiring
discussions with Prof.\ Felix Otto and Prof.\ Reiner Sch\"atzle. 
\\[5mm]
\begin{minipage}{12cm}
\centering{\bf Summary of notation}
\ \\[3mm]
\begin{small}
\begin{tabular}{lll}
  $\F_\eps(\cdot,\cdot)$         & functionals describing the mesoscale energy & \eqref{def:Fe}\\
  $\G_\eps(\cdot,\cdot)$         & rescaled functionals & \eqref{eq:def-Ge-uv}\\
  $\K_\eps$         & domain of $\F_\eps$, $\G_\eps$ & \eqref{def:Ke}\\
  $\W(\gamma)$     & bending energy of a curve $\gamma$ & \eqref{eq:def-elastica}\\
  $\W(\Gamma)$     & generalized bending energy of a system of
  curves $\Gamma$ & \eqref{eq:def-W-Gamma}\\
  $d_1(\cdot,\cdot)$             & \Wasserstein distance &\eqref{eq:def-d-1}\\
  $d_p(\cdot,\cdot)$             & $p$-Wasserstein distance & \eqref{eq:def-d-p}\\
  $\spt(\Gamma)$
                    & support of a system of
  curves $\Gamma$   & Def.~\ref{def:sys-curves}\\
  $\theta(\Gamma,\cdot)$
                    & \multiplicity of a system of
  curves $\Gamma$   & Def.~\ref{def:sys-curves}\\
  $|\Gamma|$
                    & total mass of a system of
  curves $\Gamma$   & Def.~\ref{def:sys-curves}\\
  $\Lip_1(\R^2)$ 	& Lipschitz continuous functions with Lipschitz constant 1
  					& \pref{eq:phi_eps}\\
  $\mathcal{T},\mathcal{E}$
                    & transport set and set of endpoints of rays &
  Def.~\ref{def:rays}\\
  $E$               & $\{s: \gamma(s) \text{ lies inside a transport ray}\}$ &
  Def.~\ref{def:para}\\
  $\theta(s)$          & ray direction in $\gamma(s)$ & Def.~\ref{def:para}\\
  $L^+(s),L^-(s),l^+(s)$     & positive, negative and effective ray
  length in $\gamma(s)$&
  Def.~\ref{def:para}\\
  $\psi(\cdot,\cdot)$            & parametrization & Def.~
  \ref{def:para}\\
  $E_\delta$        & boundary points with uniformly bounded ray lengths
  & \eqref{eq:def-E-delta}\\ 
  $\alpha(s),\beta(s)$    & direction of ray and difference to tangent
  at $\gamma(s)$& \eqref{eq:def-beta}\\
  $\ma(s,\cdot)$             & mass coordinates & \eqref{eq:def-mass}\\
  $\te(s,\cdot)$             & length coordinates & \eqref{eq:def-t_s}\\
  $M(s)$               & mass over $\gamma(s)$& \eqref{eq:def-Ms}\\
  $E_i,\theta_i,\psi_i$,
                    \\
  $L^+_i,L^-_i,l_i^+$,& corresponding quantities for a collection
  $\{\gamma_i\}$ & Rem.~\ref{rem:corr-lengths}\\
  $\alpha_i,\beta_i,\ma_i,\te_i,M_i$\\
  $E_{\eps,i},\theta_{\eps,i},\psi_{\eps,i}$,\\
  $L^+_{\eps,i},L^-_{\eps,i},l_{\eps,i}^+$
                    & corresponding quantities for  a collection
  $\{\gamma_{\eps,i}\}$ & Rem.~
  \ref{rem:corr-lengths}\\
  $\alpha_{\eps,i},\beta_{\eps,i},\ma_{\eps,i},\te_{\eps,i},M_{\eps,i}$
\end{tabular}
\end{small}
\end{minipage}

\section{Heuristics: the functional $\F_\eps$}
\label{sec:heuristics}

In the preceding section we postulated that the energy
functional $\F_\eps$, defined in~\pref{def:Fe}, favours
partial localization. More precisely,
we claim that for
small but fixed $\e>0$ functions $(u,v)$ with moderate $\G_\eps(u,v)$ 
will be partially localized, in the sense that
their support resembles one or more fattened curves. 

In this section we first provide some heuristic arguments to support 
this claim,  mostly to develop some intuitive
understanding of the functional~$\F_\eps$ and its properties. 

For the discussion below it is 
useful to remark that the distance $d_1(u,v)$ as defined in
Definition \ref{def:wasserstein}
scales as the mass 
$\int u = \int v$ times a length scale, and can be interpreted as
a $u$-weighted spatial distance between the mass distributions
$u$ and~$v$. 

\subsection{Fixed thickness (\boldmath $\e=1$), part 1: 
comparing disc and strip structures}
To gain some insight into the properties of $\F_\eps$ we start with the case of
fixed $\e=1$, and study the limit of large mass
\[
M = \int u  = \int v,\qquad M \longrightarrow \infty.
\]
We now compare two different specific realizations, a disc and a strip.
\begin{figure}[ht]
\centerline{\includegraphics[width=\textwidth]{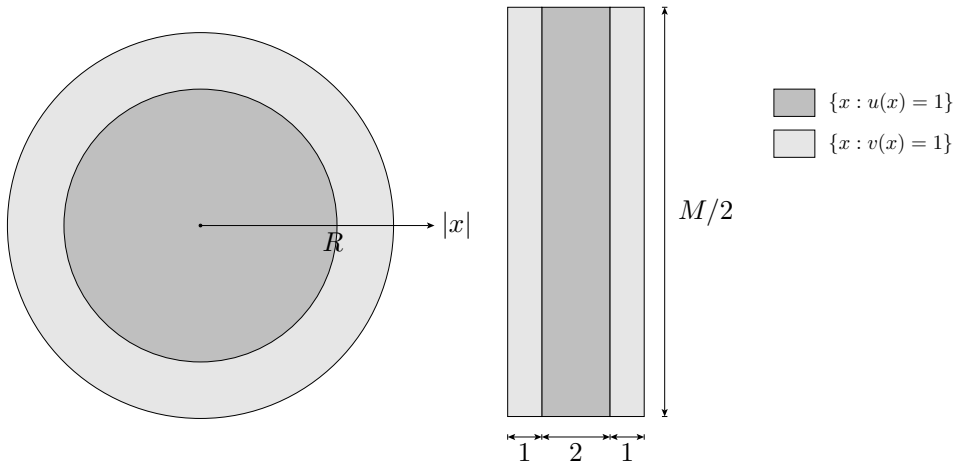}}
\caption{Disc and strip structures.}
\label{fig:disc}
\end{figure}

\begin{enumerate}
\item The \textbf{disc: } we 
concentrate all the mass of $u$ into a disc
 in $\R^2$,
of radius $R\sim M^{1/2}$, surrounded by the mass of $v$ in an
annulus (Figure~\ref{fig:disc}). In this setup mass is transported (in
the sense of the definition of $d_1(u,v)$) over
a distance  $O(R) = O(M^{1/2})$, so that
\begin{equation}
\label{sim:d1}
d_1(u,v)\sim MR\sim M^{3/2} \qquad \text{as }M\to\infty.
\end{equation}
On the other hand, the interfacial length 
$\int\mod{\nabla u}$ equals $2\pi R\sim M^{1/2}$,
and the functional $\Fone$ picks up the larger of the two:
\[
\Fone(u,v) = d_1(u,v) + \int\mod{\nabla u} \sim M^{3/2}.
\]

\item For the  \textbf{strip} we take the support of $u$ to be
a rectangle of width 2 and length $M/2$, flanked by two strips
of half this width for $\supp(v)$ (Figure~\ref{fig:disc}).
The transport distance can now be taken constant and equal to 1,
so that we find
\[
\Fone(u,v) = M + M + 4 = 2M+4.
\]
For this choice of geometry $\Fone$ scales linearly with $M$ in
the limit $M\to\infty$.
\end{enumerate}

Comparing the two we observe that the linear structure has lower energy,
for large mass, than a spherical one. This is a first
indication that $\Fone$ may favour partial localization.

\medskip

With this example we can also illustrate an additional property.
Let us take for $\supp(u)$ a strip of thickness $t$ and length $M/t$, and recalculate
the value of $\Fone$:
\[
\Fone(u,v) = \left(\frac t2  + \frac 2t\right) M + 2t.
\]
For large $M$ this expression is dominated by the value of
the prefactor $t/2+ 2/t$, suggesting two additional features:
\begin{enumerate}
\item There is a preferred thickness, which is that value of $t$ for which
$t/2+2/t$ is minimal (i.e. $t=2$);
\item For the preferred thickness, the functional $\Fone$ equals $2M$
plus `other terms'.
\end{enumerate}
Although these statements only give suggestions, not proofs, for the case of general geometry, 
we shall see below that they both are true, and that these examples
do demonstrate the general behaviour. The `other terms' in the 
example above are the single term~$2t$, a term which is associated
with the ends of the strip; we now turn to a different case,
in which there are no loose ends, but instead the curvature of the structure
creates an energy penalty.

\subsection{Fixed thickness (\boldmath $\e=1$), part 2: comparing line with
ring structures}

In the previous section we argued that the energy $\Fone$ favours 
objects of thickness~$2$ and penalizes loose ends. We now consider
\emph{ring} structures, and we will see that also the \emph{curvature} carries
an energy penalty.

\medskip

\begin{figure}[ht]
\centerline{\psfig{width=\textwidth,figure=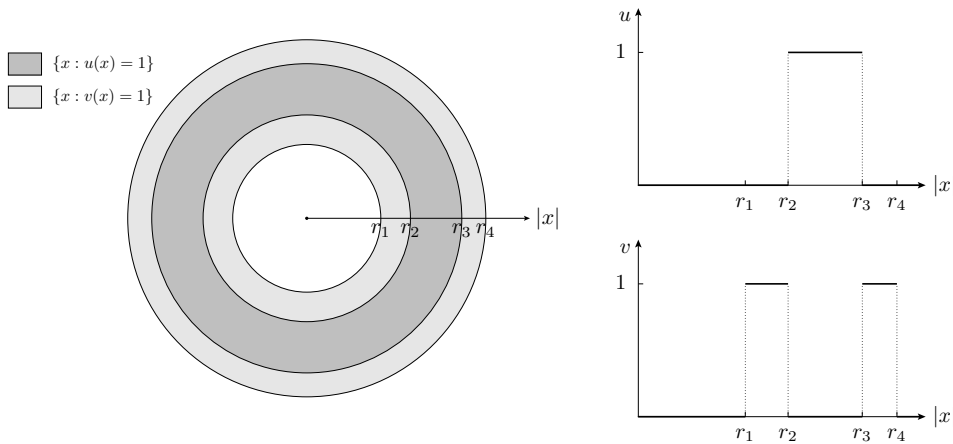}}
\caption{Ring structures; the pairs $(u,v)$ on the right-hand figure
indicate the values of $u$ and $v$ in that region in the plane.}
\label{fig:rings}
\end{figure}

Let the supports of $u$ and $v$ be the ring structures of
Fig.~\ref{fig:rings}: the support of $u$ is a single ring between 
circles of radii $r_2$ and $r_3$, and the support of $v$ consists
of two rings flanking $\supp(u)$.

We first do some direct optimization to reduce the number of 
degrees of freedom in this geometry.
For fixed outer radii $r_1$ and $r_4$ one may optimize $\Fone$ 
over variations of $r_2$ and $r_3$ that respect the mass
constraint $\int u = \int v$, and this optimization results
in the values of $r_2$ and $r_3$ given by
\begin{equation}
\label{def:r2r3}
r_2^2 = \tfrac12(R^2 + r_1^2), \qquad \text{and}\qquad
r_3^2 = \tfrac12(R^2 + r_4^2)
\end{equation}
in terms of the mean radius $R = \tfrac12(r_1+r_4)$. The functional $\Fone$
can then be computed explicitly in terms of $r_1$ and $r_4$ (see
Appendix~\ref{app:ring_solutions}). The interesting quantity is actually
the energy per unit mass, 
$\Fone/M$, as a function of the mean radius $R$ and the thickness $t$ of the
structure,
\begin{gather*}
  M\,:=\, \int u, \qquad t\,:=\, \frac{r_4-r_1}{2}.
\end{gather*}
Expanding $\Fone/M$ around $R=\infty$ and $t=2$, we find 
\begin{align}
\label{asymp:F1}
  \frac{\Fone}{M}(R,t)\,=\,& 2 +\frac{1}{4}(t-2)^2
  +\frac{1}{4}R^{-2} + O(|t-2|^3 + R^{-3}),
\end{align}
see Appendix \ref{app:ring_solutions}.

Again we recognize a preference for structures of thickness $t=2$;
we now also observe a penalization of the curvature in the term $R^{-2}/4$.
The main result of this paper indeed is to identify this 
curvature penalization for structures of arbitrary geometry, in the form of an elastica limit energy.

\subsection{Rescaling and renormalization}

In the previous sections we have seen that the functional $\F_\eps$
at $\e=1$ has a preference for structures of thickness $2$. It is easy
to see, by repeating the arguments above, that this becomes a 
preference for thickness $2\e$ in the general case. For instance,
the development~\pref{asymp:F1} generalizes to 
\[
\frac{\F_\eps}{M}(R,t) = 2 + \frac1{4}(t/\e-2)^2 + \frac14 \e^2R^{-2} 
  + O\bigl(|t/\e-2|^3+ \e^3R^{-3}\bigr).
\]

This expression provides us with a recipe for identifying the bending
energy in the limit $\e\to0$. The curvature, which is equal to $1/R$ for
this 
spherical geometry, enters the expression above as the term $\e^2R^{-2}/4$.
This suggests that the alternative functional
\[
\G_\e(u,v) := \frac1{\e^2}\bigl[\F_\eps(u,v) - 2M\big], 
\qquad \text{where}\qquad M  = \int u = \int v,
\]
may have a (Gamma-)limit similar to the Elastica functional.
This suggestion is proved in Theorem~\ref{the:main}.
\section{\Wasserstein distance, systems of curves and a generalized curve
bending energy}
\label{sec:wass}
In this section we introduce some basic definitions and concepts.
\begin{definition}
\label{def:wasserstein}
Consider $u,v\in L^1(\Rn)$ with compact support in $\Rn$ satisfying the mass balance
\begin{gather}\label{eq:fun-eq-mass}
  \int_{\Rn} u\,d\LL^2 \,=\, \int_{\Rn} v\,d\LL^2\,=\, 1.
\end{gather}
The \Wasserstein distance $d_1(u,v)$ is defined as
\begin{eqnarray}\label{eq:def-d-1}
  d_1(u,v) &:=& \min \Big( \int_{\Rn\times\Rn} |x-y| \,d\gamma(x,y)\Big)
\end{eqnarray}
where the minimum is taken over all Radon measures $\gamma$ on $\Rn\times\Rn$
with \emph{marginals}
$u\LL^2$ and $v\LL^2$, that means $\gamma$ satisfies
\begin{eqnarray}
  \int_{\Rn\times\Rn} \varphi(x) \,d\gamma(x,y) &=& \int_{\Rn}\varphi u\,d\LL^2,\label{eq:marg-u}\\
  \int_{\Rn\times\Rn} \psi(y) \,d\gamma(x,y) &=& \int_{\Rn}\psi v\,d\LL^2\label{eq:marg-v}
\end{eqnarray}
for all $\varphi,\psi\in C^0_c(\Rn)$.
\end{definition}
The \Wasserstein distance
is characterized by the optimal mass transport problem described in
\mbox{section \ref{sec:mtp}}. 
\begin{remark}\label{rem:wass}
For $\mu,\nu\in \mathcal{P}_1$, where
\begin{gather*}
  \mathcal{P}_1(\Rn)\,:=\, \big\{ \mu\text{ Radon measure on }\Rn \,:\, \int_{\Rn}d\mu\,=\,1,
  \,\,\int_{\Rn} |x|\,d\mu(x)\,<\,\infty\big\},
\end{gather*}
we can define the \Wasserstein distance $d_1(\mu,\nu)$ analogously to \eqref{eq:def-d-1},
substituting $u\LL^2$ and $v\LL^2$
in \eqref{eq:marg-u}, \eqref{eq:marg-v} by $\mu$ and $\nu$.
The minimum in \eqref{eq:def-d-1} is always attained and $d_1$ is a distance function on
$\mathcal{P}_1$~\cite{Amb04}. Moreover $d_1$ is continuous in the sense that for
$(\mu_k)_{k\in\N}\subset \mathcal{P}_1(\Rn), \mu\in \mathcal{P}_1(\Rn)$
\begin{gather*}
  d_1(\mu_k,\mu) \,\to\, 0\qquad \Longleftrightarrow\qquad
  \begin{cases}
    \mu_k\to\mu\text{ as Radon measures and}\\
    \int_{\Rn} |x|\,d\mu_k\,\to\,\int_{\Rn} |x|\,d\mu,
  \end{cases}
\end{gather*}
see \cite[Theorem 3.2]{Amb04}.

A family of related distances is given by
the $p$-Wasserstein distances
\begin{eqnarray}\label{eq:def-d-p}
  d_p(\mu,\nu) &:=& \min \Big( \int_{\Rn\times\Rn} |x-y|^p \,d\gamma(x,y)\Big)^{\frac{1}{p}},
\end{eqnarray}
where the minimum is taken over all Radon measures $\gamma$ on $\Rn\times\Rn$
with \emph{marginals} $\mu$ and $\nu$. The \Wasserstein distance 
coincides with the 1-Wasserstein distance.
\end{remark}
We will describe the limit structures in terms of \emph{systems of closed $W^{2,2}$-curves}.
The following definitions are mainly taken from \cite{BeM}.
\begin{definition}\label{def:sys-curves}
We call $\gamma:\R\to\R^2$  a closed $W^{2,2}$-curve if
\begin{align*}
	&\gamma\in W^{2,2}_{\loc}(\R,\R^2),\\
	&\gamma^\prime(r)\neq 0\text{ for all }r\in \R,\\
	&\gamma\text{ is $L$-periodic for some }0<L<\infty.
\end{align*}
If $\gamma$ is $L$-periodic and injective on $[0,L)$ the
curve 
$\gamma$ is called \emph{simple}. The length $L(\gamma)$ of a closed curve
$\gamma$ is defined by
\[
L(\gamma) := \int_0^L |\gamma'(r)|\, dr,
\]
where $L$ is the the minimal   period of $\gamma$.

A $W^{2,2}$-\emph{system of closed curves} is a finite collection of
closed $W^{2,2}$-curves. We represent a system of curves by a
\emph{multiset} $\{\gamma_i\}_{i=1,...,N}$, $N\in\N$, \emph{i.e.}
a set in which repeated elements are counted with multiplicity.

A system of curves $\Gamma=\{\gamma_i\}_{i=1,...,N}$ is called \emph{disjoint}
if each $\gamma_i$ is simple and $\gamma_i(\R)\cap\gamma_j(\R)=\emptyset$ for $i\not=j$.
$\Gamma$
has \emph{no transversal crossings} if for any $1\leq i,j\leq m$, $s_i,s_j\in\R$,
\begin{gather*}
  \gamma_i(s_i)\,=\,\gamma_j(s_j)\text{ implies that }\gamma_i^\prime(s_i)\text{ and }
  \gamma_j^\prime(s_j)\text{ are parallel}.
\end{gather*}
We define the \emph{support}, \emph{multiplicity} and \emph{total mass}
of a system of curves $\Gamma=\{\gamma_i\}_{i=1,...,N}$ by
\begin{align*}
  \spt(\Gamma) &:=\, \bigcup_{i=1}^N \gamma_i(\R),\\
    \theta(\Gamma,x) &:=\, \#\big\{(i,s) : \gamma_i(s) =x,\ 1\leq i\leq N,\ 
    0\leq s< L(\gamma_i)\big\},\\
    |\Gamma| &:=\, \sum_{i=1}^N L(\gamma_i)
\end{align*}
and define a corresponding
Radon measure $\mu_\Gamma$ on $\Rn$ to be the measure that
satisfies  
\begin{eqnarray}\label{eq:def-gamma-mu}
  \int_{\Rn} \varphi\,d\mu_\Gamma &=& \sum_{i=1}^N
  \int_0^{L(\gamma_i)}\varphi(\gamma_i(s))\,|\gamma_i^\prime(s)|\,ds 
\end{eqnarray}
for all $\varphi\in C^0_c(\Rn)$.

Two systems of curves are identified if the corresponding Radon measures
coincide. 
\end{definition}
\begin{remark}
We can represent a given system of closed curves $\Gamma$ by a multiset
$\Gamma\,=\,\{\gamma_i\}_{i=1,...,N},\,\text{ where for all }i=1,...,N$
\begin{gather}
  \gamma_i\text{ is one-periodic, with $1$ being the smallest
  possible period},\label{eq:stand-sys}\\
  \gamma_i\text{ is parametrized proportional to arclength.}
  \label{eq:curve-arclength}
\end{gather}
\end{remark}
We generalize the classical curve bending energy 
defined in \eqref{eq:def-elastica} to $W^{2,2}$-systems of closed curves.
\begin{definition}\label{def:W-Gamma}
Let $\Gamma$ be a $W^{2,2}$-system of closed
curves represented by $\Gamma=\{\gamma_i\}_{i=1,...,N}$ as in
\eqref{eq:stand-sys}-\eqref{eq:curve-arclength} and set $L_i = L(\gamma_i)$. 
Then we define
\begin{eqnarray}\label{eq:def-W-Gamma}
  \W(\Gamma) &:=& \frac{1}{2}\sum_{i=1}^N L_i^{-3}\int_0^1
  \gamma_i^{\prime\prime}(s)^2 \,ds. 
\end{eqnarray}
\end{definition}
\begin{remark}
The support and total mass of a system of curves coincide with
the support and  total mass of the corresponding Radon measure~\cite{Sim}.
The multiplicity functions of $\Gamma$ and of $\mu_\Gamma$ coincide
$\Ha^{1}$-almost everywhere on the support of $\Gamma$.\\ 
By Remark 3.8, Proposition 4.5, and Corollary 4.8 of \cite{BeM05}, 
we obtain that
$\Gamma$ is a $W^{2,2}$-system of closed curves without transversal crossings if and only if
$\mu_\Gamma$  is a \emph{Hutchinson varifold} with weak mean curvature
$\vec{H}\in L^2(\mu_\Gamma)$ 
such that in every point of $\supp(\Gamma)$ a unique tangent line exists. Moreover
\begin{eqnarray}
  \W(\Gamma) &=& \frac{1}{2}\int |\vec{H}|^2\,d\mu_\Gamma\label{eq:W-hutch}
\end{eqnarray}
holds.
\end{remark}
\section{Main results}\label{sec:main}
To investigate the limit behaviour of the functionals 
$ \F_\eps$ as $\eps\to 0$ 
we fix $M>0$ and study the rescaled functionals
\begin{eqnarray}\label{eq:def-Ge-uv}
  \G_\eps(u,v) &:=& \frac{1}{\eps^2} \Big(  \F_\eps(u,v) - 2M\Big).
\end{eqnarray}

\begin{theorem}
\label{the:main}
The curve bending energy $\W$ as defined in \eqref{eq:def-W-Gamma} is the Gamma-limit
of the functionals $\G_\eps$ in the following sense.
\begin{liste2}
\item
Let $(u_\eps, v_\eps)_{\eps>0}\subset
L^1(\Rn)\times L^1(\Rn)$, $R>0$ and a Radon measure $\mu$ on $\Rn$ be given with 
\begin{eqnarray}
  \spt(u_\eps)&\subset& B_R(0)\quad\text{ for all }\eps>0,\label{eq:spt-ue}\\
  u_\eps\LL^2 &\to& \mu\quad\text{ as Radon measures on
  }\R^2,\label{eq:conv-ue}
\end{eqnarray}
and
\begin{eqnarray}
  \liminf_{\eps\to 0} \G_\eps(u_\eps,v_\eps)&<&\infty.\label{eq:liminf-bounded}
\end{eqnarray}
Then there is a $W^{2,2}$-system of closed curves $\Gamma$
such that
\begin{gather}
  2\mu_\Gamma = \mu,\label{eq:rep-Gamma}\\
  \spt(\Gamma)\text{ is bounded},\label{eq:Gamma-0}\\
  \Gamma\text{ has no transversal crossings},\label{eq:Gamma-1}\\
  2|\Gamma|\,=\, M\label{eq:Gamma-3},
\end{gather}
and such that the \emph{liminf-estimate}
\begin{eqnarray}\label{eq:liminf-esti}
  \W(\Gamma) &\leq& \liminf_{\eps\to 0} \G_\eps(u_\eps)
\end{eqnarray}
holds.
\item
Let $\Gamma$ be a $W^{2,2}$-system of closed curves such that
\eqref{eq:Gamma-0}-\eqref{eq:Gamma-3} holds. Then there exists a
sequence $(u_\eps,v_\eps)_{\eps>0}\subset \K_\eps$ 
such that \eqref{eq:spt-ue} holds for some $R>0$,
\begin{eqnarray}\label{eq:conv-ue-Gamma}
  u_\eps\LL^2 &\to& 2\mu_\Gamma\quad\text{ as Radon measures on }\R^2,
\end{eqnarray}
 and such that the \emph{limsup-estimate}
\begin{eqnarray}\label{eq:limsup-esti}
 \W(\Gamma) &\geq& \limsup_{\eps\to 0} \G_\eps(u_\eps)
\end{eqnarray}
is satisfied.
\end{liste2}
\end{theorem}
\begin{remark}
\label{rem:convergence}
Let $X$ be the space of nonnegative Radon measures on $\R^2$. Define a concept of convergence on $X\times X$ by which 
\begin{gather*}
  (\mu_i,\nu_i) \,\to\, (\mu,\nu)\quad\text{ in }X\times X\quad\text{ iff }\quad
\begin{cases}
  \quad\mu_i\,\to\,\mu\quad\text{ as Radon measures on }\R^2\text{ and }\\
  \quad\bigcup_{i\geq i_0}\spt(\mu_i) \text{ is bounded for some }i_0\in\N.
\end{cases}
\end{gather*}
Note that no condition is placed on $\nu_i$.
We now define
\begin{eqnarray*}
  \G_\eps(\mu,\nu) &:=&
  \begin{cases}
    \G_\eps(u,v) \quad &\text{ if }\mu\,=\,u\LL^2,\,\nu\,=\,v\LL^2
       \text{ for some }(u,v)\in \K_\eps,\\
    \infty &\text{ otherwise}
  \end{cases}
\end{eqnarray*}
and
\begin{eqnarray*}
  \G_0(\mu,\nu) &:=& 
  \begin{cases}
    \W(\Gamma) &\begin{cases}
        \text{if $\mu=\nu$ is given by a $W^{2,2}$-system of closed curves}\\
        \text{$\Gamma=\{\gamma_i\}_{i=1,...,N}$ satisfying \eqref{eq:rep-Gamma}-\eqref{eq:Gamma-3}}
      \end{cases} \\[5\jot]
    \infty & \text{otherwise.}
  \end{cases}
\end{eqnarray*}
Then Theorem \ref{the:main} can be rephrased as
\begin{eqnarray}\label{eq:gamma-conv}
  \lim_{\eps\to 0} \G_\eps &=& \G_0
\end{eqnarray}
in the sense of Gamma-convergence with respect to the topology of
$X\times X$ defined above. 
\end{remark}

\begin{remark}
Implicit in the formulation of Theorem~\ref{the:main} is the following
(contrapositive) statement: if a sequence $(u_\e,v_\e)_{\e>0}\subset
\K_\e$ converges to a measure that \emph{cannot} be represented
by a system of curves satisfying~\eqref{eq:rep-Gamma}--\eqref{eq:Gamma-3},
then $\G_\e$ is unbounded along this sequence. Put differently,  
each of
the following is penalized in $\F_\e$ at an order lower (\emph{i.e.} larger) than $\eps^2$: 
\begin{enumerate}
\item a fracture in the limit structure, occurring as a nonclosed curve;
\item a total mass $|\Gamma|$ that deviates from~$M/2$;
\item a non-even \multiplicity.
\end{enumerate}

In the Introduction we conjectured the asymptotic 
development~\eqref{eq:asym-Fe}. 
Theorem~\ref{the:main} justifies the zeroth and second order of this
development in the following way: if $F_\e-2M$ is of order $O(\e^2)$,  
then the three types of degeneracy above can not occur, and the dominant term
is given in the limit by $\e^2 \W$. 
\end{remark}

\begin{remark}
By the definition of the set $\K_\eps$ and the functionals $\F_\eps$ in
\eqref{def:Fe}, \eqref{def:Ke} it is clear that for a 
sequence $(u_\eps, v_\eps)_{\eps>0}$ along which $\G_\e$ is bounded the supports of $u_\eps$ and $v_\eps$ necessarily concentrate as $\eps \to
0$. The crucial conclusions are (a) that the concentration is on curves
rather than on points or other sets and (b) that the limit measure can
be constructed as a sum of curves, each with density two. The latter
property 
implies a uniform thickness of structures with bounded energy and
paraphrases the stretching resistance of lipid bilayers.
\end{remark} 

\begin{remark}
We also obtain a compactness result for sequences with bounded $\G_\e$: Let
$(u_\eps, v_\eps)_{\eps>0}\subset L^1(\R^2)\times
L^1(\R^2)$ satisfy \eqref{eq:spt-ue}, \eqref{eq:liminf-bounded}.
Then there exists a Radon measure $\mu$ on $\R^2$ such that
\eqref{eq:conv-ue} holds for a subsequence $\eps\to 0$. Theorem
\ref{the:main} then implies that there 
is a $W^{2,2}$-system $\Gamma$ of closed 
curves such 
that the conclusions \eqref{eq:rep-Gamma}-\eqref{eq:liminf-esti} hold.
\end{remark}
\section{The Mass Transport Problem}
\label{sec:mtp}
In this section we recall the classical Monge problem of optimal mass transport and review
some results that we will use later.
\begin{definition}\label{def:trans-maps}
Fix two non-negative Borel functions $u,v\in L^1(\R^2)$ with compact support
satisfying the
mass balance \eqref{eq:fun-eq-mass}.
By $\mathcal{A}(u,v)$ we denote the set of all Borel maps $S:\R^2\to\Rn$
\emph{pushing $u$ forward to $v$}, that is
\begin{eqnarray}\label{eq:trans-map}
  \int\eta(S(x)) u(x)\,dx &=& \int_{\Rn} \eta(y)v(y)\,dy
\end{eqnarray}
holds for all $\eta\in C^0(\Rn)$.
\end{definition}
\begin{prob-mass-transport}
For $u,v$ as in Definition \ref{def:trans-maps} minimize the \emph{transport cost}
$I:\mathcal{A}(u,v)\to \R$,
\begin{eqnarray}\label{eq:transportcost}
  I(S) &:=& \int_\Omega |S(x)-x| u(x)\,dx.
\end{eqnarray}
\end{prob-mass-transport}
The following dual formulation is due to Kantorovich \cite{Kan}.
\begin{prob-Kantorovich}
For $u,v$ as in Definition \ref{def:trans-maps} maximize $K: \Lip_1(\Rn)\to\R$,
\begin{eqnarray}\label{eq:phi_eps}
  K(\phi) &:=& \int_\Omega \phi(x)(u-v)(x)dx,
\end{eqnarray}
where $\Lip_1(\Rn)$ denotes the space of Lipschitz functions on $\R^2$
with Lipschitz constant not larger than $1$.
\end{prob-Kantorovich}
There is a vast literature on the optimal mass transportation problem
and an impressive number of applications, see for example \cite{EGan,TW,CFC,Amb,Vil,JKF,Ott}.
We only list a few results which we will use later.
\begin{theorem}[\cite{CFC},\cite{FeC}]
Let $u,v$ be given as in Definition \ref{def:trans-maps}.
\begin{liste2}
\item There exists an optimal transport map $S\in \mathcal{A}(u,v)$.
\item There exists an optimal Kantorovich potential $\phi\in \Lip_1(\Rn)$.
\item The identities
\begin{gather*}
  d_1(u,v)\,=\, I(S)\,=\,K(\phi)
\end{gather*}
hold.
\item Every optimal transport map $S$ and every optimal Kantorovich potential $\phi$
satisfy
\begin{eqnarray}\label{eq:dual-crit}
  \phi(x) -\phi(S(x)) &=& |x-S(x)|\quad\text{ for almost all }x\in\spt(u).
\end{eqnarray}
\end{liste2}
\end{theorem}
The optimal transport map and the optimal Kantorovich potential are in general not
unique. We can choose $S$ and $\phi$ enjoying some additional properties.
\begin{proposition}[\cite{CFC,FeC}]\label{prop:add-prop}
There exists an optimal transport map $S\in \mathcal{A}(u,v)$ and an optimal
Kantorovich potential $\phi$ such that
\begin{eqnarray}
  \phi(x) &=& \min_{y\in\spt(v)}\big(\phi(y)+|x-y|\big)\quad\text{ for
  any }x\in\spt(u),\label{eq:add-prop-phi1}\\ 
  \phi(y) &=& \max_{x\in\spt(u)}\big(\phi(x)-|x-y|\big)\quad\text{ for
  any }y\in\spt(v),\label{eq:add-prop-phi2}
\end{eqnarray}
and such that $S$ is the unique monotone transport map in the sense of
\cite{FeC},
\begin{gather*}
  \frac{x_1-x_2}{|x_1-x_2|}
  +\frac{S(x_1)-S(x_2)}{|S(x_1)-S(x_2)|}\,\neq\, 0\quad\text{ for all
  }x_1\neq x_2\in\Rn\text{ with }S(x_1)\neq S(x_2).
\end{gather*}

\end{proposition}
We will extensively use the fact that, by~\eqref{eq:dual-crit} the optimal transport is organized along
{\itshape transport rays} which are defined as follows.
\begin{definition}[\cite{CFC}]
\label{def:rays}
Let $u,v$ be as in Definition \ref{def:trans-maps} and let $\phi\in \Lip_1(\Rn)$
be the optimal transport map as in Proposition \ref{prop:add-prop}.
A {\itshape transport ray} is a line segment in $\Rn$ with endpoints $a,b\in\Rn$ such that
$\phi$ has slope one on that segment and $a,b$ are maximal, that is
\begin{gather*}
  a\in\spt(u),\,b\in\spt(v),\quad a\neq b,\\
  \phi(a)-\phi(b)\,=\,|a-b|\\
  |\phi(a+t(a-b))-\phi(b)| \,<\, |a+t(a-b)-b|\quad\text{ for all }t>0,\\
  |\phi(b+t(b-a))-\phi(a)| \,<\, |b+t(b-a)-a|\quad\text{ for all }t>0.
\end{gather*}
We define the {\itshape transport set }$\mathcal{T}$ to consist of all points which
lie in the (relative) interior of some transport ray and define $\mathcal{E}$ to be the
set of all endpoints of rays. 
\end{definition}

Some important properties of
transport rays are given in the next proposition.
\begin{proposition}[\cite{CFC}]\label{prop:cfm-rays}
\begin{liste2}
\item Two rays can only intersect in a common endpoint.
\item
The endpoints $\mathcal{E}$ form a Borel set of measure zero.
\item If $z$ lies in the interior of a ray with endpoints $a\in\spt(u),b\in\spt(v)$
then $\phi$ is differentiable in $z$ with $\nabla\phi(z)\,=\,(a-b)/|a-b|$.
\end{liste2}
\end{proposition}
In Section \ref{subsec:para} we will use the transport rays to parametrize the support of $u$ and
to compute the \Wasserstein distance between $u$ and $v$.

\section{Overview of the proof of Theorem~\ref{the:main}}
\label{sec:overview}

\subsection{The lower bound}
The heart of the proof of the lower bound~\pref{eq:liminf-esti} is a 
parametrization of $\supp(u)\cup\supp(v)$ that allows us to rewrite the
functionals  $\F_\e$ and $\G_\e$ in terms of the geometry of $\supp(u)$ and
$\supp(v)$ and the structure of the transport rays.

The first step in the proof of~\pref{eq:liminf-esti} is to pass to
functions $u$ and $v$ with smooth boundaries (Proposition~\pref{prop:red})
so that $\partial \supp(u)$ can be represented by a collection $\{\gamma_i\}$
of smooth curves. For the discussion in this section we will pretend
that there is only one curve $\gamma:[0,L)\to\R^2$, which is closed and 
which we parametrize by arclength.

The properties of the \Wasserstein distance $d_1$ give that almost
every point $x\in \supp(u)\cup\supp(v)$ lies in the interior of exactly
one transport 
ray; the set $\supp(u)\cup \supp(v)$ can therefore be written, up to a
null set, as the disjoint union of transport rays.  

In addition, along each ray mass  is transported `from $u$ to $v$', implying 
that each ray intersects $\partial\supp(u)$. We therefore can parametrize
the collection of rays by their intersections with $\partial \supp(u)$.

The ray direction of the ray $\mathcal  R(s)$ that passes through $\gamma(s)$
defines a unit length vector $\theta(s)$;
we now introduce a parametrization $\psi$ by
\[
\psi(s,t) := \gamma(s) + t\theta(s).
\]
In terms of such a parametrization the first term of $\F_\e$ has a simple
description as the length of the curve $\gamma$,
\[
\e\int|\nabla u| = \int_0^L 1\,ds = L.
\]
To rewrite $d_1(u,v)$ we
use the fact that transport takes place along transport rays; when restricted to a single ray the transport problem becomes one-dimensional, and even explicitly solvable. 

\begin{figure}[ht]
\centering
\includegraphics[height=5.5cm]{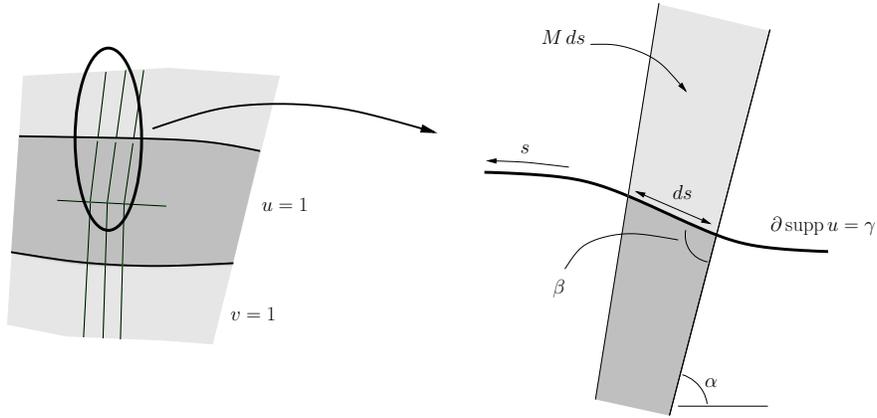}
\caption{The parametrization of $\supp(u)\cup\supp(v)$. On the left six transport rays are drawn over the two supports; on the right the functions $\alpha$, $\beta$, and $M$ are indicated. At a boundary point $\gamma(s)$, $\alpha(s)$ is the angle of the transport ray through that point, and $\beta(s)$ is the angle between $\gamma$ and the ray. Finally, $M(s)ds$ is the amount of mass contained between the rays at positions $s$ and $s+ds$.}
\label{fig:explanation}
\end{figure}

The central estimate in this part of the proof is the following lower bound:
\begin{align}
\label{est:central}
d_1(u,v) 
    &\geq \int_0^L \Biggl[\frac{\eps}{\sin\beta(s)}M(s)^2+
   \frac{\eps^3}{4\sin^5\beta(s)}\alphaprime(s)^2M(s)^4\Biggr]ds.
\end{align}
In this inequality the geometry of $\supp(u)$ is characterized by
functions $\alpha$, $\beta$, and~$M$, which are illustrated in Figure~\ref{fig:explanation}.
The functions $\alpha$ and $\beta$ are angles: $\alpha(s)$ is the angle
between the ray direction $\theta(s)$ and a reference direction and $\beta(s)$ is the angle between $\theta(s)$ and the tangent $\gamma'(s)$ to $\gamma$. The function $M(\cdot)$, finally, measures the amount of mass supported on a ray. The relation between the function $M(\cdot)$ and the scalar $M$ in~\pref{def:Ke} is
\[
M = \int_0^L M(s)\, ds.
\]

With the estimate~\pref{est:central}, the main result can readily be appreciated. Let a sequence $(u_\e,v_\e)$ be such that $\G_\e(u_\e,v_\e)$
is bounded as $\e\to0$, and let $\alpha_\e$, $\beta_\e$, and $M_\e$ be the
associated geometric quantities. With the inequality above,
\begin{align}
\G_\e(u_\e,v_\e) &= \frac{\F_\e(u_\e,v_\e)-2M_\e}{\e^2} \notag \\
&\geq \int_0^L \Bigl[\frac1{\e^2} - \frac{2M_\e(s)}{\e^2}+\frac{M_\e(s)^2}{\e^2\sin\beta_\e(s)}+
   \frac{\alphaprime_\e(s)^2M_\e(s)^4}{4\sin^5\beta_\e(s)}\Bigr]ds\notag\\
&= \int_0^L \Biggl[\frac1{\e^2}\bigl(1-M_\e(s)\bigr)^2\
   +\frac1{\e^2}\left(\frac1{\sin\beta_\e(s)}-1\right) M_\e(s)^2+
   \frac{\alphaprime_\e(s)^2M_\e(s)^4}{4\sin^5\beta_\e(s)}\Biggr]ds
   \label{est:informal}
\end{align}
Note that each of the three terms above is non-negative.
If $\G_\e(u,v)$ is bounded, then one concludes that as $\e\to0$ the mass $M_\eps(s)$ tends to one for almost all $s$. Similarly, $\beta_\e$ converges to $\pi/2$, implying that 
the ray angle $\alpha_\eps$ converges to the angle
 of the normal to $\gamma_\eps$, and in a weak sense therefore $\alpha^\prime_\eps$ converges to the curvature $\kappa_\eps=\gamma_\eps^{\prime\prime}$.
The inequality above suggests 
a $W^{2,2}_{\mathrm{loc}}$ bound for curves $\gamma_\eps$ with the last term approximating 
\begin{gather*}
		\frac{1}{4}\int_0^L \kappa_\eps(s)^2\,ds
\end{gather*}
This integral is the bending energy that we expect in the limit.

Based on the inequality~\pref{est:informal} we show that the boundary curves of $\spt(u_\eps)$ are compact, that the limit is given
by a $W^{2,2}$ system of closed curves as in \eqref{eq:rep-Gamma}-\eqref{eq:Gamma-3}
 and that the lim-inf estimate is satisfied. We finally conclude
that the limit of the mass distributions $u_\eps$ is identical to the limit of the boundary curves.

The main reasons why the proof in Section~\ref{sec:liminf} is more involved than this brief explanation are related to the gaps in the reasoning above:
\begin{itemize}
\item A ray can intersect $\partial\supp(u)$ multiple times, and care must be taken to ensure that the parametrization is a bijection.
\item The ray direction $\theta$  is not necessarily smooth; to be precise, when the ray length tends to zero (which is equivalent to $M(\cdot)$ vanishing), $\theta$ may vary wildly. 
\item Similarly, when $M(\cdot)$ vanishes, the $L^2$-estimate on $\alpha'$ in~\pref{est:informal} degenerates,
resulting in a compactness problem  for the boundary curves.
\end{itemize}

\subsection{The upper bound}
With the machinery of the lower bound in place, and with the insight that is provided, the upper bound becomes a relatively simple construction. Around a given limit curve tubular neighbourhoods are constructed for the supports of $u$ and $v$. The thickness of these neigbourhoods can be chosen just right, and the calculation that leads to~\pref{est:central} now gives an exact value for $\G_\e$.

The main remaining difficulty of the proof is the fact that a $W^{2,2}$-curve need not have a non-self-intersecting tubular neighbourhood of any thickness, and therefore an approximation argument is needed.

\section{Proof of the lim-inf estimate}
\label{sec:liminf}
In this section we prove the first part of Theorem~\ref{the:main}. 
The main idea is to use the optimal Kantorovich potential
of the mass transport problem of $u_\eps$ to $v_\eps$ to construct a
parametrization of the support of $u_\eps$ 
and $v_\eps$ and to derive a lower bound for the functionals
$\G_\eps(u_\eps,v_\eps)$. This lower bound yields the compactness of the
systems of curves 
which describe the boundary of the support of $u_\eps$ as well as the
desired lim-inf estimate.

We first show that we can restrict ourselves to a class of `generic
data'. 
\begin{proposition}\label{prop:red}
It is sufficient to prove the \emph{liminf} part of Theorem
\ref{the:main} under the additional assumptions that
\begin{align}
  &M=1,\quad\text{ that is } \int u_\e \,=\, \int v_\e \,=\,1\quad\text{ for 
  }(u_\e,v_\e)\in\K_\eps, \label{eq:M=1}\\[2mm]
  &\sup_{\eps>0}\G_\eps(u_\eps,v_\eps)\,\leq
  \,\Lambda\quad\text{ for some }\quad 0<\Lambda<\infty,\label{eq:bound-G-eps}\\[2mm]  
  &(u_\eps, v_\eps)\,\in\, \K_\eps\quad\text{ for all
  }\eps>0,\label{eq:ass-K}\\[2mm]
  &\lim_{\eps\to 0} \G_\eps(u_\eps,v_\eps)\,=\,\liminf_{\eps\to 0}
  \G_\eps(u_\eps,v_\eps)\quad\text{ exists,} \label{eq:ex-lim-G}\\[2mm]
  &\partial\spt(u_\eps)\text{ is given by a disjoint system of
  smooth, simple curves } \Gamma_{\eps}. \label{eq:reg-u-eps}
\end{align}
\end{proposition}
\begin{proof}
To prove that we can restrict ourselves to $M=1$ we do a
spatial rescaling and set
\begin{gather*}
  \tilde{\eps}\,:=\,\frac{\eps}{M},\quad
  \tilde{u}_{\tilde{\eps}}(x)\,:=\, M u_\eps(Mx),\quad
  \tilde{v}_{\tilde{\eps}}(x)\,:=\, M v_\eps(Mx).
\end{gather*}
We then obtain
\begin{gather}
  \int_{\R^2} \tilde{u}_{\tilde{\eps}}\,=\,\int_{\R^2} \tilde{v}_{\tilde{\eps}}\,=\,1,\qquad\qquad 
  {\tilde{\eps}}\int |\nabla \tilde{u}_\eps|\,=\,
  \frac{\eps}{M}\int|\nabla\label{eq:tilde-u-bdry} 
  u_\eps|.
\end{gather}
Moreover, if $S$ is the optimal transport map from $u_\eps$ to
$v_\eps$, then $\tilde{S}(x):=
M^{-1}S(Mx)$ is the optimal transport map from $\tilde{u}_{\tilde{\eps}}$
to $\tilde{v}_{\tilde{\eps}}$ and
\begin{gather}
  d_1(\tilde{u}_{\tilde{\eps}},\tilde{v}_{\tilde{\eps}})\,=\, M^{-2}
  d_1(u_\eps,v_\eps). \label{eq:tilde-d1} 
\end{gather}
We obtain from \eqref{eq:tilde-u-bdry},
\eqref{eq:tilde-d1} that
\begin{align}
  {\G}_{\tilde{\eps}}(\tilde{u}_{\tilde{\eps}},\tilde{v}_{\tilde{\eps}})
  \,=\,&\frac{1}{\tilde{\eps}^2}  
  \Big(\frac{1}{\tilde{\eps}}d_1(\tilde{u}_{\tilde{\eps}},\tilde{v}_{\tilde{\eps}})
  +\tilde{\eps}\int|\nabla\tilde{u}_{\tilde{\eps}}| -2\Big)\notag\\
  \,=\,& \frac{M^2}{\eps^2}\Big( \frac{1}{M\eps}d_1(u_\eps,v_\eps) +
  \frac{\eps}{M}\int|\nabla u_\eps| - \frac{2}{M}\int u_\eps\Big)\notag\\
  =&\, M \G_\eps(u_\eps,v_\eps).\label{eq:tilde-Ge}
\end{align}
We observe that $(\tilde{u}_{\tilde{\eps}},\tilde{v}_{\tilde{\eps}})\in
\K_{\tilde{\eps}}$ iff 
$(u_\eps,v_\eps)\in \K_\eps$ and that
$(\tilde{u}_{\tilde{\e}})_{\tilde{\e}>0}$ satisfies 
\eqref{eq:spt-ue}, \eqref{eq:conv-ue} for \mbox{$\tilde{R}:= R/M$} and
$\tilde{\mu}$ defined by
\begin{gather}
  \int \eta(x)\,d\tilde{\mu}(x)\,=\, \int
  \frac{1}{M}\eta\big(\frac{x}{M}\big)\,d\mu(x)\quad\text{ for }\eta\in
  C^0_c(\R^2). \label{eq:tilde-mu}
\end{gather}
If Theorem \ref{the:main} holds for $M=1$ then there exists a
$W^{2,2}$-system of closed curves $\tilde{\Gamma}$ with
\begin{gather}
  2\mu_{\tilde{\Gamma}}\,=\,\tilde{\mu},\qquad
  2|\tilde{\Gamma}|\,=\,1,\label{eq:the-mu-tilde}
\end{gather}
such that \eqref{eq:Gamma-0}, \eqref{eq:Gamma-1} holds and
\begin{gather}
  \W(\tilde{\Gamma}) \,\leq\, \liminf_{\tilde{\eps}\to
    0}\G_{\tilde{\eps}}(\tilde{u}_{\tilde{\eps}},\tilde{v}_{\tilde{\eps}}). \label{eq:liminf-tilde}
\end{gather}
We deduce from \eqref{eq:tilde-mu} and \eqref{eq:the-mu-tilde}
that $\mu$ is
given as system of closed curves:
Let $\tilde{\Gamma}\,=\,\{\tilde{\gamma}_i\}_{i=1,...,N}$ with
\begin{gather*}
  \tilde{\gamma}_i : [0,\tilde{L}_i)\,\to\,\R^2, \qquad
    |\tilde{\gamma}_i^\prime|\,=\,1 
\end{gather*}
and define a
system of curves $\Gamma=\{\gamma_i\}_{i=1,...,N}$ by
\begin{gather*}
  \gamma_i : [0,L_i)\,\to\,\R^2,\quad L_i:= M\tilde{L}_i,\\
    \gamma_i(s):= M\tilde{\gamma}\big(\frac{s}{M}\big)\quad\text{ for }s\in
    [0,L_i). 
\end{gather*}
Then $|\gamma_i^\prime|=1$ and $\Gamma$ satisfies \eqref{eq:Gamma-0},
\eqref{eq:Gamma-1}.
We obtain from \eqref{eq:tilde-mu} that
\begin{align*}
  \int \eta \,d\mu\,&=\, M\int
  \eta(Mx)\,d\tilde{\mu}(x)\,=\,2M\sum_{i=1}^m\int_0^{\tilde{L}_i} 
  \eta(M\tilde{\gamma}_i(s))\,ds \\
  \,&=\, 2\sum_{i=1}^m\int_0^{L_i}\eta(\gamma_i(s))\,ds
  \,=\, 2\int \eta\,d\mu_{\Gamma},
\end{align*}
which yields $\mu=2\mu_\Gamma$ and, by taking $\eta=1$ inside $B_R(0)$,
that \eqref{eq:Gamma-3} holds.
Finally we deduce from \eqref{eq:tilde-mu}
\begin{align*}
  \W(\Gamma)\,&=\,\frac{1}{2}\sum_{i=1}^m \int_0^{L_i}
  |\gamma_i^{\prime\prime}(s)|^2\,ds \\
  \,&=\,
  \frac{1}{2}\sum_{i=1}^m \int_0^{L_i}
  \frac{1}{M^2}\Big|\tilde{\gamma}_i^{\prime\prime}\Big(\frac{s}{M}\Big)\Big|^2\,ds
  \,=\, 
  \frac{1}{2}\sum_{i=1}^m \int_0^{\tilde{L}_i}
  \frac{1}{M}|\tilde{\gamma}_i^{\prime\prime}(s)|^2\,ds \\
  \,&=\,
  \frac{1}{M}\W(\tilde{\Gamma})
\end{align*}
and by \eqref{eq:tilde-Ge}, \eqref{eq:liminf-tilde} this yields that
\eqref{eq:liminf-esti} 
holds. Therefore we have shown that it is sufficient to prove the
lim-inf part of Theorem \ref{the:main} for $M=1$.

Eventually restricting ourselves to a subsequence $\eps\to 0$ we can
assume \eqref{eq:ex-lim-G}. The existence of a $0<\Lambda<\infty$
such that \eqref{eq:bound-G-eps} holds follows from
\eqref{eq:liminf-bounded}. By the definition of $\G_\eps$ this in
particular implies \eqref{eq:ass-K}.

It remains to show that it is sufficient to consider functions $u_\eps$
such that $\partial\spt(u_\eps)$ is smooth. We already have seen that we
can assume that $(u_\eps,v_\eps)\in\K_\eps$. This implies that
$\spt(u_\eps)$ is a set of finite perimeter. Moreover
$\spt(u_\eps)\subset B_R(0)$ and by
\cite[Theorem~3.42]{AFP} there exists a sequence of
open sets $(E_k)_{k\in\N}$ with smooth boundary such that
\begin{eqnarray}
  \Chi_{E_k} &\to& \eps u_\eps\quad\text{ in }L^1(\Rn),\label{eq:E-k-1}\\
  |\nabla\Chi_{E_k}|(\Rn) &\to& \eps|\nabla u_\eps|(\Rn)\label{eq:E-k-2}\\
  E_k &\subset& B_{2R}(0).\label{eq:E-k-3}
\end{eqnarray}
We set
\begin{eqnarray}\label{eq:def-r-k}
  r_k &:=& \eps^{\frac{1}{2}}|E_k|^{-\frac{1}{2}}
\end{eqnarray}
and define functions $\hat{u}_k$,
\begin{eqnarray}
  \hat{u}_k(x) &:=& \frac{1}{\eps}\Chi_{E_k}(r_k^{-1}x).\label{eq:def-u-hat}
\end{eqnarray}
This yields
\begin{gather}
  \int_{\R^2} \hat{u}_k \,=\, \frac{r_k^2}{\eps}|E_k|\,=\,1\label{eq:mass-hat-u}
\end{gather}
and
\begin{gather}
  \int_{\R^2} |\nabla\hat{u}_k| \,=\, \frac{r_k}{\eps}\int_{\Rn}
  |\nabla E_k|.\label{eq:surf-hat-u} 
\end{gather}
Moreover, from \eqref{eq:E-k-1}, \eqref{eq:def-r-k} and
$|\spt(u_\eps)|=\eps$ we 
observe that
\begin{gather}
  r_k\,\to\, 1\quad\text{ as }k\to\infty.\label{eq:conv-k}
\end{gather} 
We compute that
\begin{eqnarray*}
  \int_{\R^2} |\hat{u}_k -u_\eps| &=&
  \frac{1}{\eps}\int_{\R^2} \Big|\Chi_{E_k}(r_k^{-1}x)-\eps
  u_\eps(x)\Big|\,dx\\ 
  &\leq& \frac{1}{\eps}\int_{\R^2} \big|\Chi_{E_k}(r_k^{-1}x)
  -\eps u_\eps(r_k^{-1}x)\big|\,dx\\
  &&+\frac{1}{\eps}\int_{\R^2} \big|\eps u_\eps(r_k^{-1}x)-\eps
  u_\eps(x)|\,dx\\ 
  &=& \frac{r_k^2}{\eps}\int_{\R^2} \big|\Chi_{E_k}-\eps
  u_\eps(x)\big|\,dx 
  +\int_{\R^2} \big|u_\eps(r_k^{-1}x)-u_\eps(x)|\,dx.
\end{eqnarray*}
Since the right-hand side converges to zero as $k\to\infty$ by
\eqref{eq:E-k-1}, \eqref{eq:conv-k} we deduce that
\begin{eqnarray}
  \hat{u}_k &\to& u_\eps\quad\text{ in }L^1({\R^2})\label{eq:conv-u-hat-L1}
\end{eqnarray}
as $k\to\infty$. Equations \eqref{eq:E-k-2}, \eqref{eq:surf-hat-u} and
\eqref{eq:conv-k} yield that 
\begin{eqnarray}
  \int_{\R^2} |\nabla\hat{u}_k| &\to& \int_{\R^2} |\nabla
  u_\eps|\quad\text{ as }k\to\infty.\label{eq:conv-u-hat-area} 
\end{eqnarray}
We construct approximations $\hat{v}_k$ of $v_\eps$ by setting
\begin{gather}
  \hat{v}_k(x) \,:=\, v_\eps(x)\big(1-\eps \hat{u}_k(x)\big) +w_k(x),
  \label{eq:def-hat-v} 
\end{gather}
for a suitable $w_k\in L^1\big(B_{2R}(0),\{0,\eps^{-1}\}\big)$ with
\begin{gather}
  \int_{\R^2} ( w_k -\eps v_\eps\hat{u}_k)\,=\, 0. \label{eq:def-wk}
\end{gather}
We then deduce from \eqref{eq:def-hat-v}, \eqref{eq:def-wk} that
$(\hat{u}_k,\hat{v}_k )\in\K_\eps$. It follows from
\eqref{eq:conv-u-hat-L1}, \eqref{eq:def-wk} and $u_\eps v_\eps=0$ that
$w_k\to 0$ as $k\to \infty$. Together with \eqref{eq:def-hat-v} we
obtain that 
\begin{gather}
  \lim_{k\to\infty}\hat{v}_k\,=\, v_\eps\quad\text{ in
  }L^1(\R^2). \label{eq:conv-hat-vk} 
\end{gather}

By \eqref{eq:E-k-3}, \eqref{eq:conv-u-hat-L1} and \eqref{eq:conv-hat-vk}
we deduce from Remark \ref{rem:wass} the continuity of the \Wasserstein
distance term in $\F_\eps(\hat{u}_k,\hat{v}_k)$. Together with
\eqref{eq:conv-u-hat-area} this yields
\begin{eqnarray*}
  \F_\eps(\hat{u}_k,\hat{v}_k) &\to& \F_\eps(u_\eps,v_\eps)\quad\text{ as }k\to\infty.
\end{eqnarray*}
Therefore we can choose $k(\eps)\in\N$ such that for
$\tilde{u}_\eps:=\hat{u}_{k(\eps)}$, $\tilde{v}_\eps:=\hat{v}_{k(\eps)}$
satisfy the estimate
\begin{eqnarray*}
  \|u_\eps -\tilde{u}_\eps\|_{L^1({\R^2})} +
  |\G_\eps(u_\eps,v_\eps)-\G_\eps(\tilde{u}_\eps,\tilde{v}_\eps)| &\leq&
  \eps. 
\end{eqnarray*}
This yields a sequence $(\tilde{u}_\eps,\tilde{v}_\eps)_{\eps>0}$ that satisfies
\eqref{eq:reg-u-eps} and has the same limit as
$(u_\eps,v_\eps)_{\eps>0}$ in \eqref{eq:conv-ue}, \eqref{eq:liminf-esti}.
\end{proof}

From now on, for the rest of this section, we assume that
\eqref{eq:M=1}-\eqref{eq:reg-u-eps} hold.
\subsection{Parametrization by rays}
\label{subsec:para}
Due to \eqref{eq:reg-u-eps} the boundary $\partial\spt(u_\eps)$ is smooth.
We are free to choose a suitable representation.
\begin{remark}\label{rem:arclength-curves}
We can represent $\partial\spt(u_\eps)$ by a disjoint system
$\Gamma_\eps$ of closed curves such that for all $\gamma\in \Gamma_\eps$
\begin{gather}
  \gamma \text{ is parametrized by arclength}, |\gamma^\prime|=1,\notag
  \\ 
  \spt(u_\eps)\text{ is `on the left hand side' of
  }\gamma. \label{eq:spt-ue-lhs} 
\end{gather}
In particular $\gamma$ is $L(\gamma)$-periodic and
\begin{gather*}
  \det (\gamma^\prime(s),\nu(s)) \,\leq\, 0,
\end{gather*}
where $\nu(s)$ denotes the outward unit normal of $\spt(u_\eps)$ at
$\gamma(s)$. 
\end{remark}
Let $\phi\in \Lip_1(\R^2)$ be an optimal Kantorovich potential for
the mass transport from $u_\eps$ to $v_\eps$ as in Proposition
\ref{prop:add-prop}, with 
$\mathcal{T}$ being the set of transport rays as in Definition
\ref{def:rays}. 
Recall that $\phi$ is differentiable, with $|\nabla \phi| = 1$, in the
relative interior of any ray. 

\begin{definition}[Parametrization by rays]\label{def:para}
For $\gamma\in \Gamma_\eps$ we define
\begin{liste}
\item
a set $E$ of `inner points' with respect to the transport set
$\mathcal{T}$,
\begin{eqnarray*}
  E &:=& \{s\in \R :\gamma(s) \in \mathcal{T}\},
\end{eqnarray*}
\item
a direction field
\begin{eqnarray*}
  \theta: E\to \sphere^1, \qquad \theta(s) := \nabla\phi(\gamma(s)),
\end{eqnarray*}
\item
the positive and negative total ray length $L^+,L^-:E\to \R$,
\begin{eqnarray}
  L^+(s) &:=& \sup\{t>0 : \phi(\gamma(s)+t\theta
  (s))-\phi(\gamma(s))\,=\,t\},\label{eq:def-L+}\\ 
  L^-(s) &:=& \inf\{t<0 : \phi(\gamma(s)+t\theta
  (s))-\phi(\gamma(s))\,=\,t\},\label{eq:def-L-} 
\end{eqnarray}
\item the \emph{effective} positive ray length $l^+: E\to\R$,
\begin{gather}
  l^+(s) := \sup\ \{t\in [0,L^+(s)] : \gamma(s) + \tau\theta(s)\in
  \Int\bigl(\spt(u_\eps)\bigr)\text{ for all }0<\tau<
  t\}.\label{eq:def-l+} 
\end{gather}
\end{liste}
Finally we define a map $\psi$ which will serve as a parametrization of
$\spt(u_\eps)\cup\spt(v_\eps)$ by 
\begin{gather}
  \psi: \big[0,L(\gamma)\big)\times\R\,\to\,\R^2,\quad
    \psi(s,t) \,:=\, \gamma(s) + t\theta(s).\label{eq:def-psi}
\end{gather}
\end{definition}
\begin{remark}\label{rem:corr-lengths}
All objects defined above are properties of $\gamma$ even if we
do not denote this dependence explicitely. When dealing with a
collection of curves $\{\gamma_i :i=1,...,N\}$ or $\{\gamma_{\eps,i} : \eps>0,
i=1,...,N(\eps)\}$ then $E_i$, $l_{\eps,i}^+$ etc. refer to the
objects defined for the corresponding curves.
\end{remark}

\begin{figure}[ht]
\centerline{\psfig{width=\textwidth,figure=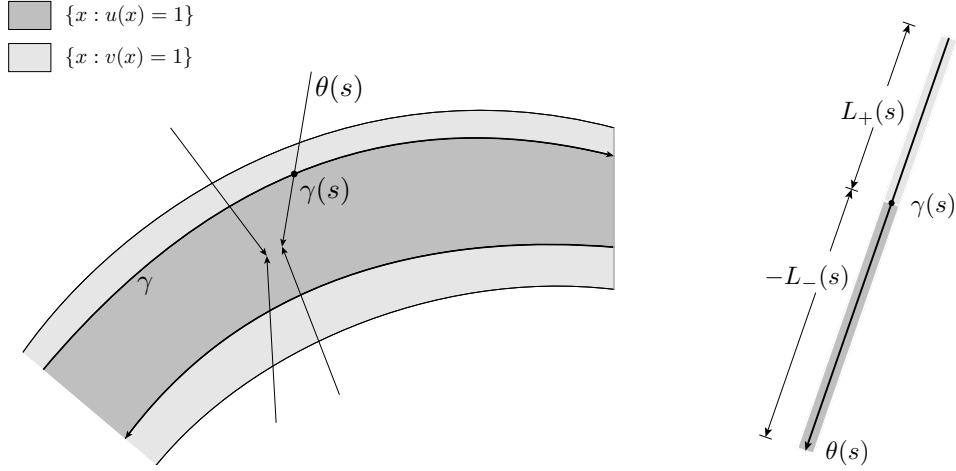}}
\caption{The parametrization in a `simple' situation. Here
  $l^+(s)=L^+(s)$ coincide.}
\label{fig:para}
\end{figure}

\begin{figure}[ht]
\centerline{\psfig{width=\textwidth,figure=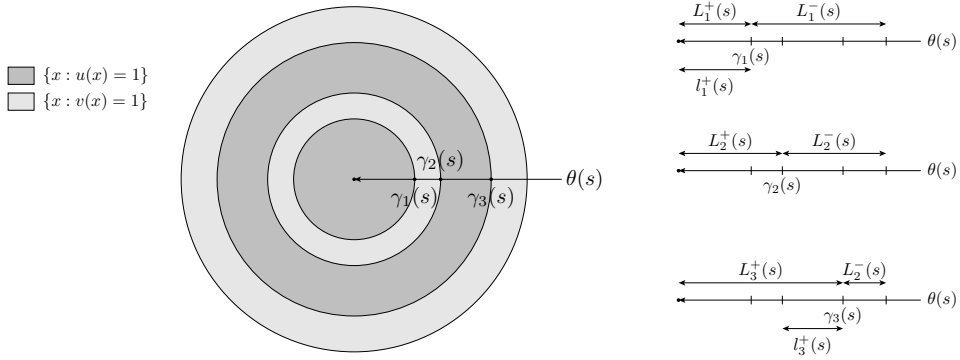}}
\caption{A situation where a ray crosses several times the boundary of
  $\spt(u_\eps)$, given by curves $\gamma_1,\gamma_2,\gamma_3$. On the
  right for each of the curves $\gamma_i$ the positive, negative
  and effective ray lengths $L^+_i,L^-_i,l^+_i$ is depicted. Observe that
  $l^+_2(s)=0$.} 
\label{fig:para2}
\end{figure}

See the Figures
\ref{fig:para} and \ref{fig:para2} for an illustration of the
parametrization defined above. The {\itshape effective ray length} is
introduced to obtain the injectivity of the parametrization 
in the case that a ray crosses several times the boundary
of $\spt(u_\eps)$ (as it is the case in the situation depicted in Figure
\ref{fig:para2}).
The set $\{\gamma(s): l^+(s)>0\}$ represents the points of the boundary
where mass is transported in the `right
direction'.

We will first prove some results which are analogous to those in \cite{CFC}.
\begin{lemma}\label{lem:usc}
The ray direction $\theta: E\to\R^2$ is continuous.
The positive, negative, and effective positive ray lengths $L^+, L^-,l^+ : E\to\R$
are measurable.
\end{lemma}

\begin{proof}
Consider a sequence $(s_k)_{k\in\N}\subset E$, $s\in E$, with $s_k\to s$.
Choose $\tilde{t}_k\leq t_k$ such that
\begin{align}\label{eq:t_n}
  L^+(s_k)-\frac{1}{k} \,&<\, t_k \, <\, L^+(s_k),\\
  l^+(s_k)-\frac{1}{k} \,&<\, \tilde{t}_k\, < \, l^+(s_k).\label{eq:tilde-t_n}
\end{align}
It follows from \eqref{eq:def-L+} and \eqref{eq:t_n} that
\begin{eqnarray}
  \phi(\gamma(s_k)+t_k\theta(s_k)) &=& \phi(\gamma(s_k))+t_k.\label{eq:theta_n}
\end{eqnarray}
Since $(t_k)_{k\in\N},(\tilde{t}_k)_{k\in\N}$ are uniformly bounded by
$2R$ and $|\theta(s_k)|=1$ there exists a subsequence $k\to\infty$ and
$\tilde{t}$,  
$t\in\R,\theta\in\R^2$
with $\tilde{t}\leq t$, $|\theta|=1$ such that
\begin{gather*}
  t_k \,\to\, t,\quad \tilde{t}_k\,\to\, \tilde{t},\quad\text{and} \quad
  \theta(s_k)\,\to\, \theta.
\end{gather*}
We deduce from \eqref{eq:t_n}-\eqref{eq:theta_n} that
\begin{eqnarray}\label{eq:t}
  \limsup_{k\to\infty} L^+(s_k) &\leq& t,\\\label{eq:tilde-t}
  \limsup_{k\to\infty}\, l^+(s_k) &\leq& \tilde{t},\\\label{eq:theta}
  \phi(\gamma(s)+t\theta) &=& \phi(\gamma(s)) + t.
\end{eqnarray}
Since $\phi$ has Lipschitz constant one it follows from
\eqref{eq:theta} that
\begin{gather}
  \theta \,=\, \nabla\phi(\gamma(s))\,=\,\theta(s).\label{eq:equal-theta}
\end{gather}
We therefore deduce by \eqref{eq:theta} that $t\leq
L^+(s)$ and by \eqref{eq:t} we find that 
\begin{eqnarray*}
 \limsup_{k\to\infty} L^+(s_k) &\leq&  L^+(s),
\end{eqnarray*}
which proves the upper-semicontinuity and therefore the measurability 
of $L^+$.  By analogous arguments one proves that $L^-$
is lower-semicontinuous and measurable.\\
Let now $0<\tau<\tilde{\tau}$. Since $\tilde{t}_k\to\tilde{t}$  we
obtain from \eqref{eq:tilde-t_n} that 
\begin{gather*}
  \phi\big(\gamma(s_{k_i}) +\tau\theta(s_{k_i})\big) \,\in\, \spt
  (u_\eps)\quad\text{ for all sufficiently large }k\in\N.
\end{gather*}
Since $\spt(u_\eps)$ is closed and since $\phi$ is continuous this yields
\begin{eqnarray*}
  \phi\big(\gamma(s)+\tau\theta(s)\big) &\in& \spt(u).
\end{eqnarray*}
But $0<\tau<\tilde{\tau}$ was arbitrary and we deduce that
$\tilde{t}\leq l^+(s)$. 
By \eqref{eq:tilde-t} this shows the upper-semicontinuity
and measurability of $l^+$.
\end{proof}

The next result is similar to Lemma~16 of \cite{CFC}:
since rays can only intersect in a common endpoint, the ray directions vary
Lipschitz continuously.

\begin{proposition}\label{prop:lip-l>c}
Let $\gamma\in \Gamma_\eps$ be given as in Remark
\ref{rem:arclength-curves} and set for 
$\delta>0$
\begin{gather}
  E_\delta \,:=\, \{s\in E :
  L^+(s),|L^-(s)|\geq\delta\}. \label{eq:def-E-delta} 
\end{gather}
Then $\theta$ is Lipschitz continuous on $E_\delta$ with
\begin{eqnarray}
  \big|\theta(s_1)-\theta(s_2)\big|
  &\leq& \frac{2}{\delta}\, |s_1-s_2|\quad\text{ for all }s_1,s_2\in
  E_\delta. \label{eq:lipconst-theta} 
\end{eqnarray}
\end{proposition}
\begin{proof}
To prove \eqref{eq:lipconst-theta} it is sufficient
to consider the case $\theta(s_1)\neq\theta(s_2)$. Then there exist
$t_1,t_2\in\R$ such that
\begin{eqnarray}
  \gamma(s_1)+t_1 \theta(s_1) &=& \gamma(s_2) +t_2\theta(s_2).
  \label{eq:triangle}
\end{eqnarray}
By the definition of $L^+,L^-$ and since rays can only intersect in
common endpoints we obtain
\begin{gather}
  t_i\,\in\, \R\setminus\big(L^-(s_i),L^+(s_i)\big),\quad
  i=1,2. \label{eq:bounds-triangle} 
\end{gather}
By taking the determinant with $\theta(s_2)$ in
\eqref{eq:triangle} we compute
\begin{gather}
  \det\big(\gamma(s_1)-\gamma(s_2),\theta(s_2)\big)\,=\,
  -t_1\det\big(\theta(s_1),\theta(s_2)\big).\label{eq:eq-det}
\end{gather}
From \eqref{eq:bounds-triangle}
and since $s_1,s_2\in E_\delta$ we deduce that
  $t_1^2\,\geq\,\delta^2$. 
Taking squares in \eqref{eq:eq-det} we therefore obtain
\begin{align}
  (s_1-s_2)^2 \,&\geq\,
  \delta^2\det\big(\theta(s_1),\theta(s_2)\big)^2\notag\\
  &=\,  \delta^2\Big[ 1-(\theta(s_1)\cdot\theta(s_2))^2\Big] \notag\\
  &=\, \frac{\delta^2}{4}\big|\theta(s_1)-\theta(s_2)\big|^2
  \big|\theta(s_1)+\theta(s_2)\big|^2. \label{eq:est-rays-1}
\end{align}
Next we observe that
\begin{gather}
  \big|\theta(s_1)+\theta(s_2)\big|\,\geq\, 1 \quad\text{ if }
  |s_1-s_2|\leq \delta.\label{eq:small-sdiff}
\end{gather}
Indeed we compute
\begin{align*}
  4\delta\,&=\, \phi(\gamma(s_1)+\delta \theta(s_1)) -
  \phi(\gamma(s_1)-\delta\theta(s_1)) + \phi(\gamma(s_2)+\delta
  \theta(s_2)) - \phi(\gamma(s_2)-\delta\theta(s_2))\\
  &=\, \phi(\gamma(s_1)+\delta \theta(s_1))-
  \phi(\gamma(s_2)-\delta\theta(s_2))+  \phi(\gamma(s_2)+\delta
  \theta(s_2))-\phi(\gamma(s_1)-\delta\theta(s_1))\\
  &\leq\, 2|\gamma(s_1)-\gamma(s_2)| +2\delta|\theta(s_1)+\theta(s_2)|\\
  &\leq\, 2|s_1-s_2| +2\delta|\theta(s_1)+\theta(s_2)|
\end{align*}
which proves \eqref{eq:small-sdiff}.\\
By \eqref{eq:est-rays-1}, \eqref{eq:small-sdiff} we deduce that
\eqref{eq:lipconst-theta} holds for
$s_1,s_2\in E_\delta$ 
with $|s_1-s_2|\leq \delta$. In the case $|s_1-s_2|>\delta$
we obtain \eqref{eq:lipconst-theta} since $|\theta(s_1)-\theta(s_2)|\leq
2$. 
\end{proof}
\begin{definition}\label{def:alpha}
Define two functions $\alpha,\beta: E\,\to\, \R\modulo 2\pi$ by
requiring that 
\begin{gather}
  \theta(s) \,=\,
  \begin{pmatrix}
    \cos \alpha(s)\\
    \sin \alpha(s)
  \end{pmatrix},
  \qquad\det\big(\gamma^\prime(s),\theta(s)\big) \,=\, \sin\beta(s).\label{eq:def-beta}
\end{gather}
\end{definition}
We note that
\begin{gather}
  \sin\beta(s)\,\geq\, 0\quad\text{ for }s\in E\text{ with }l^+(s)>0
  \label{eq:sin-beta} 
\end{gather}
by \eqref{eq:spt-ue-lhs} and the definition of $l^+$.

Proposition \ref{prop:lip-l>c} yields the Lipschitz continuity of
$\theta,\alpha$ on sets where the positive and negative ray lengths are
strictly positive. By Kirszbraun's Theorem and Rademacher's Theorem we
obtain an extension which is differentiable almost everywhere. To
identify these derivatives as a property of $\alpha$ we prove
the existence of $\alpha^\prime$ in the sense of \emph{approximate
  derivatives}~\cite[section 6.1.3]{EG}.  

\begin{lemma}\label{lem:alpha-prime}
The function $\alpha: E\to \R\modulo 2\pi$ as defined in Definition
\ref{def:alpha} is  almost everywhere approximately differentiable and
its approximate differential 
satisfies
\begin{eqnarray}
  \sin \beta(s) - t \alpha^\prime(s) &\geq& 0\quad\text{ for all }\quad
  L^-(s)\,\leq t\, \leq\, L^+(s) 
  \label{eq:det>0}
\end{eqnarray}
for almost all $s\in E$ with $l^+(s)>0$.
\end{lemma}
\begin{proof}
Consider for $\delta>0$ the set $E_\delta$ defined in \eqref{eq:def-E-delta}.
By Proposition \ref{prop:lip-l>c} the restriction of $\theta$ to
$E_\delta$ is Lipschitz continuous. 
From Kirszbraun's Theorem we deduce the existence of a Lipschitz
continuous extension $\tilde{\alpha}_\delta:\R\to\R\modulo 2\pi$ of $\alpha$, see for example the construction in \cite{Gan}.
This extension is almost everywhere differentiable by Rademacher's Theorem.
By \cite[Corollary 1.7.3]{EG} almost all points of $E_\delta$ have
density one in $E_\delta$ and 
we deduce that $\alpha$ is approximately differentiable in these points
and that the approximate differential coincides with
$\tilde{\alpha}_\delta^\prime$.
Since \mbox{$E=\cup_{k\in\N}E_{1/k}$} this proves that $\alpha$ is
approximately differentiable almost everywhere in $E$.

Since a measurable function is almost everywhere approximately continuous
by \cite[Theorem 1.7.3]{EG}, it is sufficient to prove \eqref{eq:det>0}
for $s\in E$ with $l^+(s)>0$ such that 
\begin{itemize}
\item $\alpha$ is approximately differentiable in $s$, and 
\item there exists a sequence
$(s_k)_{k\in\N}\subset E$ with $s_k\to s$ as $k\to\infty$ and 
\begin{gather*}
  L^+(s_k)\to L^+(s),\quad L^-(s_k)\to L^-(s),\quad l^+(s_k)\to l^+(s),\\
  L^+(s_k),\, -L^-(s_k),\, l^+(s_k)\,>\,0\quad\text{ for all }k\in \N.
\end{gather*}
\end{itemize}
Since \eqref{eq:det>0} is satisfied for $\alpha'(s)=0$ we can
also assume that
\begin{itemize}
\item $\alpha'(s)>0$ (the other case is analogous);
\item $(\alpha(s_k)-\alpha(s))/(s_k-s)\to \alpha'(s)$, or equivalently, that
$(\theta(s_k)-\theta(s))/(s_k-s)\to\theta'(s)$;
\item for all $k\in\N$ there exist $t_k,t^*_k$ such that
\begin{gather}
\label{eq:gamma-theta}
  \gamma(s_k)+t_k\theta(s_k) \,=\, \gamma(s)+t^*_k\theta(s).
\end{gather}
\end{itemize}
Analogously to \eqref{eq:bounds-triangle}, \eqref{eq:eq-det} we find
from~\pref{eq:gamma-theta}
\begin{gather}
   t_k\,\in\, \R\setminus\big(L^-(s_k),L^+(s_k)\big),\quad
   t^*_k\,\in\, \R\setminus\big(L^-(s),L^+(s)\big),
   \label{eq:length-t-k}\\ 
  \det\big(\gamma(s_k)-\gamma(s),\theta(s)\big)\,=\,
  -t_k\det\big(\theta(s_k),\theta(s)\big).\label{eq:eq-det-k}
\end{gather}
Writing equation~\pref{eq:eq-det-k} as
\[
\det\Big(\frac{\gamma(s_k)-\gamma(s)}{s_k-s},\theta(s)\Big)\,=\,
  -t_k\det\Big(\frac{\theta(s_k)-\theta(s)}{s_k-s},\theta(s)\Big),
\]
we observe that the left-hand side converges to
$\det(\gamma'(s),\theta(s)) = \sin \beta(s)\geq0$
and the determinant on the right-hand side to $\det(\theta'(s),\theta(s)) = -\alpha'(s)<0$.
For sufficiently large  $k$ we
can therefore eliminate the negative range for $t_k$ in~\pref{eq:length-t-k} 
and obtain
\[
t_k\,\geq\, L^+(s_k).
\]
Again using the negative sign of $\det\bigl((\theta(s_k)-\theta(s))/(s_k-s),\theta(s)\bigr)$ we then find 
\begin{gather*}
  \det\Big(\frac{\gamma(s_k)-\gamma(s)}{s_k-s},\theta(s)\Big)\,\geq\,
  -L^+(s_k)\det\Big(\frac{\theta(s_k)-\theta(s)}{s_k-s},\theta(s)\Big)
\end{gather*}
Taking the limit $k\to\infty$ we
deduce
\begin{gather*}
  \sin\beta(s) \,\geq\,
  L^+(s)\,\alpha'(s).
\end{gather*}
This proves the Lemma.
\end{proof}

For each $\gamma\in\Gamma_\eps$ we define analogously to
\eqref{eq:def-psi}  a map $\psi$. Restricting these maps
suitably we obtain a parametrization of $\spt(u_\eps)$ which is
essentially 
one-to-one. 
\begin{proposition}\label{prop:parametrization}
Let $\Gamma_\eps=\{\gamma_i:i=1,...,N\}$ be chosen as in Remark
\ref{rem:arclength-curves}. For all $i=1,...,N$ let $L_i=L(\gamma_i)$.
Then, with
\begin{eqnarray}\label{eq:def-E^i}
  D_i &:=& \big\{(s,t): s\in E_i\cap [0,L_i), 0\leq t<l_i^+(s)\},
\end{eqnarray}
the restrictions $\psi_i: D_i\to\R^2$ give, up to a Lebesgue
nullset, an injective map onto $\spt(u_\eps)$:
for almost all $x\in \spt(u_\eps)$ there exists a unique
$i\in\{1,...,N\}$ and a unique 
$(s,t)\in D_i$ such that $\psi_i(s,t)=x$.
\end{proposition}
\begin{proof}
For both the injectivity and the surjectivity it
is sufficient to consider only interior points of
$\spt(u_\eps)$ that are not ray ends, since the boundary of
$\spt(u_\eps)$ and the sets 
of ray ends form an $\LL^2$-nullset (Proposition \ref{prop:cfm-rays}). 

For the surjectivity, let $x$ be such a point in $\spt(u_\eps)$. By
\eqref{eq:add-prop-phi1} there exists $y\in \spt(v)$ such
that 
\begin{eqnarray*}
  \phi(x)-\phi(y) &=& |x-y|
\end{eqnarray*}
and it follows that $x$ is on the interior of a ray with direction
$\nabla\phi(x)=\frac{x-y}{|x-y|}$.
Define
\begin{eqnarray*}
  t &:=& \min\ \{t>0: x-t\nabla\phi(x)\in\partial\spt(u_\eps)\}.
\end{eqnarray*}
Then there exists $i\in\{1,...,N\}$ and $s\in E_i$ such that $l^+(s)\geq
t>0$ and
\[
  x \,=\,\gamma_i(s)+t\theta_i(s) \,=\, \psi_i(s,t). 
\]

To prove the injectivity part, let $x$ be the same point again, and
assume that there exists $j\in\{1,...,N\}$, 
$(\sigma,\tau)\in D_j$ such that
\begin{gather*}
  x \,=\, \psi_i(s,t)\,=\,\psi_j(\sigma,\tau).
\end{gather*}
Since different rays cannot intersect in interior points such as $x$, 
the three points $x$, $\gamma_i(s)$, and 
$\gamma_j(\sigma)$ are on the same ray, and since $t>0$ and $\tau>0$
we have
\begin{gather*}
  \phi(\gamma_i(s))\, <\, \phi(x) \qquad\text{and}\qquad\phi(\gamma_j(\sigma)) \,<\, \phi(x).
\end{gather*}
This implies that $\gamma_i(s)$ and $\gamma_j(\sigma)$ are on the same side of the ray with respect to $x$. 
On the other hand by the definition of $l^+$ and since $t<l^+(s)$,
$\tau<l^+(\sigma)$, 
neither the part of the ray between $x$ and $\gamma_i(s)$
nor the part between $x$ and $\gamma_j(\sigma)$ contains a point of
$\partial\spt(u_\eps)$. This implies that $\gamma_i(s)=\gamma_j(\sigma)$
which proves the injectivity part of the proposition.
\end{proof}

Using Proposition \ref{prop:lip-l>c} and Proposition \ref{prop:parametrization}
we can justify the following transformation formula for integrals.

\begin{lemma}\label{lem:trafo-formula}
Let $\Gamma_\eps=\{\gamma_i\}_{i=1,...,N}$ be as in Remark
\ref{rem:arclength-curves} and $D_i$ as defined in \eqref{eq:def-E^i}. 
Then for all \mbox{$g\in L^1(\R^2)$}
\begin{gather}\label{eq:trafo}
  \int g(x)u_\eps(x)\,dx =
  \sum_{i=1}^N \int_{D_i} g(\psi_i(s,t))\frac{1}{\eps}
  \big(\sin\beta_i(s)-t\alphaprime_i(s)\big)\,dt\,ds
\end{gather}
holds.
\end{lemma}
\begin{proof}
We deduce from Proposition \ref{prop:lip-l>c} that for all $i\in
\{1,...,N\}$ the functions $\theta_i$ and $\psi_i$ are approximately
differentiable on $D_i$ with approximate differentials
\begin{align*}
  \theta_i^\prime(s)\,&=\, \alphaprime_i(s)
  \begin{pmatrix}
    -\sin \alpha_i(s)\\ \cos \alpha_i(s)
  \end{pmatrix},\\
  \partial_s\psi_i(s,t)\,&=\,
    \gamma_i^{\prime}(s)+t\theta_i^\prime(s),\qquad
   \partial_t\psi_i(s,t)\,=\, \theta_i^\prime(s).
\end{align*}
This yields that
\begin{align*}
  |\det D\psi_i(s,t)|
  \,=\,& \big|\det \big(\gamma_i^\prime(s),\theta_i(s)\big)+
  t \det\big(\theta_i^\prime(s),\theta_i(s)\big)\big|\\
  =\,& | \sin\beta_i(s) - t\alphaprime_i(s)|\\
  =\,& \sin\beta_i(s) -
  t\alphaprime_i(s)
\end{align*}
where we have used \eqref{eq:det>0} in the last equality.\\
We then deduce from the generalized transformation formula~\cite[Remark 5.5.2]{AGS} that
\begin{gather*}
  \int_{D_i} g(\psi_i(s,t))\frac{1}{\eps}
  \big(\sin\beta_i(s)-t\alphaprime_i(s)\big)\,dt\,ds\,=\,
  \int_{\psi_i(D_i)} g(x)u_\eps(x)\,dx. 
\end{gather*}
Summing these equalities over $i=1,...,N$ we deduce by Proposition
\ref{prop:parametrization} that \eqref{eq:trafo} holds.
\end{proof}
Often it is more convenient to work not in {\itshape length coordinates} but rather in
{\itshape mass coordinates} which are defined as follows.

\begin{definition}\label{def:mass-coordinates}
For $\gamma\in\Gamma_\eps$ and $s\in \R$ we define a map
$\mathfrak{m}_s:\R\to \R$ and a map $M: \R\to\R$ by
\begin{eqnarray}\label{eq:def-mass}
  \mathfrak{m}(s,t) &:=&
  \begin{cases}
    \frac{t}{\eps}\sin\beta(s) -\frac{t^2}{2\eps}\alphaprime(s)&\text{ if }l^+(s)>0,\\
    0 &\text{ otherwise.}
  \end{cases}\\
  \label{eq:def-Ms}
  M(s) &:=& \mathfrak{m}(s,l^+(s)).
\end{eqnarray}
\end{definition}

\begin{figure}[ht]
\centerline{\includegraphics[height=6cm]{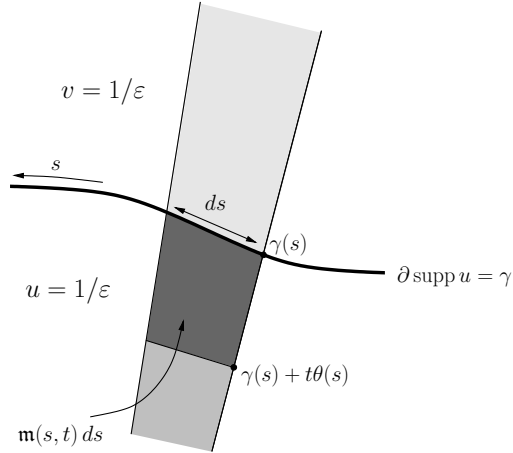}}
\caption{The function $\mathfrak m$. Two rays delimit a section of $u$-mass (below the curve $\gamma$, in this figure) that is transported to the associated section of $v$-mass. For given $t>0$, $\mathfrak m(s,t)\,ds$ is the amount of $u$-mass contained between the two rays and stretching from $\gamma(s)$ to the point $\gamma(s)+t\theta(s)$. For $t<0$,  $\mathfrak m(t,s)$ is the corresponding amount of $v$-mass, with a minus sign.}
\end{figure}

\begin{lemma}\label{lem-mass}
The map $\mathfrak{m}(s,\cdot)$ is strictly monotone on $\big(L^-(s),L^+(s)\big)$ with
inverse
\begin{eqnarray}\label{eq:def-t_s}
  \te(s,m) &=& \frac{\sin\beta(s)}{\alphaprime(s)}
    \Big[1- \Big(1-\frac{2\alphaprime(s)\eps} 
    {\sin^2\beta(s)}m\Big)^\frac{1}{2}\Big].
\end{eqnarray}
Let $\Gamma_\eps=\{\gamma_i\}_{i=1,...,N}$ be given as in Remark
\ref{rem:arclength-curves}, set $L_i=L(\gamma_i)$, and let $M_i$ be the
mass function~\pref{eq:def-Ms} for curve $\gamma_i$. We obtain for
$g\in L^1(\R^2)$ 
that 
\begin{gather}\label{eq:trafo2}
  \int g(x)u_\eps(x)\,dx =
  \sum_{i=1}^N \int_0^{L_i}\int_0^{M_i(s)}
  g(\psi_i(s,\te_i(s,m)))\,dm\,ds. 
\end{gather}
In particular, the total mass of $u_\eps$ is given by
\begin{eqnarray}
  \int u_\eps(x)\,dx &=&
    \sum_{i=1}^N \int_0^{L_i} M_i(s)\,ds.\label{eq:tot-mass}
\end{eqnarray}
\end{lemma}

\begin{proof}
The calculation of the inverse is straightforward. For the transformation formula~\pref{eq:trafo2} we combine~\pref{eq:trafo} with the remark that
\[
\frac{\partial}{\partial t}\,\mathfrak m(s,t)= \frac{1}{\eps}
  \big(\sin\beta_i(s)-t\alphaprime_i(s)\big).
\]
Note that the integration interval $0<t<l^+_i(s)$ in the definition~\pref{eq:def-E^i} of $D_i$ transforms to $0<m<M_i(s)$.
\end{proof}

Typically, the optimal transport map is built by gluing solutions of
one-dimensional transport problems on single rays together.
We observe a similar structure here.

\begin{proposition}\label{prop:mass-balance}
Let $\gamma\in \Gamma_\eps$ be as in Remark \ref{rem:arclength-curves}
and $S$ be 
the optimal transport map from $u_\eps$ to $v_\eps$, as in Proposition
\ref{prop:add-prop}. Define
for $s\in E$ with $l^+(s)>0$ 
an interval $I(s)\subset\R$ and
measures $f_s^+,f_s^-$ on $I(s)$,
\begin{align*}
  I(s) &:=
  \Big(\mathfrak{m}_s\big(L^-(s)\big),
  \mathfrak{m}_s\big(L^+(s)\big)\Big),\\  
  df_s^+ &:= u_\eps\big(\psi(s,\te(s,m))\big)\,dm,\qquad
  df_s^- := v_\eps\big(\psi(s,\te(s,m))\big)\,dm.
\end{align*}
Moreover, define a map
\begin{eqnarray*}
  \hat{S}:I(s) \cap\spt(f_s^+) &\to& I(s)\cap\spt(f_s^-),
\end{eqnarray*}
by the equation 
\begin{gather}\label{eq:def-s-hat}
  S\big(\psi(s,\te(s,m))\big) \,=\,
  \psi\big(s,\te(s,\hat{S}(m))\big)\quad\text{ for }m\in I\cap\spt(f_s^+).
\end{gather}
Then for almost all $s\in E$ with $l^+(s)>0$
the map $\hat{S}$
is the unique monotone transport map pushing $f_s^+$ forward to $f_s^-$. 
\end{proposition}

\begin{proof}
It is sufficient to prove the claim for almost all $s\in \{l^+>k^{-1}\}$,
$k\in\N$ arbitrary.
Let $A\subset E$ be any set such that
\begin{gather}
  A\,\subset\, \{s\in E: l^+(s)>k^{-1}\},\quad \diam(A)\,<\,k^{-1}\label{eq:setA}
\end{gather}
We define rays and a set of rays
\begin{gather*}
  \mathcal{R}(s)\,:=\,\{\gamma(s)+ \tau \theta(s) : \tau \in (L^-(s),
  L^+(s))\},\\
  \mathcal{R}(A)\,:=\, \bigcup_{s\in A} \mathcal{R}(s).
\end{gather*}
Since different rays can only intersect in a common endpoint, two rays
$\mathcal{R}(s), \mathcal{R}(\sigma)$ with $s,\sigma\in A$, $s\neq
\sigma$ can only be disjoint or identical. The latter case is excluded
by \eqref{eq:setA}: since $l^+(s),l^+(\sigma)>k^{-1}$ the distance between
$\gamma(s)$ and $\gamma(\sigma)$ has to be at least $k^{-1}$ in
contradiction to the assumption on the diameter of $A$ in \eqref{eq:setA}.
Therefore the union $\{ \mathcal{R}(s):s\in A\}$ is
pairwise disjoint.

Let now $(g_j)_{j\in\N}$ be a dense subset of $C^0(\R)$. Since $S$ pushes 
$u_\eps|_{\mathcal{R}(A)}$
forward to $v_\eps|_{\mathcal{R}(A)}$ we obtain for all $j\in\N$
\begin{eqnarray}\label{eq:loc-trans}
  \int_{\mathcal{R}(A)}g_j(\phi(S(x)))u_\eps(x)\,dx &=&
  \int_{\mathcal{R}(A)}g_j(\phi(x)) v_\eps(x)\, dx. 
\end{eqnarray}
Repeating the changes of variables of Lemmas~\ref{lem:trafo-formula} and~\ref{lem-mass}, and using the fact that $\{
\mathcal{R}(s):s\in A\}$ is 
pairwise disjoint we obtain 
\begin{align*}
  &\int_A \int_{I(s)}
  g_j\Big(\phi\bigl(S(\psi(s,\te(s,m)))\bigr)\Bigr)u_\eps\big(\psi(s,\te(s,m))\big)\,dm\,ds\\
  &\qquad =\,
  \int_A \int_{I(s)}  g_j\Bigl(\phi\bigl(\psi(s,\te(s,m))\bigr)\Bigr)v_\eps\big(\psi(s,\te(s,m))\big)\,dm\,ds,
\end{align*}
and by using the definition of $\hat S$~\pref{eq:def-s-hat} and the linearity of
$\phi$ along rays we write this
as
\begin{align*}
  &\int_A \int_{I(s)}
  g_j\big(\phi(\gamma(s)) + \te(s,\hat S(m))\bigr)u_\eps\big(\psi(s,\te(s,m))\big)\,dm\,ds\\
  &\qquad =\,
  \int_A \int_{I(s)}  g_j\big(\phi(\gamma(s))
  +\te(s,m)\big)v_\eps\big(\psi(s,\te(s,m))\big)\,dm\,ds,
\end{align*}
Since $A$ as above was arbitrary we deduce that for almost all $s\in \{l^+>0\}$
and all $j\in\N$
\begin{eqnarray*}
  \int_{I(s)} g_j\big(\phi(\gamma(s))+\te(s,\hat{S}(m))\big)\,df_s^+(m) &=&
  \int_{I(s)} g_j\big(\phi(\gamma(s))+\te(s,m)\big)\,df_s^-(m).
\end{eqnarray*}
Since $(g_j)_{j\in\N}$ is dense in $C^0(\R)$ and since for fixed $s$ the
function $m\mapsto
\phi(\gamma(s)) + \te(s,m)$ is a homeomorphism it follows that
\begin{eqnarray*}
  \int_{I(s)} g(\hat{S}(m))\,df_s^+(m) &=&
  \int_{I(s)} g(m)\,df_s^-(m)
\end{eqnarray*}
for all $g\in C^0(\R)$. Therefore $\hat{S}$ pushes $f_s^+$ forward to $f_s^-$.
The monotonicity of $\hat{S}$ follows from the monotonicity of $S$. By \cite[Theorem 3.1]{Amb}
the map $\hat{S}$ is the unique monotone transport map from $f_s^+$ to
$f_s^-$.
\end{proof}

\begin{lemma}\label{lem:eff-mass-balance}
Let $\gamma\in \Gamma_\eps$ as in Remark \ref{rem:arclength-curves}.
Then for almost all $s\in E$ with $l^+(s)>0$ we
obtain that
\begin{gather}
  \big|\psi\big(s,\te(s,m)\big)-
  S\big(\psi\big(s,\te(s,m)\big)\big)\big|
  \,\geq\, \te(s,m)-\te(s,m-M(s))
  \label{eq:esti-t-diffs} 
\end{gather}
for all $0<m<M(s)$.
\end{lemma}
\begin{proof}
We let $\hat{S}$ be as in \eqref{eq:def-s-hat}.
Using the monotonicity of $\te(s,.)$ we obtain
\begin{align}
  \big|\psi\big(s,\te(s,m)\big)-
  S\big(\psi\big(s,\te(s,m)\big)\big)\big|
  &= \big|\psi\big(s,\te(s,m)\big)-
  \psi\big(s,\te(s,\hat{S}(m))\big)\big|\notag\\
  &= \te(s,m)-\te(s,\hat{S}(m)).\label{eq:t-diff}
\end{align}
By Proposition \ref{prop:mass-balance} the map $\hat{S}$ is the unique monotone
transport map from $f_s^+$ to $f^-_s$. In particular
\begin{gather}
  f^-_s\big(\big[\hat{S}(m),\hat{S}(M(s))\big]\big)
  \,=\,   f^+_s\big([m,M(s)]\big)\label{eq:eq-opt-fs}
\end{gather}
holds. Since $f^+_s$ restricted to the set $[0,M(s)]$ coincides
with the Lebesgue measure and since $0\leq f^-_s,f^+_s\leq \LL^1$ we deduce from
\eqref{eq:eq-opt-fs} that
\begin{gather}
  \hat{S}(M(s))-\hat{S}(m)\,\geq M(s)-m\label{eq:eq-opt-fs2}
\end{gather}
for any $0\leq m\leq M(s)$. The value $\hat{S}(M(s))$ is
less or equal than zero since $\hat{S}$ is decreasing and maps onto the
support of 
$f^-_s$. This implies by \eqref{eq:eq-opt-fs2} that \mbox{$\hat{S}(m)\leq
m-M(s)$}. By the montonicity of $\te_s$ we deduce
\begin{gather*}
  \te(s,m)-\te(s,\hat{S}(m))\,\geq\,
  \te(s,m)-\te(s,m-M(s)).
\end{gather*}
Together with \eqref{eq:t-diff} this proves \eqref{eq:esti-t-diffs}.
\end{proof}

By Lemma \ref{prop:lip-l>c} the ray directions vary Lipschitz
continuously on sets where the positive and negative ray lengths are
bounded from below. We obtain a Lipschitz bound also on sets
where $M(\cdot)$ is bounded from below.
\begin{lemma}\label{lem:lip-mass>c}
Let $\gamma\in \Gamma_\eps$ be as in Remark \ref{rem:arclength-curves}.
Then the inequality
\begin{eqnarray}\label{eq:est-l-m}
  \sin\beta(s) \min\big( L^+(s),|L^-(s)|\big) &\geq& \frac{\eps}{2} M(s)
\end{eqnarray}
holds. In particular, for all $0<\kappa<1$ the function $\alpha$ is
Lipschitz continuous on the set 
$\{s: M(s)\geq (1-\kappa)\}$ and
\begin{eqnarray}\label{eq:lip-bound-m>0}
  |\alphaprime(s)| &\leq& \frac{2}{\eps(1-\kappa)}
\end{eqnarray}
holds for almost all $s\in \R$ with $M(s)\geq (1-\kappa)$.
\end{lemma}
\begin{proof}
By \eqref{eq:det>0} we get that $\sin\beta(s)\geq 0$ for $s\in E$ with
$l^+(s)>0$ and that $\alphaprime(s)=0$ if $\sin\beta(s)=0$. Since in the
latter case \eqref{eq:est-l-m} holds it is sufficient to assume that
$\sin\beta(s)>0$.
We consider first the case $\alphaprime(s)\leq 0$.
From \eqref{eq:def-mass} and \eqref{eq:def-Ms} we obtain
\begin{eqnarray}
  M(s) &=& \frac{1}{\eps}\Big(l^+(s)\sin\beta(s)
  -\frac{1}{2}l^+(s)^2\alphaprime(s)\Big)\notag\\ 
  &\leq&
  \frac{\sin\beta(s)}{\eps}l^+(s)-\frac{\sin\beta(s)}{\eps}
  \cdot\frac{l^+(s)^2}{2L^-(s)}.\label{eq:eff-mass-est1}   
\end{eqnarray}
By Proposition \ref{prop:mass-balance} the estimate
\begin{gather*}
  M(s)\,=\, f_s^+((0,M(s))\,\leq\,  f^-_s\big(I(s)\cap (L^-(s),0]\big)
\end{gather*}
holds and we deduce that
\begin{eqnarray}
  M(s) &\leq&
  \int_{L^-(s)}^0\frac{1}{\eps}\big(\sin\beta(s)-t\alphaprime(s)\big)\,dt\notag\\ 
  &\leq& \frac{|L^-(s)|}{\eps}\sin\beta(s)\notag\\
  &=&
  -\frac{\sin\beta(s)}{\eps}L^-(s).\label{eq:eff-mass-est5} 
\end{eqnarray}
Together with \eqref{eq:eff-mass-est1} we deduce
\begin{eqnarray*}
  M(s) &\leq& \frac{\sin\beta(s)}{\eps}l^+(s) + \frac{1}{2M(s)}\Big(\frac{\sin\beta(s)}{\eps}l^+(s)\Big)^2.
\end{eqnarray*}
Therefore $\xi=\frac{\sin\beta(s)}{\eps}l^+(s)$ satisfies the inequality
\begin{eqnarray*}
 \xi^2 +2M(s)\xi-2M(s)^2 &\geq & 0
\end{eqnarray*}
and we obtain that
\begin{eqnarray}\label{eq:eff-mass-est2}
  \frac{\sin\beta(s)}{\eps}l^+(s)&\geq&(\sqrt{3}-1)M(s).
\end{eqnarray}
Together with \eqref{eq:eff-mass-est5} and $l^+(s)\leq L^+(s)$
this proves \eqref{eq:est-l-m} in the case that $\alphaprime(s)\leq0$.\\
Next we assume that $\alphaprime(s)>0$. From
\eqref{eq:def-mass}, \eqref{eq:def-Ms} we observe that
\begin{eqnarray}\label{eq:eff-mass-est4}
  M(s) &\leq& \frac{1}{\eps}L^+(s)\sin\beta(s).
\end{eqnarray}
To prove \eqref{eq:est-l-m} also for $L^-(s)$ let us assume---without loss 
of generality---that
$|L^-(s)|\leq \frac{\eps}{\sin\beta(s)}M(s)$. We then deduce that
\begin{gather}
  |L^-(s)|\,\leq\, L^+(s)\label{eq:ass-l+-}
\end{gather}
and compute
\begin{eqnarray*}
  M(s) &\leq& \int_{L^-(s)}^0 v_\eps(\psi(s,t))\big(\sin\beta(s)
  -t\alphaprime(s)\big)\,dt\notag\\
  &\leq& \int_{L^-(s)}^0 \frac{1}{\eps}\big(\sin\beta(s)
  -L^-(s)\alphaprime(s)\big)\,dt\,=\,\frac{|L^-(s)|}{\eps}\big(\sin\beta(s)
  -L^-(s)\alphaprime(s)\big).
\end{eqnarray*}
By \eqref{eq:ass-l+-} and \eqref{eq:det>0} we deduce that
\begin{gather*}
   M(s) \,\leq\, \frac{|L^-(s)|}{\eps}\big(\sin\beta(s)
  +L^+(s)\alphaprime(s)\big)\,\leq\,2|L^-(s)|\frac{\sin\beta(s)}{\eps}.
\end{gather*}
Together with \eqref{eq:eff-mass-est4} this proves \eqref{eq:est-l-m} in
the case $\alphaprime(s)>0$. 

The estimate \eqref{eq:lip-bound-m>0} follows from \eqref{eq:est-l-m}
and \eqref{eq:det>0}.
\end{proof}
\subsection{Estimate from below}
\label{subsec:est-below}
Let $S$ be the optimal transport map and $\phi\in \Lip_1(\R^2)$
be an 
optimal Kantorovich potential for the mass transport from $u_\eps$ to
$v_\eps$ as in Proposition 
\ref{prop:add-prop}. 

Using Lemma \ref{lem-mass} and Lemma \ref{lem:eff-mass-balance}
we obtain for the
\Wasserstein distance between $u$ to $v$ that
\begin{eqnarray}
  d_1(u,v) &=&\int_{\R^2} |x-S(x)|u_\eps(x)\,dx\notag\\
  &=& \sum_{i=1}^N\int_0^{L_i}\int_0^{M_i(s)}
  \big|\psi_i(s,\te_i(s,m))-S
  \big(\psi_i(s,\te_i(s,m))\big|\,dm\,ds\notag\\  
  &\geq& \sum_{i=1}^N\int_0^{L_i}\int_0^{M_i(s)}\bigl[
  \te_i(s,m) -
  \te_i(s,m-M_i(s))\bigr]\,dm.\label{eq:cost-1} 
\end{eqnarray}
We further estimate the right-hand side of this inequality.

\begin{lemma}\label{lem:stand-trans-est}
Let $\Gamma_\eps$ be a disjoint system of curves parametrized by
arclength as in Remark \ref{rem:arclength-curves} and let $\gamma\in
\Gamma_\eps$.
Then we obtain for all $s\in E$ with $l^+(s)>0$ and
$\sin\beta(s)>0$ that
\begin{align}
  &\int_0^{M(s)} \bigl[\te(s,m) -
  \te(s,m-M(s))\bigr]\,dm\notag\\ 
  =\,& \frac{\eps}{\sin\beta(s)}M(s)^2+
   \frac{\eps^3}{4\sin^5\beta(s)}\alphaprime(s)^2M(s)^4 +
  R(s)\eps^5,\label{eq:stand-trans-est}
\end{align}
where
\begin{gather}
  0\,\leq\, R(s)\,\leq\,
  \frac{7}{9}\frac{M(s)^6}{\sin\beta(s)^9}\alphaprime(s)^4.\label{eq:est-R}
\end{gather}
\end{lemma}

\begin{proof}
We compute for $r\in\R$
\begin{eqnarray}
  \int_0^r \te(s,m)\,dm &=& \int_0^r
    \frac{\sin\beta(s)}{\alphaprime(s)}\Big[1- \Big(1-\frac{2\alphaprime(s)\eps} 
    {\sin^2\beta(s)}m\Big)^\frac{1}{2}\Big]\, dm\notag\\
  &=& \frac{\sin^3\beta(s)}{3\alphaprime(s)^2\eps}\Big[\Big(
    1-\frac{2\alphaprime(s)\eps}{\sin^2\beta(s)}r\Big)^{\frac{3}{2}} 
    +\frac{3\eps\alphaprime(s)}{\sin^2\beta(s)}r -1\Big]\label{eq:cost-4}
\end{eqnarray}
and thus
\begin{align}
  &\int_0^{M(s)} \bigl[\te(s,m)-
  \te(s,m-M(s))\bigr]\,dm\notag\\ 
  =\,&\int_0^{M(s)} \te(s,m)\, dm+ \int_0^{-M(s)}
  \te(s,m)\,dm\notag\\ 
  =\,&\frac{\sin^3\beta(s)}{3\alphaprime(s)^2\eps}
  \Bigg[\Big(
  1+\frac{2\alphaprime(s)\eps}{\sin^2\beta(s)}M(s)\Big)^{\frac{3}{2}} 
    +\Big(
  1-\frac{2\alphaprime(s)\eps}{\sin^2\beta(s)}M(s)\Big)^{\frac{3}{2}} 
    -2\,\Bigg]\label{eq:cost-5}
\end{align}
By a Taylor expansion we observe that for all $\xi>0$
\begin{eqnarray}
\label{eq:Taylor_xi}
  (1+\xi)^{\frac{3}{2}} + (1-\xi)^{\frac{3}{2}} -2 &=& \frac{3}{4}\xi^2
  +\frac{3}{64}\xi^4 + \frac{7}{9}2^{-6}\zeta(\xi)^6
\end{eqnarray}
with $0\leq \zeta(\xi)\leq \xi$.
Using this expansion in \eqref{eq:cost-5} we obtain
\eqref{eq:stand-trans-est}, \eqref{eq:est-R}.
\end{proof}

Putting all information together we derive the following
estimate. 

\begin{proposition}\label{prop:lower-bound}
Let $\partial\spt(u_\eps)$ be given by a disjoint system $\Gamma_\eps$
of closed curves parametrized by arclength as in Remark
\ref{rem:arclength-curves}. Then the lower bound
\begin{eqnarray}\label{eq:lower-bound}
  \G_\eps(u_\eps,v_\eps) &\geq& \sum_{i=1}^N\int_0^{L_i}
  \left[\frac{1}{\eps^2}\Big(\frac{1}{\sin\beta_i(s)}-1\Big)M_i(s)^2 
  +\frac{1}{\eps^2}\big(M_i(s)-1\big)^2\right]\,ds \notag\\
  &&+\sum_{i=1}^N\int_0^{L_i}\frac{1}{4\sin\beta_i(s)}
  \Big(\frac{M_i(s)}{\sin\beta_i(s)}\Big)^4\alphaprime_i(s)^2\,ds  
\end{eqnarray}
holds.
\end{proposition}

\begin{proof}
Since $L_i=L(\gamma_i)$ and by \eqref{eq:tot-mass} we obtain that
\begin{gather*}
  \eps\int_{\R^2}|\nabla u_\eps| \,=\, \sum_{i=1}^N L_i\qquad\text{and}\qquad
  1\,=\, \int_{\R^2} u_\eps \,=\, \sum_{i=1}^N\int_0^{L_i} M_i(s)\,ds.
\end{gather*}
From \eqref{eq:cost-1} and Lemma \ref{lem:stand-trans-est} we therefore obtain
\begin{align*}
  \G_\eps(u_\eps,v_\eps)
  \;\geq\;& \sum_{i=1}^N\int_0^{L_i}\Big(\frac{1}{\eps^2\sin\beta_i(s)}M_i(s)^2
   -\frac{2}{\eps^2}M_i(s)
  +\frac{1}{\eps^2}\Big) \\
  & +\sum_{i=1}^N\int_0^{L_i}
  \frac{1}{4\sin^5\beta_i(s)} 
  \alphaprime_i(s)^2M_i(s)^4,
\end{align*}
and observing that
\begin{align*}
  \frac{1}{\sin\beta_i(s)}M_i(s)^2 -2M_i(s) + 1 \,=\,
  \Big(\frac{1}{\sin\beta_i(s)}-1\Big)M_i(s)^2
  +\big(M_i(s)-1\big)^2 
\end{align*}
the estimate \eqref{eq:lower-bound} follows.
\end{proof}
\subsection{Compactness of the boundary curves and proof of the lim-inf
  estimate} 
\label{subsec:conv}
In this subsection we prove that the boundaries of $\spt(u_\eps)$ 
converge
to a system of closed curves which satisfies the
lim-inf estimate. We later identify this limit with the limit of
$u_\eps\LL^2$.

Let $\partial\spt(u_\eps)$ be given by a system of closed curves
$\Gamma_{\eps}=\{\gamma_{\eps,i}\}_{i=1,...,N(\eps)}$ 
parametrized by arclength as in Remark \ref{rem:arclength-curves}. In
particular we recall that \mbox{$L_{\eps,i}=L(\gamma_{\eps,i})$}.

The total length of the boundary of $\spt(u_\eps)$ is given by
$|\Gamma_{\eps}|$ and we obtain from
\eqref{eq:bound-G-eps} and \eqref{eq:lower-bound}
\begin{gather}
  |\Gamma_{\eps}|\,=\,\sum_{i=1}^{N(\eps)} L_{\eps,i}\,=\, \eps\int_{\R^2}|\nabla
  u_\eps|\,\leq\, 2+\eps^2 \G_\eps(u_\eps,v_\eps)\,\leq\, 2+\eps^2\Lambda.
  \label{eq:bound-boundaries}
\end{gather}
This implies the convergence of the measures $\eps |\nabla u_\eps|
=\mu_{\Gamma_\eps}$ as Radon measures on $\R^2$ for a subsequence
$\eps\to 0$. 

We first define modified curves $\tilde{\gamma}_{\eps,i}$ which are
uniformly bounded in 
$W^{2,2}\big(0,L_{\eps,i}\big)$.

\begin{definition}\label{def:mod-curves}
Let $\gamma\in \Gamma_\eps$ and $0<\lambda<1$.
Choose a periodic Lipschitz continuous function
$\tilde{\alpha}_{\eps,i}:\R\to\R\modulo 2\pi$ with period $L_{\eps,i}$ such that
\begin{align}
  \tilde{\alpha}_{\eps,i}\,&=\,\alpha_{\eps,i}\quad\text{ on }\quad 
  \{s\in \R\,:\, M_{\eps,i}(s)\geq 1-\lambda\},\label{eq:ext-equal}\\
  |\tilde{\alpha}_{\eps,i}^\prime| \,&\leq\,
  \frac{2}{\eps(1-\lambda)}\label{eq:ext-curv-bound}
\end{align} 
and set
\begin{gather*}
  \tilde{\theta}_{\eps,i} \,=\, \begin{pmatrix}
    \cos \alpha_{\eps,i}\\
    \sin \alpha_{\eps,i}
  \end{pmatrix},
  \qquad
  \tilde{\theta}_{\eps,i}^\perp \,=\, \begin{pmatrix}
    \sin \alpha_{\eps,i}\\
    -\cos \alpha_{\eps,i}
  \end{pmatrix}.
\end{gather*}
We then define $\tilde{\gamma}_{\eps,i}:\R\to\R^2$ to be the curve which satisfies
\begin{eqnarray}
  \tilde{\gamma}_{\eps,i}(0) &=&
  \gamma_{\eps,i}(0),\label{eq:new-curve-0}\\
  \tilde{\gamma}_{\eps,i}^\prime(s) &=& \tilde{\theta}_{\eps,i}^\perp(s)
  \quad\text{ for all }s\in\R.\label{eq:new-curve-tan}
\end{eqnarray}
\end{definition}
\begin{remark}
By Proposition \ref{prop:lip-l>c} the
function $\alpha_{\eps,i}$ is Lipschitz continuous on the set
$\{M_{\eps,i}>1-\lambda\}$. By Kirszbraun's Theorem and
\eqref{eq:lip-bound-m>0} 
an extension $\tilde{\alpha}_{\eps,i}$ as in Definition
\ref{def:mod-curves} exists.

The curves $\tilde{\gamma}_{\eps,i}$ are not necessarily closed (or, equivalently, the functions $\tilde{\gamma}_{\eps,i}$ are not necessarily periodic);
the restriction $\tilde{\gamma}_{\eps,i}|_{[0,L_{\eps,i})}$ may also
have self-intersections. However the \emph{tangents} of these
curves, the functions
$\tilde{\gamma}_{\eps,i}^\prime$, are periodic with period~$L_{\eps,i}$.
\end{remark}

\begin{lemma}\label{lem:mod-curves}
Let $\eps>0$, $0<\lambda<1$, $1\leq p\leq 2$, and consider for $i\in \{1,...,N(\eps)\}$
functions
$\tilde{\alpha}_{\eps,i}$ and modified curves $\tilde{\gamma}_{\eps,i}$
as in Definition \ref{def:mod-curves}. Then there exists a constant
$C(\lambda,\Lambda)$, which is independent of $\eps$, such that
\begin{align}
  \sum_{i=1}^{N(\eps)} \int_0^{L_{\eps,i}}
  |\gamma_{\eps,i}^\prime(s)-\tilde{\gamma}_{\eps,i}^\prime(s)|\, ds &\leq
  \, \eps C(\lambda,\Lambda)
  \label{eq:mod-curves-1}\\
  \sum_{i=1}^{N(\eps)} \int_0^{L_{\eps,i}}
  |\tilde{\gamma}_{\eps,i}^{\prime\prime}(s)|^2\,ds
  &\leq\, C(\lambda,\Lambda),
  \label{eq:mod-curves-2}\\
  \sum_{i=1}^{N(\eps)} \int_0^{L_{\eps,i}}
  |\tilde{\gamma}_{\eps,i}^{\prime\prime}(s)|^p\,ds
  &\leq\, \Big[\sum_{i=1}^{N(\eps)}L_{\eps,i}\Big]^{1-\frac{p}{2}}
  \Big[4(1-\lambda)^{-4} \G_\eps(u_\eps,v_\eps)\Big]^{\frac{p}{2}}\notag\\
  &\qquad{}+  4\lambda^{-2}(1-\lambda)^{-2}\eps^{2-p}\G_\eps(u_\eps,v_\eps).
  \label{eq:mod-curves-3}
\end{align}
\end{lemma}
\begin{proof}
Dropping the indexes $\eps,i$ for a moment and letting
$\theta(s)^\perp = (\sin \alpha(s),-\cos \alpha(s))^T$ we compute
that 
\begin{eqnarray}
  \int_0^L |\gamma^\prime(s)-\tilde{\gamma}^\prime(s)|\,ds &\leq&
  \int_0^L \Chi_{\{M\geq 1-\lambda\}}(s) \Big|\gamma^\prime(s) -
  \theta(s)^\perp  \Big|\, ds\notag\\
  &&+ 2\int_0^L \Chi_{\{M< 1-\lambda\}}(s)\,ds\label{eq:diff-gamma-prime}
\end{eqnarray}
and 
\begin{align*}
  & |\gamma^\prime(s) - \theta(s)^\perp|^2
  \,=\, 2-2\gamma^\prime(s)\cdot \theta(s)^\perp\\
  =\,& 2- 2\det \big(\gamma^\prime(s),\theta(s)\big)\,=\,
  2\sin\beta(s)\Big(\frac{1}{\sin\beta(s)} -1\Big),
\end{align*}
which implies
\begin{gather}
\int_0^L \Chi_{\{M\geq 1-\lambda\}}(s) \Big|\gamma^\prime(s) -
  \theta(s)^\perp  \Big|^2\, ds  \leq\,
  2(1-\lambda)^{-2}\int_0^L \Big(\frac{1}{\sin\beta(s)}
  -1\Big)M(s)^2\,ds.\label{eq:diff-g-p1}
\end{gather}
Moreover we calculate
\begin{eqnarray}\label{eq:diff-g-p2}
  \int_0^L \Chi_{\{M< 1-\lambda\}}(s)\,ds &\leq&
  \eps^2\lambda^{-2}\int_0^L \frac{1}{\eps^2}\big(1-M(s)\big)^2\,ds.
\end{eqnarray}
By \eqref{eq:diff-g-p1}, \eqref{eq:diff-g-p2} we obtain from
\eqref{eq:diff-gamma-prime} that
\begin{eqnarray*}
  \int_0^L |\gamma^\prime(s)-\tilde{\gamma}^\prime(s)|\,ds &\leq&
  \sqrt{2L}\eps(1-\lambda)^{-1}\Big( \int_0^L \frac{1}{\eps^2}\Big(\frac{1}{\sin\beta(s)}
  -1\Big)M(s)^2\,ds\Big)^{1/2}\\
  &&+ \eps^2\lambda^{-2}\int_0^L \frac{1}{\eps^2}\big(1-M(s)\big)^2\,ds.
\end{eqnarray*}
Summing the corresponding inequalities for $i=1,...,N(\eps)$, and using
H\"older's inequality, we deduce
that
\begin{align}
  \sum_{i=1}^{N(\eps)}&\int_0^{L_{\eps,i}}
 |\gamma_{\eps,i}^\prime(s)-\tilde{\gamma}_{\eps,i}^\prime(s)|\,ds\notag\\
 &\leq
  \sqrt{2}\eps(1-\lambda)^{-1}\Big(\sum_{i=1}^{N(\eps)}L_i\Big)^{1/2}
  \Big( \sum_{i=1}^{N(\eps)}\int_0^{L_{\eps,i}} \frac{1}{\eps^2}\Big(\frac{1}{\sin\beta_{\eps,i}(s)}
  -1\Big)M_{\eps,i}(s)^2\,ds\Big)^{1/2}\notag\\
  &\qquad+ \eps^2\lambda^{-2}\sum_{i=1}^{N(\eps)}\int_0^{L_{\eps,i}}
 \frac{1}{\eps^2}\big(1-M_{\eps,i}(s)\big)^2\,ds.\\ 
 &\leq
  \sqrt{2}\eps(1-\lambda)^{-1}\Big(\sum_{i=1}^{N(\eps)}L_i\Big)^{1/2}
  \Lambda^{\frac{1}{2}} + \eps^2\lambda^{-2}\Lambda,\label{eq:aux-mod-curves}
\end{align}
where we have used \eqref{eq:bound-G-eps} and \eqref{eq:lower-bound}. By \eqref{eq:bound-boundaries}
the inequality \eqref{eq:mod-curves-1} follows from \eqref{eq:aux-mod-curves}.\\
Dropping indexes $\eps,i$ again we obtain that
\begin{eqnarray}
  \int_0^L |\tilde{\gamma}^{\prime\prime}(s)|^p\,ds
  &=& \int_0^L \Chi_{\{M\geq 1-\lambda\}}(s) |\alphaprime(s)|^p\, ds\notag\\
  &&\qquad+ \int_0^L \Chi_{\{M<
  1-\lambda\}}(s)|\tilde{\alpha}^\prime(s)|^p\,ds\notag\\
  &\leq& L^{1-\frac{p}{2}}\Big[4(1-\lambda)^{-4}\int_0^L
  \frac{1}{4\sin\beta(s)}\Big(\frac{M(s)}{\sin\beta(s)}\Big)^4|\alphaprime(s)|^2\,ds\Big]^{\frac{p}{2}}
  \notag\\
  && \qquad{}+4(1-\lambda)^{-2}\frac{1}{\eps^p}\int_0^L \Chi_{\{M<
  1-\lambda\}}(s)\,ds,\notag
\end{eqnarray}
where we have used \eqref{eq:ext-curv-bound} in the last inequality. We
can further estimate the second term on the right-hand side by
\eqref{eq:diff-g-p2},
\begin{gather*}
  4(1-\lambda)^{-2}\frac{1}{\eps^p}\int_0^L \Chi_{\{M<
  1-\lambda\}}(s)\,ds\,\leq\, 4\lambda^{-2}(1-\lambda)^{-2}\eps^{2-p}\int_0^L
  \frac{1}{\eps^2}(1-M(s))^2\,ds.
\end{gather*}
Putting everything together and summing over $i=1,...,N(\eps)$ we obtain
\begin{eqnarray*}
  \sum_{i=1}^{N(\eps)}\int_0^{L_{\eps,i}} |\tilde{\gamma}_{\eps,i}^{\prime\prime}(s)|^p\,ds
  &\leq& \Big[\sum_{i=1}^{N(\eps)}L_{\eps,i}\Big]^{1-\frac{p}{2}}
  \Big[4(1-\lambda)^{-4} \G_\eps(u_\eps,v_\eps)\Big]^{\frac{p}{2}}\\
  &&\qquad {}+  4\lambda^{-2}(1-\lambda)^{-2}\eps^{2-p}\G_\eps(u_\eps,v_\eps)
\end{eqnarray*}
which is \eqref{eq:mod-curves-3}. Setting  $p=2$
and using \eqref{eq:bound-G-eps} the estimate \eqref{eq:mod-curves-2} follows.
\end{proof}

The next result shows that those curves $\gamma_{\eps,i}$ that
are `too short' do not contribute to the limit.

\begin{lemma}\label{lem:density}
There exists $\delta=\delta(\Lambda)$ and $C=C(\Lambda)$ independent of
$\eps$ such that
for the index set
\begin{eqnarray*}
  J(\eps) &:=& \big\{i\in \{1,...,N(\eps)\} : L_{\eps,i} \leq\delta\big\}
\end{eqnarray*}
the estimate
\begin{eqnarray}\label{eq:density}
  \sum_{i\in J(\eps)} L_{i,\eps} &\leq& \eps C(\Lambda)
\end{eqnarray}
holds.
\end{lemma}

\begin{proof}
Let $\eps>0$ and $i\in J(\eps)$ be fixed and consider
$\tilde{\alpha}_{\eps,i}$ and $\tilde{\gamma}_{\eps,i}$ as in Definition
\ref{def:mod-curves} with $\lambda=1/2$. We drop for a moment the indexes $\eps,i$.
We then have
\begin{eqnarray}
  L &=& \int_0^L |\tilde{\gamma}^\prime(s)|^2\,ds\notag\\
  &=& \int_0^L \tilde{\gamma}^\prime(s)\cdot \big(\tilde{\gamma}(s)-\tilde{\gamma}(0)\big)^\prime\,ds\notag\\
  &=& -\int_0^L \tilde{\gamma}^{\prime\prime}(s)\big(\tilde{\gamma}(s)-\tilde{\gamma}(0)\big)\,ds+
  \tilde{\gamma}^\prime(L)\cdot \big(\tilde{\gamma}(L)-\tilde{\gamma}(0)\big)\notag\\
  &\leq& \int_0^L s |\tilde{\gamma}^{\prime\prime}(s)|\,ds+
  \Big|\int_0^L \bigl(\tilde{\gamma}^\prime(s)-\gamma^\prime(s)\bigr)\,ds \Big|\notag\\
  &\leq& \frac{1}{\sqrt{3}}L^{3/2}\Big(\int_0^L |\tilde{\gamma}^{\prime\prime}(s)|^2\,ds\Big)^{1/2}
  + \int_0^L \big| \tilde{\gamma}^\prime(s)-\gamma^\prime(s)\big|\,ds. \label{eq:proof-dens-1}
\end{eqnarray}
Summing the corresponding inequalities over $i\in J(\eps)$ and using
\eqref{eq:mod-curves-1} and \eqref{eq:mod-curves-2} we obtain with the
H\"older inequality
\begin{eqnarray*}
  \sum_{i\in J(\eps)} L_{\eps,i} &\leq& 
  C(\Lambda)\Big(\sum_{i\in J(\eps)} L_{\eps,i}^3\Big)^{1/2}\Lambda +
  \eps C(\Lambda)\\
  &\leq& {\delta} \Big(\sum_{i\in J(\eps)} L_{\eps,i}\Big)C(\Lambda) +
  \eps C(\Lambda)\\
  &\leq& \frac{1}{2}\Big(\sum_{i\in J(\eps)} L_{\eps,i}\Big) +
  \eps C(\Lambda)
\end{eqnarray*}
for $\delta=\delta(\Lambda)$ small enough. This proves \eqref{eq:density}.
\end{proof}

\begin{proposition}\label{prop:conv-interface}
There exists a
subsequence $\eps\to 0$ and a $W^{2,2}$-system
$\Gamma=\{c_i, i=1,...,N_1\}$ of closed curves without transversal crossings 
such that
\begin{eqnarray}
  \eps|\nabla u_\eps| &\to& \mu_{\Gamma}\label{eq:conv-nabla-ue}
\end{eqnarray}
as measures on $\R^2$. Moreover $\Gamma$ has even \multiplicity and satisfies
\begin{eqnarray}\label{eq:willmore-1}
 \W(\Gamma) &\leq& 2\liminf_{\eps\to 0} \G_\eps(u_\eps,v_\eps).
\end{eqnarray}
\end{proposition}

\begin{proof}
By Lemma \ref{lem:density} we can, without
loss of generality, assume that $L_{\eps,i}> \delta(\Lambda)$ for
all $\eps,i$. We then deduce that the number of curves $N(\eps)$ is
bounded uniformly in $\eps$.
Next we consider the reparametrized one-periodic functions
$c_{\eps,i}:\R\to\R^2$ defined by 
\begin{gather*}
  c_{\eps,i}(r)\,:=\, \gamma_{\eps,i}(L_{\eps,i}r)\quad\text{ for
  }r\in\R.
\end{gather*}
Since the functions $c_{\eps,i}$ are Lipschitz continuous with uniformly
bounded Lipschitz constant, and since $L_{\eps,i}$ are bounded from above 
and away from zero, there exists a subsequence
$\eps\to 0$, a number $N_1\in\N$ and
\begin{gather*}
  L_i>0,\, 
  c_i:\R\to\R^2,\quad i=1,...,N_1,
\end{gather*}
such that for all $i\in\{1,...,N_1\}$ the functions $c_i$ are Lipschitz
continuous and
\begin{eqnarray}
  N(\eps) &=& N_1\quad\text{ for all }\eps>0,\notag\\
  L_{\eps,i} &\to& L_i\quad\text{ as }\eps\to 0,\label{eq:conv-curve-org-L}\\
  c_{\eps,i} &\to& c_i\quad\text{ in } C^{0,\sigma}(\R;\R^2)\quad\text{ as
  }\eps\to 0\label{eq:conv-curve-org}
\end{eqnarray}
holds for all $0<\sigma<1$.
In particular, the functions $c_i$ are
one-periodic and  $\eps|\nabla u_\eps|$ converge as measures to
$\mu_{\Gamma}$. 

We next turn to the $W^{2,2}$-bound on the limit curves $c_i$. 
We fix
$0<\lambda<1$ and consider for $\gamma_{\eps,i}$ modified functions
$\tilde{\gamma}_{\eps,i}$ as in Definition \ref{def:mod-curves} and the
reparametrizations $\tilde{c}_{\eps,i}$,
\begin{eqnarray*}
  \tilde{c}_{\eps,i}(r) &=&
  \tilde{\gamma}_{\eps,i}(L_{\eps,i}r)\quad\text{ for }r\in\R.
\end{eqnarray*}
By Lemma \ref{lem:mod-curves} these curves are uniformly bounded in
$W^{2,2}_{\loc}(\R;\R^2)$ and we deduce that there exists a subsequence
$\eps\to 
0$ of the sequence in \eqref{eq:conv-curve-org} such that the curves
$\tilde{c}_{\eps,i}$ converge weakly in $W^{2,2}_{\loc}(\R;\R^2)$ and
strongly in 
$C^{1,\sigma}([-K,K];\R^2)$ for all $0 <\sigma<1/2$, $K>0$. Since
$c_{\eps,i}(0)=\tilde{c}_{\eps,i}(0)$ we obtain for any $r\in \R$
\begin{eqnarray*}
  \lim_{\eps\to 0} |c_{\eps,i}(r) -\tilde{c}_{\eps,i}(r)| &\leq&
  \lim_{\eps\to 0} \Big|\int_0^r |c_{\eps,i}^\prime(\varrho)-
  \tilde{c}_{\eps,i}^\prime(\varrho)|\,d\varrho\Big|\\
  &\leq& \lim_{\eps\to 0}\Big(1+\frac{|r|}{L_{\eps,i}}\Big) \int_0^{L_{\eps,i}}
  |\gamma_{\eps,i}^\prime(s)-
  \tilde{\gamma}_{\eps,i}^\prime(s)|\,ds\\
  &=& 0,
\end{eqnarray*}
where we have used \eqref{eq:mod-curves-1}, the periodicity of
$c_{\eps,i}$, $\tilde{c}^\prime_{\eps,i}$ and
$L_{\eps,i}\geq\delta(\Lambda)$. Therefore we can identify 
the limits of $c_{\eps,i},\tilde{c}_{\eps,i}$ and deduce that
\begin{alignat}{3}
  \tilde{c}_{\eps,i} \,&\schwto\, c_i\quad&&\text{ weakly in }
  W^{2,2}_{\loc}(\R;\R^2)\quad&&\text{ as 
  }\eps\to 0,\label{eq:conv-curve-mod-W22}\\
  \tilde{c}_{\eps,i} \,&\to\, c_i\quad&&\text{ in }
  C^{1,\sigma}([-K,K];\R^2)\quad&&\text{ as 
  }\eps\to 0,\label{eq:conv-curve-mod-C1}
\end{alignat}
for all $0\leq\sigma<1/2$, $K>0$. In particular $c_i\in W^{2,2}_{\loc}(\R;\R^2)$.
The lower-semicontinuity of the norm under weak
convergence and \eqref{eq:mod-curves-3} yield that
\begin{eqnarray*}
  \sum_{i=1}^{N_1} L_i^{1-2p}\int_0^1 c_i^{\prime\prime}(r)^p \,dr &\leq&
  \liminf_{\eps\to 
  0}\sum_{i=1}^{N_1}  L_{\eps,i}^{1-2p}\int_0^1
  \tilde{c}_{\eps,i}^{\prime\prime}(r)^p \,dr\\ 
  &=& \liminf_{\eps\to
  0}\sum_{i=1}^{N_1}  \int_0^{L_{\eps,i}}
  \tilde{\gamma}_{\eps,i}^{\prime\prime}(s)^p \,ds\\
  &\leq& \Big[\sum_{i=1}^{{N_1}}L_{i}\Big]^{1-\frac{p}{2}}
  \Big[4(1-\lambda)^{-4} \liminf_{\eps\to
  0}\G_\eps(u_\eps,v_\eps)\Big]^{\frac{p}{2}}
\end{eqnarray*}
holds for all
$1\leq p<2$. Since $0<\lambda<1$ and $1\leq p<2$ are arbitrary we
obtain \eqref{eq:willmore-1}.

Since the curves $c_{\eps,i}$
are pairwise disjoint and have no self-intersections, by
\eqref{eq:conv-curve-org} we deduce that  the curves $c_i$ have no
transversal crossings.
Finally we obtain for any $\eta\in C^1_c(\R^2)$ by the divergence theorem 
that
\begin{eqnarray*}
  0 &=& \lim_{\eps\to 0} \int_{\spt(u_\eps)} \nabla\cdot\eta(x)\,dx\\
  &=& \lim_{\eps\to 0} \sum_{i=1}^{N_1} \int_0^{1}
  \eta(c_{\eps,i}(s))\cdot c_{\eps,i}^\prime(s)^\perp\,ds\\
  &=& \sum_{i=1}^{N_1} \int_0^1 \eta(c_{i}(s))\cdot
  c_i^\prime(s)^\perp\,ds\\
  &=& \int_{\spt(\Gamma)} \eta(x)\cdot \Big[ \sum_{\{(i,s): c_i(s)=x\}}
  \frac{1}{L_i}c_i^\prime(s)^\perp\Big]\,d\Ha^{1}(x).
\end{eqnarray*}
We deduce that
\begin{eqnarray*}
   \sum_{(i,s): c_i(s)=x} \frac{1}{L_i}c_i^\prime(s)^\perp &=& 0\quad\text{ for
   }\Ha^1-\text{almost all }x\in \spt(\Gamma)
\end{eqnarray*}
and since $|c_i^\prime|=L_i$ and the vectors
$\{c_i^\prime(s)\,:\,c_i(s)=x\}$ are collinear, this implies that
$\theta_\Gamma(x)=\#\{(i,s): c_i(s)=x\}$ is even.  
\end{proof}

Using results from \cite{BeM05} we deduce that $\mu_\Gamma$ as in
Proposition \ref{prop:conv-interface} is given by an alternative system of curves,
where each curve is passed twice.
\begin{lemma}\label{lem:alt-sys}
Let $\Gamma=\{c_i, i=1,...,N\}$ be a $W^{2,2}$-system of closed curves without
transversal crossings and with an even \multiplicity function $\theta_{\Gamma}$. Then
there exists a system of curves $\overline{\Gamma}=\{\overline\gamma_i :
i=1,...,\overline{N}\}$ such that
\begin{align}
  \mu_\Gamma\,&=\, 2\mu_{\overline{\Gamma}},\label{eq:alt-sys-1}\\
 \W(\Gamma) \,&=\, 2\W(\overline{\Gamma}),\label{eq:alt-sys-2}\\
  \theta_{\Gamma} \,&=\, 2\theta_{\overline{\Gamma}}.\label{eq:alt-sys-3}
\end{align}
\end{lemma}

\begin{proof}
This Lemma requires some machinery from geometric measure
theory. For the relevant definitions of Sobolev type submanifolds,
Hutchinson varifolds and the various definitions for tangential lines
and \multiplicity functions see \cite{BeM05}.

By \cite{BeM05} Remark 4.9 we obtain that
\begin{gather*}
  f\,:=\, \theta(\Gamma,\cdot)\Chi_{\spt(\Gamma)}
\end{gather*}
belongs to $\mathcal{S}^2_{tg}(\R^2)$, that is the set of Sobolev-type
submanifold with 
the additional property that a unique tangential line exists in all points of
the support.
By \cite[Proposition 4.5]{BeM05} this implies that $f\in HV^2(\R^2)$ is a
Hutchinson varifold such that a unique tangential line exists in all points of
the support. Since the \multiplicity is even we conclude that
also $\frac{1}{2}f$ belongs to $HV^2(\R^2)$ and that a unique tangential
line exists in all points of the support. Using again \cite[Proposition 4.5]{BeM05}
we obtain that $\frac{1}{2}f\in \mathcal{S}^2_{tg}(\R^2)$. Now
\cite[Corollary 4.8]{BeM05} gives that $\frac{1}{2}f$ is given by a
$W^{2,2}$-system $\overline{\Gamma}=\{\overline\gamma_i :
i=1,...,\overline{N}\}$ of closed curves without transversal crossings. This implies
\eqref{eq:alt-sys-1}. Equations \eqref{eq:alt-sys-2} and
\eqref{eq:alt-sys-3} are an immediate consequence.
\end{proof} 

We next show that the limit of the boundaries of the support of
$u_\eps$ and the limit of the mass distributions $u_\eps$ are identical.

\begin{proposition}\label{prop:lim-mass}
For a system of curves $\Gamma$ as obtained in Proposition
\ref{prop:conv-interface} and the measure $\mu$ in \eqref{eq:conv-ue} we have
\begin{gather*}
  \mu_\Gamma\,=\, \mu.
\end{gather*}
In particular, \eqref{eq:conv-nabla-ue} holds for the whole sequence
$\eps\to 0$.
\end{proposition}

\begin{proof}
Let $\partial\spt(u_\eps)$ be given by a system of closed curves
$\Gamma_\eps=\{\gamma_{\eps,i}\}_{i=1,...,N(\eps)}$ as in Remark
\ref{rem:arclength-curves}. 
defined in Definition \ref{def:para} we obtain from Lemma \ref{lem:trafo-formula} that
for all $\eta\in C^1_c(\R^{2};R^+_0)$
\begin{align}
  &\Big|\int_{\R^2} \eta u_\eps -\int_{\R^2} \eta \eps|\nabla u_\eps|\,\Big|\notag\\
  =\,& \Bigg|
  \sum_{i=1}^{N(\eps)}\int_0^{L_{\eps,i}}\Big( \int_0^{M_{\eps,i}(s)}
  \eta\big(\gamma_{\eps,i}(s)+\te_{\eps,i}(s,m)\theta_{\eps,i}(s)\big)\,dm
  -\eta\big(\gamma_{\eps,i}(s)\big)\Big)\,ds\;\Bigg|\notag\\
  \leq\,& \sum_{i=1}^{N(\eps)}\int_0^{L_{\eps,i}}
  \big|M_{\eps,i}(s)-1\big|\eta\big(\gamma_{\eps,i}(s)\big)\,ds\notag\\
  &+ \sum_{i=1}^{N(\eps)}\int_0^{L_{\eps,i}}\int_0^{M_{\eps,i}(s)}
  \Big|\eta\big(\gamma_{\eps,i}(s)+\te_{\eps,i}(s,m)\theta_{\eps,i}(s)\big)-
  \eta\big(\gamma_{\eps,i}(s)\big)\Big|\,dm\,ds\notag\\  
  \leq\,& \Big(\sum_{i=1}^{N(\eps)}\int_0^{L_{\eps,i}}
  \big(M_{\eps,i}(s)-1\big)^2\,ds\Big)^{\frac{1}{2}}\Big(\sum_{i=1}^{N(\eps)}\int_0^{L_{\eps,i}}
  \eta\big(\gamma_{\eps,i}(s)\big)^2\,ds\Big)^{\frac{1}{2}}\notag\\
  &+\|\eta\|_{C^1(\R^{2})}\sum_{i=1}^{N(\eps)}\int_0^{L_{\eps,i}}\int_0^{M_{\eps,i}(s)}
  \te_{\eps,i}(s,m)\,dm\,ds\notag\\
  \leq\,& \eps \G_\eps(u_\eps,v_\eps)^{\frac{1}{2}}2\|\eta\|_{C^0(\R^{2})}+
  \|\eta\|_{C^1(\R^{2})}d_1(u_\eps,v_\eps), 
  \label{eq:proof-conv-20}
\end{align}
where we have used \eqref{eq:cost-1}. We observe that by
\eqref{eq:bound-G-eps}
\begin{gather*}
  \frac{1}{\eps} d_1(u_\eps,v_\eps)\,\leq\, 2 +\eps^2 \Lambda
\end{gather*}
and we deduce from \eqref{eq:proof-conv-20} that
\begin{gather*}
  \Big|\int_{\R^{2}} \eta u_\eps -\int_{\R^2} \eta \eps|\nabla u_\eps|\,\Big|
  \,\leq\, \eps \Big(2\Lambda^{\frac{1}{2}}\|\eta\|_{C^0(\R^{2})}+
  (2+\eps^2\Lambda)\|\eta\|_{C^1(\R^{2})}\Big) 
\end{gather*}
This shows that
\begin{gather*}
  \mu\,=\, \lim_{\eps\to 0}u_\eps\LL^2\,=\,\lim_{\eps\to 0}\eps|\nabla
  u_\eps| \,=\, \mu_\Gamma
\end{gather*}
in the sense of convergence as measures.
\end{proof}
The proof of the lim-inf estimate now follows: 
by  Proposition \ref{prop:conv-interface} there exists a system
of curves $\Gamma$ of even \multiplicity such that 
\[
\frac{1}{2}\W(\Gamma) \leq \liminf_{\eps\to 0} \G_\eps(u_\eps,v_\eps).
\]
By Lemma \ref{lem:alt-sys} there is an alternative system of curves
$\tilde \Gamma$, of integer \multiplicity, such that $\mu_\Gamma =
2\mu_{\tilde\Gamma}$ 
and $\W(\Gamma) = 2\W(\tilde \Gamma)$. This second system $\tilde \Gamma$
is the system referred to in Theorem~\ref{the:main} as $\Gamma$. Finally, 
Proposition~\ref{prop:lim-mass} guarantees that 
$\mu_\Gamma = 2\mu_{\tilde\Gamma}$ is the limit of the sequence 
$(u_\e,v_\e)$ in the appropriate sense (see \eqref{eq:conv-ue}).

\section{The limsup estimate}
\label{sec:limsup}
\subsection{The heart of the construction}
Let us first perform the lim-sup construction in the case that $\mu=2\mu_{\Gamma}$,
where $\Gamma$ consists of one simple curve.
\begin{lemma}\label{lem:single-curve}
Let $\gamma: [0,L)\to\R^2$ be a closed $C^2$-curve without self-intersections
and with $|\gamma^\prime|=1$ and let $\Gamma=\{\gamma\}$.
Then there exists $\eps_0>0$,
$\eps_0=\eps_0(\|\gamma^{\prime\prime}\|_{C^0([0,L))})$ and sequences
$(u_\eps, v_\eps)_{0<\eps<\eps_0}$ with
\begin{gather}
  u_\eps,v_\eps\,\in\, BV(\R^2;\{0,1/\eps\}),\quad u_\eps v_\eps\,=\,0,
  \label{eq:ls-reg}\\
  \int_{\R^2} u_\eps \,=\, \int_{\R^2} v_\eps \,=\,
  2L,\label{eq:ls-mass}\\
  \spt(u_\eps)\cup\spt(v_\eps)\,\subset\, \{x: \dist(x,\Gamma)\leq 3\eps\},
  \label{eq:ls-spt-uv} 
\end{gather}
such that
\begin{align}
  2\mu_{\Gamma} &\,=\, \lim_{\eps\to 0}u_\eps\,\LL^2,\label{eq:ls-conv}\\ 
  \frac{1}{2}\int_0^{L} \gamma^{\prime\prime}(s)^2\,ds &\,\geq\, \limsup_{\eps\to
  0} \frac{1}{\eps^2}\Big(\frac{1}{\eps} d_1(u_\eps,v_\eps)
  +\eps\int_{\R^2}|\nabla u_\eps| -4L\Big).\label{eq:ls-ls} 
\end{align}
\end{lemma}
\begin{proof}
There exists a $\eps_0>0$ such that
\begin{gather}
  \sup_{s\in [0,L)} |\gamma^{\prime\prime}(s)| \,\leq\,
  \frac{1}{4\eps_0},\label{eq:ls-eps0-curv}\\
  x \mapsto \dist(x,\Gamma)\text{ is of class }C^2\text{ in }\{x:\dist(x,\Gamma)< 4\eps_0\}.
  \label{eq:ls-eps0-dist}
\end{gather}
Let $\nu:[0,L)\to S^1$ be the $C^1$ unit normal field of $\gamma$ that satisfies
\begin{gather*}
  \det (\gamma^\prime,\nu)\,=\,-1
\end{gather*}
and let $\kappa(s)$ be the curvature
in direction of $\nu(s)$,
\begin{gather*}
  \kappa(s) \,=\, -\nu^\prime(s)\cdot\gamma^\prime(s) \,=\,
  \nu(s)\cdot\gamma^{\prime\prime}(s). 
\end{gather*}
For $\eps<\eps_0$ we set
\begin{eqnarray*}
  u_\eps(x) &:=&
  \begin{cases}
    \frac{1}{\eps} &\text{ if }\dist(x,\Gamma)<\eps,\\
    0 &\text{ elsewise,}
  \end{cases}
\end{eqnarray*}
and define two curves $\hat{\gamma}_+,\hat{\gamma}_- :[0,L)\to\R^2$ by
\begin{eqnarray*}
  \hat{\gamma}_+ (s)&:=& \gamma(s) +\eps\nu(s),\\
  \hat{\gamma}_- (s)&:=& \gamma(s) -\eps\nu(s),
\end{eqnarray*}
see Figure \ref{fig:limsup}.
Using the parametrization $\Phi(s,r):=\gamma(s)+r\nu(s)$ we calculate
that
\begin{gather*}
  |\det D\Phi(s,r)|\,=\, 1-r\kappa(s)\quad\text{ for }0<
   s<L,\,-\eps<r<\eps.
\end{gather*}
Therefore for any $\eta\in C^0(\R^2)$
\begin{gather*}
  \lim_{\eps\to 0}\int_{\R^2} \eta(x)u_\eps(x)\,ds\,=\, \lim_{\eps\to 0}
  \int_0^L\int_{-\eps}^\eps
  \eta(\Phi(s,r))\frac{1}{\eps}(1-r\kappa(s))\,dr\,ds\\
  =\, \int_0^L
  2\eta(\Phi(s,0))\,ds\,=\, 2\int_0^L \eta(\gamma(s))\,ds
\end{gather*}
holds and we deduce \eqref{eq:ls-conv}.

The arc length functions corresponding to $\hat{\gamma}_+, \hat{\gamma}_-$ are given by
\begin{eqnarray*}
  \vartheta_{+}(s) &=& \int_0^s|\hat{\gamma}_+^\prime(\sigma)|\,
  d\sigma\,=\,\int_0^s (1-\eps\kappa(\sigma))\,d\sigma,\\ 
  \vartheta_{-}(s) &=& \int_0^s|\hat{\gamma}_-^\prime(\sigma)|\,
  d\sigma\,=\,\int_0^s (1+\eps\kappa(\sigma))\,d\sigma. 
\end{eqnarray*}
Let $\tilde{\gamma}_+,\tilde{\gamma}_-$ be the corresponding
parametrizations with respect to arclength, 
\begin{gather*}
  \tilde{\gamma}_{+} \,=\, \hat{\gamma}_{+}\circ \vartheta_{+}^{-1},\qquad
  \tilde{\gamma}_+: [0,L_+)\to\R^2,\quad L_+\,=\, L
  -\eps\int_0^L\kappa(\sigma)\,d\sigma,\\ 
  \tilde{\gamma}_{-} \,=\, \hat{\gamma}_{-}\circ
  \vartheta_{-}^{-1},\qquad 
  \tilde{\gamma}_-: [0,L_-)\to\R^2,\quad L_-\,=\, L
  +\eps\int_0^L\kappa(\sigma)\,d\sigma.\\ 
\end{gather*}
The curves $\tilde{\gamma}_+,\tilde{\gamma}_-$ parametrize the boundary of
$\{u_\eps=1/\eps\}$ and we obtain that
\begin{eqnarray}\label{eq:boundary-limsup}
  \int_{\R^2} \eps|\nabla u_\eps| &=& L_+ + L_- \,=\, 2L.
\end{eqnarray}
Corresponding to $\tilde{\gamma}_+,\tilde{\gamma}_-$ we define unit
normal fields
\begin{eqnarray*}
  \tilde{\nu}_+ &:=& \nu\circ\vartheta_+^{-1},\\
  \tilde{\nu}_- &:=& \nu\circ\vartheta_-^{-1}
\end{eqnarray*}
and we compute the curvature of $\tilde{\gamma}_+,\tilde{\gamma}_-$ in
direction of $\nu_{\pm}$ 
\begin{eqnarray*}
  \tilde{\kappa}_+ &=&
  \frac{\kappa\circ\vartheta_+^{-1}}{1-\eps\kappa\circ\vartheta_+^{-1}},\\  
  \tilde{\kappa}_- &=&
  \frac{\kappa\circ\vartheta_-^{-1}}{1+\eps\kappa\circ\vartheta_-^{-1}}. 
\end{eqnarray*}
Observe that by \eqref{eq:ls-eps0-curv} for $\eps<\eps_0$
\begin{gather}
  |\tilde{\kappa}_+|,|\tilde{\kappa}_-|\,\leq\,
   \frac{1}{3\eps_0}.\label{eq:ls-est-kappa} 
\end{gather}
As in subsection \ref{subsec:para} we define \emph{mass coordinates} $\ma_+,\ma_-$ by
\begin{eqnarray*}
  \ma_+(r,t) &:=& \frac{1}{\eps}\big( t-\frac{t^2}{2}\tilde{\kappa}_+(r)\big),\\
  \ma_-(r,t) &:=& \frac{1}{\eps}\big( t-\frac{t^2}{2}\tilde{\kappa}_-(r)\big)
\end{eqnarray*}
and the inverse mappings $\te_+(r,\cdot)=\ma_+(r,\cdot)^{-1}$, $\te_-(r,\cdot)=\ma_-(r,\cdot)^{-1}$, 
\begin{eqnarray*}
  \te_+(r,m) &=& \frac{1}{\tilde{\kappa}_+(r)}\Big(1-\sqrt{1-2\eps\tilde{\kappa}_+m}\Big),\\
  \te_-(r,m) &=& \frac{1}{\tilde{\kappa}_-(r)}\Big(1-\sqrt{1-2\eps\tilde{\kappa}_-m}\Big).
\end{eqnarray*}
By \eqref{eq:ls-est-kappa} the expressions
$1-2\eps\tilde{\kappa}_\pm m$ are positive for $|m|\leq 1$. Using that
\begin{alignat*}{2}
  \sqrt{1+2z}\,&\leq\, 1+z,\quad&&\text{ for }z\geq -\frac{1}{2},\\
  \sqrt{1+2z}\,&\geq\, 1+2z,\quad&&\text{ for } -\frac{1}{2}\leq z\leq 0
\end{alignat*}
we deduce that
\begin{gather}
  0\,\leq\, \frac{1}{m}\te_+(r,m)\,\leq\,
  2\eps,\qquad 0\,\leq\, \frac{1}{m}\te_-(r,m)\,\leq\,
  2\eps.\label{eq:range-l} 
\end{gather} 
We construct parametrizations $\psi_+,\psi_-$ in the same way as
we did in subsection \ref{subsec:para}. Considering 
$\tilde{\gamma}_+$,  the unit
normal field $\tilde{\nu}_+$ takes the role of the direction field
$\theta$ in subsection \ref{subsec:para}. Therefore $\tilde{\kappa}_+$ and
$1$ appear in place of $\alphaprime$ and $\sin\beta$. Note that,
with respect to $\tilde{\gamma}_+$, $\spt(u_\eps)$ is  in the direction of
$-\tilde{\nu}_+$ and, with respect to $\tilde{\gamma}_-$, in direction of
$\tilde{\nu}_-$.\\
\begin{figure}[ht]
\centerline{\psfig{width=\textwidth,figure=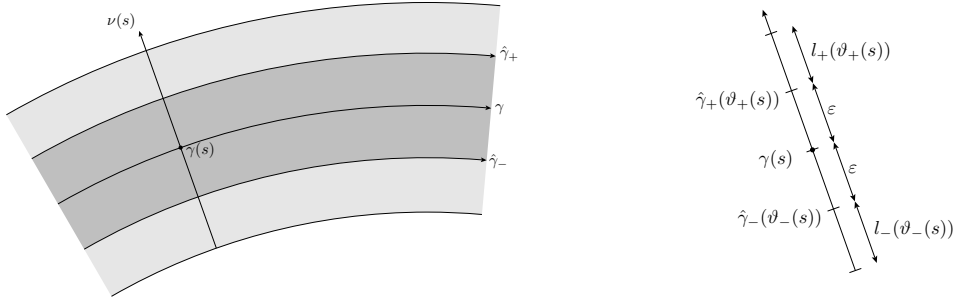}}
\caption{The limsup construction.}
\label{fig:limsup}
\end{figure}
For $\tilde{\gamma}_+, \tilde{\gamma}_-$ respectively we define `ray
lengths' $l_+(r),l_-(r)$ such that their values in mass
coordinates become $-1$ and $1$, 
\begin{align}
  l_+(r) &:= \te_+(r,-1)\,=\,
  \frac{1}{\tilde{\kappa}_+(r)}\Big(1-\sqrt{1+2\eps\tilde{\kappa}_+}\Big),
  \label{eq:ls-def-l+}\\
  l_-(r) &:= \te_-(r,1)\,=\,
  \frac{1}{\tilde{\kappa}_-(r)}\Big(1-\sqrt{1-2\eps\tilde{\kappa}_-}\Big).
  \label{eq:ls-def-l-}
\end{align}
To parametrize $\spt(u_\eps)$ we define sets
\begin{alignat*}{1}
  D_+ &:=\{ (r,t) : 0\leq r< L_+, l_+(r)<t<0\}\\
  D_- &:=\{ (r,t) : 0\leq r< L_-, 0<t<l_-(r)\}.
\end{alignat*}
and the maps
\begin{xalignat*}{2}
  \psi_+(r,t) &:=
  \tilde{\gamma}_+(r)+ t\tilde{\nu}_+(r)&&\text{ for }0<r<L_+, -2\eps<t<2\eps\\
  \psi_-(r,t) &:=
  \tilde{\gamma}_-(r)+ t\tilde{\nu}_-(r)&&\text{ for }0<r<L_-, -2\eps<t<2\eps,
\end{xalignat*}
which are, for
$\eps<\eps_0$, by \eqref{eq:ls-eps0-dist} injective on their
domains. Analogous to Lemma~\ref{lem-mass} we observe that for any 
$\eta\in C^0_c(\R^2)$
\begin{align}
  \int_{\psi_+(D_+)} \eta(x)u_\eps(x)\,dx \,&=\,
  \int_0^{L_+}\int_{-1}^0 \eta\big(\psi_+(r,\te_+(r,m))\big)\,dm\,dr,
  \label{eq:ls-trafo} \\
  \int_{\psi_-(D_-)} \eta(x)u_\eps(x)\,dx \,&=\,
  \int_0^{L_-}\int_0^1
  \eta\big(\psi_-(r,\te_-(r,m))\big)\,dm\,dr. \label{eq:ls-trafo2} 
\end{align}
Moreover we deduce from \eqref{eq:ls-def-l+}, \eqref{eq:ls-def-l-} that
\begin{align*}
  l_+(\vartheta_+(s)) &= \frac{1}{\kappa(s)}\Big(1-\eps\kappa(s)
  -\sqrt{1-(\eps\kappa(s))^2} \Big),\\
  l_-(\vartheta_-(s)) &= \frac{1}{\kappa(s)}\Big(1+\eps\kappa(s)
  -\sqrt{1-(\eps\kappa(s))^2} \Big).
\end{align*}
which implies that
\begin{gather*}
  l_-(\vartheta_-(s)) - l_+(\vartheta_+(s))\,=\, 2\eps.
\end{gather*}
We conclude that the sets $\psi_+(D_+)$ and $\psi_-(D_-)$
are pairwise disjoint and cover almost all of $\spt(u_\eps)$. We
therefore obtain by \eqref{eq:ls-trafo}, \eqref{eq:ls-trafo2} that
\begin{align*}
  &\int \eta(x)u_\eps(x)\,dx \\
  &\,=
  \int_0^{L_+}\int_{-1}^0 \eta\big(\psi_+(r,\te_+(r,m))\big)\,dm\,dr +
  \int_0^{L_-}\int_0^1 \eta\big(\psi_-(r,\te_-(r,m))\big)\,dm\,dr
\end{align*}
holds.
In particular we obtain
\begin{gather}
  \int_{\R^2} u_\eps \,=\, L_+ +L_- \,=\, 2L. \label{eq:mass-u-eps}
\end{gather}
We now define an injective transport map $S:\spt(u_\eps)\to\R^2$
by
\begin{alignat*}{2}
  S(\psi_+(r,t))\,:=&\,
  \psi_+\big(r,\te_+\big(r,\ma_+(r,t)+1\big)\big)
  &&\text{ for }(r,t)\in D_+,\\
  S(\psi_-(r,t))\,:=&\,
  \psi_-\big(r,\te_-\big(r,\ma_-(r,t)-1\big)\big)
  &&\text{ for }(r,t)\in D_-
\end{alignat*}
and set
\begin{gather*}
 v_\eps \,:=\, \frac{1}{\eps}\Chi_{S(\spt(u_\eps))}.
\end{gather*}
This gives
\begin{gather*}
  \spt(v_\eps)\,=\, \{\psi_+(r,t): 0\leq t\leq \te_+(r,1)\} \cup
  \{\psi_-(r,t): \te_-(r,-1)\leq t\leq 0\}.
\end{gather*}
By \eqref{eq:range-l} we obtain that $\te_+(r,1),|\te_-(r,-1)|\leq
2\eps$ and 
since the distance of $\hat{\gamma}_+$ and $\hat{\gamma}_-$ to $\gamma$
equals $\eps$ the equation \eqref{eq:ls-spt-uv} follows.
We compute for an arbitrary $\eta\in C^0_c(\R^2)$ that
\begin{align*}
  &\int\eta(S(x))u_\eps(x)\,dx\\
  =\,& \int_0^{L_+}\int_{-1}^0
  \eta\big(S(\psi_+(r,\te_+(r,m)))\big)\,dm\,dr
  + \int_0^{L_-}\int_0^1
  \eta\big(S(\psi_-(r,\te_-(r,m)))\big)\,dm\,dr\\
  =\,& \int_0^{L_+}\int_{-1}^0
  \eta\big(\psi_+(r,\te_+(r,m+1))\big)\,dm\,dr
  + \int_0^{L_-}\int_0^1
  \eta\big(\psi_-(r,\te_-(r,m-1))\big)\,dm\,dr\\
  =\,& \int_0^{L_+}\int_0^1
  \eta\big(\psi_+(r,\te_+(r,m))\big)\,dm\,dr
  + \int_0^{L_-}\int_{-1}^0
  \eta\big(\psi_-(r,\te_-(r,m))\big)\,dm\,dr\\
  =\,& \int \eta(x)v_\eps(x)\,dx,
\end{align*}
where we have used the change-of-variables formula in the last equality. 
This shows that $S$ indeed is a transport map. In particular it follows
from \eqref{eq:mass-u-eps} that \mbox{$\int_{\R^2} v_\eps =2L$} holds,
which 
proves \eqref{eq:ls-mass}. Moreover
we obtain
\begin{eqnarray}
 d_1(u_\eps,v_\eps) &\leq& \int_{\R^2} |x-S(x)| u_\eps(x)\,dx\notag\\
 &=& \int_0^{L_+}\int_{-1}^0  \big(
 \te_+(r,m+1) -\te_+(r,m)\big)\,dm\,dr \notag\\
 && +\int_0^{L_-}\int_0^1 \big( \te_-(r,m)
 -\te_-(r,m-1))\big)\,dm\,dr \label{eq:ls-d1}
\end{eqnarray}
By \eqref{eq:ls-def-l+} we have $\ma_+(r,l_+(r))=-1$ and we deduce for
the inner integral of the first term on the right-hand side of
\eqref{eq:ls-d1} that
\begin{align*}
 & \int_{-1}^0  \big(
 \te_+(r,m+1) -\te_+(r,m)\big)\,dm\notag\\
 =\,& \int_0^{\ma_+(r,l_+(r))}\te_+(r,m)
 -\te_+\bigl(r,m-\ma_+(r,l_+(r))\bigr)\,dm.
\end{align*}
Therefore Lemma \ref{lem:stand-trans-est} yields that
\begin{gather}
 \int_{-1}^0  \big(
 \te_+(r,m+1) -\te_+(r,m)\big)\,dm
 \,=\, \eps +\frac{1}{4}\eps^3
 \tilde{\kappa}_+(r)^2 +R_+(r)\eps^5,\label{eq:ls-d1+}
\end{gather}
where 
\begin{gather}
  0\,\leq\, R_+(r)\,\leq\, \tilde{\kappa}_+(r)^4\,\leq\, C(\eps_0)
  \label{eq:ls-R+} 
\end{gather}
by \eqref{eq:est-R} and \eqref{eq:ls-eps0-curv}.
For the inner integral of the second term in \eqref{eq:ls-d1} we deduce
by similar arguments 
\begin{align}
 \int_0^1 \te_-(r,m)-\te_-(r,m-1)\,dm
  =\,& \eps +\frac{1}{4}\eps^3
 \tilde{\kappa}_-(r)^2 +R_-(r)\eps^5,\label{eq:ls-d1-}
\end{align}
with
\begin{gather}
  0\,\leq\, R_-(r)\,\leq\, \tilde{\kappa}_-(r)^4\,<\, C(\eps_0).
  \label{eq:ls-R-} 
\end{gather}
By \eqref{eq:ls-d1} and \eqref{eq:ls-d1+}-\eqref{eq:ls-R-} we obtain
that 
\begin{align}
  &\frac{1}{\eps} d_1(u_\eps,v_\eps)\notag\\
  \leq\,& L_+ +L_- +\frac{1}{4}\eps^2
  \int_0^{L_+} \tilde{\kappa}_+(r)^2\,dr +\frac{1}{4}\eps^2
  \int_0^{L_-} \tilde{\kappa}_-(r)^2\,dr +\eps^4 L
  C(\eps_0).\label{eq:ls-d1-sum} 
\end{align} 
We compute that
\begin{align*}
  \int_0^{L_+} \tilde{\kappa}_+(r)^2\,dr +
  \int_0^{L_-} \tilde{\kappa}_-(r)^2\,dr \,&=\, \int_0^L\Big(
  \frac{\kappa(s)^2}{1-\eps\kappa(s)} +
  \frac{\kappa(s)^2}{1+\eps\kappa(s)}\Big)\,ds\\
  &=\, 2\int_0^L \kappa(s)^2\,ds + 2\eps^2\int_0^L
  \frac{\kappa(s)^2}{1-\eps^2\kappa(s)^2}\,ds\\ 
  &\leq\, 2\int_0^L \kappa(s)^2\,ds + 2\eps^2 LC(\eps_0)
\end{align*}
which gives with \eqref{eq:ls-d1-sum} and $2L=L_++L_-$ that
\begin{gather}
  \frac{1}{\eps}d_1(u_\eps,v_\eps) \,\leq\, 2L
  +\frac{\eps^2}{2}\int_0^L\kappa(s)^2\,ds +\eps^4LC(\eps_0).
  \label{eq:ls-d1-sum2}
\end{gather}
Since $|\kappa|=|\gamma^{\prime\prime}|$ we obtain from
\eqref{eq:boundary-limsup} and \eqref{eq:ls-d1-sum2} that
\begin{gather*}
  \frac{1}{\eps^2}\Big(\frac{1}{\eps} d_1(u_\eps,v_\eps)
  +\eps\int_{\R^2}|\nabla u_\eps| -4L\Big) \,\leq\,
  \frac{1}{2}\int_0^L\gamma^{\prime\prime}(s)^2\,ds +\eps^2 LC(\eps_0), 
\end{gather*}
which proves \eqref{eq:ls-ls}.
\end{proof}
\subsection{The general case}
\label{subsec:ls-gen}
We prove now the limsup estimate stated in Theorem \ref{the:main}.
First we need a technical Lemma which follows from
results in \cite{BMP} and \cite{BeM} and approximates a given system of
curves as in Theorem \ref{the:main} strongly in $W^{2,2}$ by a sequence
of more regular systems of curves.
\begin{lemma}\label{lem:detach}
Let $\Gamma$ be a $W^{2,2}$-system of closed curves without
transversal crossings. Then there exists a sequence $(\Gamma^j)_{j\in \N}$
of $W^{2,2}$-systems of closed curves and a number
$m\in\N$ such that the following holds:
\begin{liste(a)}
\item
For all $j\in\N$ the system of curves
$
  \Gamma^j = \{\gamma^j_k\}_{k=1,...,m}
$
is given by a pairwise disjoint family of simple closed curves and satisfies
\begin{gather}
  |\Gamma^j| = |\Gamma|.\label{eq:equal-length}
\end{gather}
\item
For all $1\leq k\leq m$ holds
\begin{align}\label{eq:conv-strong-gamma}
  & \gamma^j_k \,\to\, \gamma_k\quad\text{ in }W^{2,2}(0,1)\quad\text{ as }
    j\to\infty.
\end{align}
\item $\tilde{\Gamma}:=\{\gamma_k\}_{k=1,...,m}$ is a
$W^{2,2}$-system of closed curves without transversal crossings.
\item
$\tilde{\Gamma}$ is equivalent to $\Gamma$ in the sense that
\begin{gather*}
  \mu_{\Gamma}\,=\, \mu_{\tilde{\Gamma}}.
\end{gather*}
\end{liste(a)}
In particular we have
\begin{align}
  \W(\Gamma^j)\,&\to \W(\Gamma)\quad\text{ as }
    j\to\infty.\label{eq:conv-G}
\end{align}
\end{lemma}

\begin{proof}
By \cite[Corollary 5.2]{BeM} there exists a sequence
$(\tilde{\Gamma}^j)_{j\in\N}$ and a $W^{2,2}$-system of closed curves
$\tilde{\Gamma}$ such that
\begin{liste}
\item
for all $j\in\N$ $\tilde{\Gamma}^j$ is an oriented parametrization of a bounded
  open smooth set $E_j\subset\R^2$,\label{enum:i}
\item
$\tilde{\Gamma}^j$ converge weakly in $W^{2,2}$ to $\tilde{\Gamma}$ as
  $j\to\infty$,\label{enum:ii}  
\item
$\tilde{\Gamma}^j$ converge `in energy' to $\tilde{\Gamma}$ as
  $j\to\infty$.\label{enum:iii}
\item
$\tilde{\Gamma}$ and $\Gamma$ are equivalent. 
\end{liste}
Property \eqref{enum:i} implies that $\tilde{\Gamma}^j$ is a disjoint system of
simple closed curves parametrized on $(0,1)$ proportional to arclength.

By property \eqref{enum:ii} there exists a number
$m\in\N$ such that
\begin{gather*}
  \tilde{\Gamma}^j\,=\, \{\tilde{\gamma}^j_k\}_{k=1,...,m},\quad
  \tilde{\Gamma}\,=\, \{\tilde{\gamma}_k\}_{k=1,...,m},\\
  \tilde{\gamma}^j_k,\,\tilde{\gamma}_k : (0,1)\,\to\,\R^2,\\
  |(\tilde{\gamma}^j_k)^\prime|\,=\, L^j_k,\quad
  |(\tilde{\gamma}_k)^\prime|\,=\, L_k\text{ on }(0,1),\notag\\
\end{gather*}
with
\begin{gather}
  \tilde{\gamma}^j_k\,\to\, \tilde{\gamma}_k\quad\text{ weakly in
  }W^{2,2}(0,1)\text{ as }j\to\infty.\label{eq:conv-gamma}
\end{gather}
In particular we deduce that $L^j_k\to\ L_k$ as $j\to\infty$ for all
$k=1,...,m$ 
and
\begin{gather}
  |\tilde{\Gamma}^j|\,\to\, |\tilde{\Gamma}|\label{eq:conv-length}
\end{gather}
holds.
\\
The `convergence in energy' stated in \eqref{enum:iii} gives that, as
$j\to\infty$, 
\begin{gather}
  |\tilde{\Gamma}^j| + \sum_{k=1}^m \int_0^1 (L^j_k)^{-3}
  |(\tilde{\gamma}^j_k)^{\prime\prime}|^2\,\to\,
  |\tilde{\Gamma}|+\sum_{k=1}^m \int_0^1 (L_k)^{-3}
  |(\tilde{\gamma}_k)^{\prime\prime}|^2.\label{eq:conv-energy}
\end{gather}
Together with \eqref{eq:conv-length} and the weak convergence
\eqref{eq:conv-gamma} this 
yields
\begin{gather}
  \tilde{\gamma}^j_k\,\to\, \tilde{\gamma}_k\quad\text{ strongly in
  }W^{2,2}(0,1)\text{ as }j\to\infty,\label{eq:conv-tilde-gamma}\\
  \W(\tilde{\Gamma}^j)\,\to
  \W(\Gamma)\text{ as }j\to\infty.\label{eq:conv-tilde-G} 
\end{gather}
Finally we define for $j\in\N$ a modified system of curves
$\Gamma^j_k=\{\gamma^j_k\}_{k=1,...,m}$ by
\begin{gather*}
  \tilde{\gamma}^j_k(s)\,:=\,
  \frac{|\Gamma|}{|\tilde{\Gamma}^j|}\tilde{\gamma}^j_k(s).
\end{gather*}
Then \eqref{eq:equal-length} is satisfied. By \eqref{eq:conv-length}
we see that $\tilde{\gamma}^j_k$ and $\gamma^j_k$ have for all $1\leq
k\leq m$ the same limits as $j\to\infty$. We therefore obtain from
\eqref{eq:conv-tilde-gamma}, \eqref{eq:conv-tilde-G} that 
\eqref{eq:conv-strong-gamma}, \eqref{eq:conv-G}
hold.
\end{proof}
\ \\
{\bf Proof of Theorem \ref{the:main} (lim-sup part):}\\
Let $\Gamma$ be given as in Theorem \ref{the:main} and
$(\Gamma^j)_{j\in\N}$ be a sequence of systems of closed curves as in 
Lemma \ref{lem:detach}, given by $\Gamma^j=\{\gamma^j_k\}_{k=1,...,m}$. 
We now parametrize the curves $\gamma^j_k$ by arclength,
\begin{gather*}
  \gamma^j_k: (0,L^j_k)\to \R^2,\quad |(\gamma^j_k)^\prime|=1.
\end{gather*}
Choose $\eps^j_0$ such that the distance functions $\dist(\cdot,\Gamma^j)$
is smooth in the set $\{x:\dist(x,\Gamma^j)<4\eps^j_0\}$.\\
For $0<\eps<\eps^j_0$ we let $u^{j,k}_\eps$, $v^{j,k}_\eps$ be the
approximations constructed in Lemma \ref{lem:single-curve} and define
\begin{gather*}
 u^j_{\eps}\,:=\, \sum_{k=1}^m u^{j,k}_\eps,\qquad
 v^j_{\eps}\,:=\, \sum_{k=1}^m v^{j,k}_\eps.
\end{gather*}
By \eqref{eq:ls-spt-uv} all the functions $u^{j,k}_\eps,
v^{j,k}_\eps$ have pairwise disjoint supports.
We compute that
\begin{align*}
 \int_{\R^2} u^j_\eps\,d\LL^2\,=\, \sum_{k=1}^m \int_{\R^2}
 u^{j,k}_\eps\,d\LL^2\,=\, 
 \sum_{k=1}^m 2L^k_j\,=\, 2|\Gamma^j|\,=\, 2|\Gamma|\,=\, M
\end{align*}
and, by similar calculations, $\int v^j_\eps =M$.
Moreover, by \eqref{eq:ls-conv}, \eqref{eq:conv-strong-gamma},
\begin{align*}
 \lim_{j\to\infty}\lim_{\eps\to 0} u^j_\eps\LL^2\,=\, \lim_{j\to\infty}
 \sum_{k=1}^m \lim_{\eps\to 0} u^{j,k}_\eps\LL^2 \,=\,  2\lim_{j\to\infty}
 \mu_{\Gamma^j}\,=\, 2\mu_{\Gamma} 
\end{align*}
and, by \eqref{eq:conv-G}, \eqref{eq:ls-ls},
\begin{align*}
  \W(\Gamma)\,&=\,\lim_{j\to\infty} \W(\Gamma^j)\\
  &=\, \lim_{j\to\infty}\frac{1}{2}\sum_{k=1}^m \int_0^{L^j_k}
  \big|(\gamma^j_k)^{\prime\prime}\big|^2 \,ds\\
  &\geq\, \limsup_{j\to\infty}\sum_{k=1}^m \limsup_{\eps\to 0} \frac{1}{\eps^2}\Big(
  \frac{1}{\eps}d_1(u^{j,k}_\eps,v^{j,k}_\eps) + \eps\int |\nabla u^{j,k}_\eps| -
  2\int u^{j,k}_\eps\,d\LL^2\Big)\\
  &\geq\, \limsup_{j\to\infty}\limsup_{\eps\to 0} \frac{1}{\eps^2}\Big(
  \frac{1}{\eps}d_1(u^j_\eps,v^{j}_\eps) + \eps\int |\nabla u^j_\eps| -
  2\int u^j_\eps\,d\LL^2\Big).
\end{align*}
The final inequality follows from constructing an admissible joint transport map for $(u_\eps^j,v_\eps^j)$ from the transport maps of $(u_\eps^{j,k},v_\eps^{j,k})$.

Therefore there exist subsequences $(j_l)_{l\in\N}, (\eps_l)_{l\in\N}$
such that
\begin{gather*}
 \lim_{l\to\infty} u^{j_l}_{\eps_l}\LL^2\,=\,2\mu_\Gamma
\end{gather*}
as Radon measures on $\R^2$ and
\begin{gather*}
    \W(\Gamma) \,\geq\, \limsup_{l\to\infty} \G_{\eps_l}(u_{\eps_l},v_{\eps_l}).
\end{gather*}
This proves the lim-sup estimate in Theorem \ref{the:main}.\hfill $\Box$
\section{Discussion}
\label{sec:discussion}

\subsection{General remarks}
In the Introduction we formulated two basic questions about
self-aggregating, partially localized structures: first, why do they
exist at all (and why are they stable), and secondly, how can we
understand their resistance to stretching, bending, and fracture, that
is observed experimentally. 

The fact that the functional $\F_\eps$ favours structures in which $u$
and $v$ alternate on a length scale $\e$ is well illustrated by the
special cases of Section~\ref{sec:heuristics}. Simply put, for
$d_1(u,v)$ to be moderate, the masses represented by $u$ and $v$ have to
be close; the fact that $\Mod u_{L^\infty}$ and $\Mod v_{L^\infty}$ are
limited implies that these masses occupy a certain amount of volume; and
finally the interface penalization prevents the masses of $u$ and $v$
from being too finely interspersed. Together these restrictions impose a
length scale on structures with moderate $\F_\eps$. 

This argument does not determine, however, the \emph{geometry} of
structures with moderate (or even lowest) energy. The fact that the
straight lamellar geometry has lowest energy is not obvious, as is
demonstrated by the `wriggled lamellar' patterns of Ren and Wei in a
strongly related system~\cite{RenWei05}. The reason for the penalization
of curvature in the current context can be traced back to a property of
strict convexity of the function $\mathfrak t(s,m)$ defined in
\eqref{eq:def-t_s}. Since this convexity property lies at the 
heart of the stability of these partially localized structures, we take
some time to investigate it further. 

\subsection{The geometric basis for stability}

An essential observation in the proof is that it is more convenient to
use 
\emph{mass} coordinates along rays instead of \emph{length} coordinates,
since the \Wasserstein distance measures spatial distances between pairs
of infinitesimal portions of mass. The function $\mathfrak t(s,m)$
(see~\eqref{eq:def-t_s}) connects the two descriptions by giving the
length coordinate $t$ as a function of mass coordinate $m$.  

\begin{figure}[ht]
\centering
\centerline{\vc{\psfig{figure=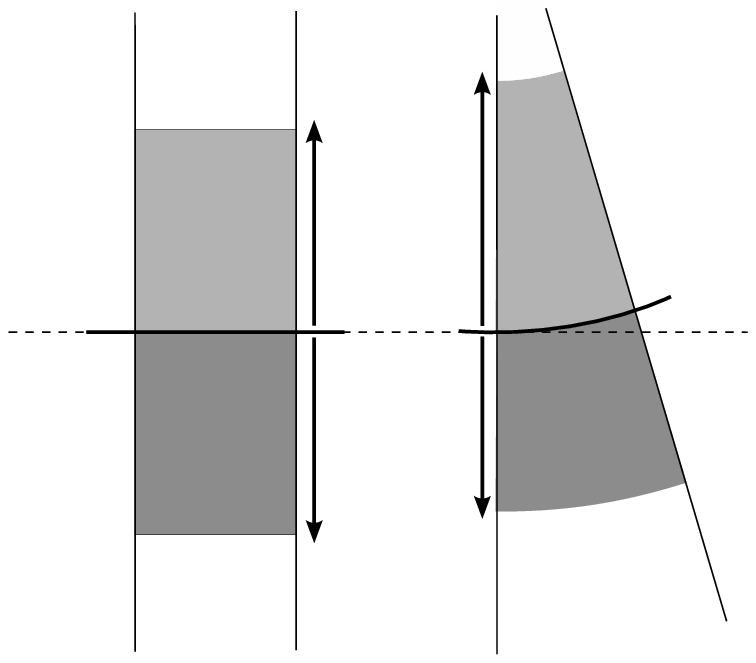,height=5cm}}
  \hskip5mm \vc{\psfig{figure=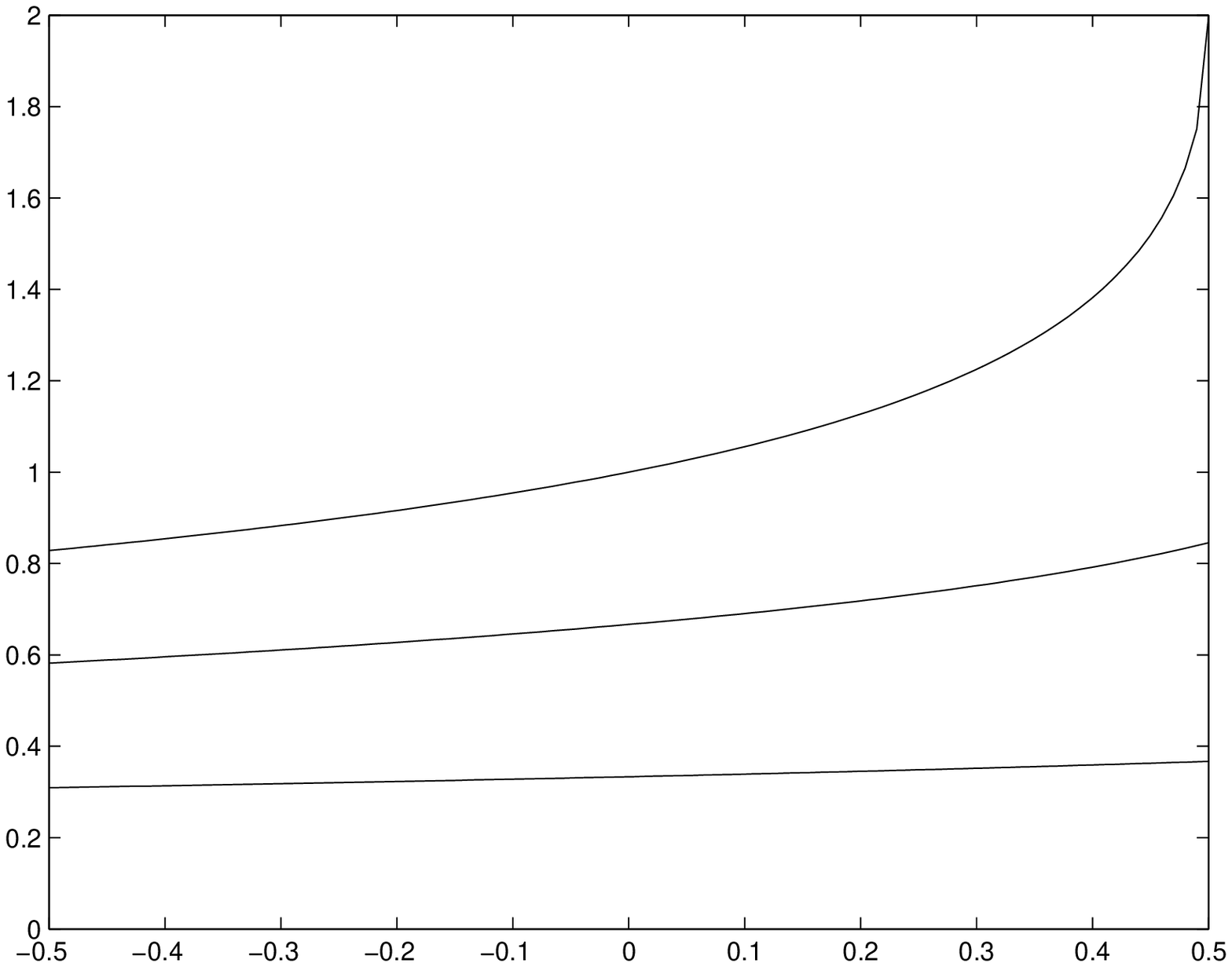,height=5cm}}}
\caption{Left: the function $\te$ maps the mass coordinate $m$ to a
  length coordinate $t$, in the following way. Fixing an infinitesimal
  section $ds$ of interface, a section of mass $mds$, delimited by $ds$
  and by two rays, extends away from the interface by a distance of
  $\te(s,m;\alpha'(s))$.  
Right: 
the function $\alpha'(s)\mapsto\mathfrak t(s,m;\alpha'(s))$
  (see~\eqref{eq:def-t_s}), plotted for $m\in\{1/3,2/3,1\}$ (curves
  (1--3)). Here we have taken $\e=1$ and $\sin \beta(s)=1$.} 
\label{fig:effectcurvature}
\end{figure}

For fixed $m$, the value of $\te(s,m)$ depends on the value of
$\alpha'(s)$, as is obvious both from the definition and from a
geometric point of view (see Fig.~\ref{fig:effectcurvature} (left)).  
For the purpose of this discussion we write the dependence on
$\alpha'(s)$ explicitly, as  
$\mathfrak t(s,m;\alpha'(s))$. The pertinent observation is that for
fixed $m\not=0$, the function $\alpha'(s)  \mapsto \mathfrak
t(s,m;\alpha'(s))$ is strictly convex. 
This convexity can be seen in a simple plot of the function $\te$, as in
Fig.~\ref{fig:effectcurvature} (right), but can also be recognized in
the geometry of rays that are rotated with respect to each other
(Fig.~\ref{fig:effectcurvature} (left)).

With this convexity we can understand why curvature is penalized. 
In~\eqref{eq:cost-5} the relevant expression is
\[
\int_0^{M(s)} \te(s,m;\alpha'(s))\, dm+ \int_0^{-M(s)}  \te(s,m;\alpha'(s))\,dm
\]
which can be written as
\[
\int_0^{M(s)} \bigl[\te(s,m;\alpha'(s)) - \te(s,-m;\alpha'(s))\bigr]\, dm.
\]
This integral describes the transport cost for the transport along an
individual ray at position $s$.  
Using a symmetry property of $\te$ and the strict convexity mentioned above, 
\begin{align*}
\mathfrak t(s,m;\alpha'(s)) - \mathfrak t(s,-m;\alpha'(s)) 
&= \mathfrak t(s,m;\alpha'(s)) + \mathfrak t(s,m;-\alpha'(s))  \\
&> 2\mathfrak t(s,m;0)\\
&= \mathfrak t(s,m;0) - \mathfrak t(s,-m;0).
\end{align*}
It is in this inequality that we see how curving the interface (or, more
precisely, rotating the rays) creates a higher transport cost and
therefore a larger value of the energy.

\subsection{Connections to other work: continuum models and phase separation}

Beyond the direct context of lipid bilayers, this work is at the
intersection of a variety of different lines of research. There is of
course a long-running tradition of slender-body limits in solid
mechanics, resulting in a wide variety of simplified models (see e.g.\
\cite{FrieseckeJamesMueller05TR} for an overview of the case of elastic
plates and shells). Although our final result is rather similar---the
Gamma-limiting energy $\W$ is the classical elastica bending
energy---the current work is fundamentally different, in that the
material at the $\e>0$ level is closer to a fluid than a solid:
particles may be arbitrarily rearranged without incurring an energy
penalty. Indeed, the penalization of curvature that we find in the limit
is an effect of \emph{global} geometry rather than of (local)
deformation. One might compare this with the difference between an
elastic ball and a drop of water; both resist deformation, but for the
ball this arises from local intermolecular forces, while the form of the
drop results from the global effect of surface area minimization.  

Turning to fluid-like systems, the simple fact that the mass represented
by $u$ and $v$ concentrates onto a curve is in itself remarkable. Phase
separation naturally leads to penalization of interfacial length, as
reflected by the term $\int \mod{\nabla u}$ in $\F_\eps$, and the distance
penalization by the term $d_1(u,v)$ is necessary to prevent bulk-scale
separation. To our knowledge the current work is the first example of a
phase-separating system that concentrates on low-dimensional sets.  

Concentration onto low-dimensional sets is a commonly observed feature
in many (often non-conservative, non-variational) reaction-diffusion
systems. For domains with sufficient symmetry a soliton- or pulse-like
solution in one dimension can be interpreted as a solution in higher
dimensions that concentrates on a hyperplane. In recent years less
trivial examples of low-dimensional concentration have been
uncovered~\cite{DAprile00,BadialeDAprile02,AmbrosettiMalchiodiNi02,AmbrosettiMalchiodiNi03,AmbrosettiMalchiodiNi04,DoelmanVanderPloeg02,MalchiodiMontenegro02,MalchiodiMontenegro03,Malchiodi05}.
Those concentrated solutions are often unstable, however, and sometimes
even highly unstable
(e.g.\cite{MalchiodiMontenegro02,MalchiodiMontenegro03,Malchiodi05}),
thus strongly contrasting with the stable nature of the structures of
this paper. 


\subsection{Connections to other work: The elastica functional}

Originally the elastica functional was introduced by Daniel Bernoulli and
Euler as the bending energy of an elastic rod, but it has many other
applications in different fields. In variational methods for image
reconstruction it is widely used~\cite{Mum}. In 
the theory of phase separations the elastica functional appears
as the two dimensional reduction of the Willmore functional
\eqref{eq:willmore}. 

It is often natural to consider the elastica functional as acting on 
boundaries of subsets of the ambient space. In \cite{BMP}
the lower semicontinuity of the elastica functional under 
$L^1$-convergence of sets is investigated; see
also \cite{Sch} 
for a generalization to arbitrary dimensions.
The Gamma-convergence of a diffuse interface approximation
of the elastica, and 
more generally the Willmore functional, was conjectured by De Giorgi
\cite{DG} and proved in a modified form in \cite{RS} for space
dimensions two and three. 

Our approach is not restricted to curves which are
boundaries of sets, and therefore open curves can be represented in this framework. 
A related approach can be found in \cite{BrM}, where
functionals which act on phase-field approximations of `thickened
curves' are considered and the Gamma-convergence of these functionals
is proved in a topology based on the Haussdorf metric. The
elastica functional is part of the limit, however the approximation is
completely different. 

\subsection{Relevance to the understanding of lipid bilayers}

The derivation that leads from a self-consistent mean-field theory for
two-bead copolymers to the energy $\F_\eps$  (Appendix~\ref{app:derivation})
contains a number of highly suspicious assumptions. The most glaring one
is the severing of the head-tail bond in individual polymers: under this
assumption a given head does not keep track of to \emph{which} tail it
is connected, as long as it is connected to exactly one tail. The
\Wasserstein distance $d_1(u,v)$ is the mathematical implementation
of this assumption, and this term is all that keeps heads and tails from
separating to arbitrarily large distances. 

The most interesting conclusion, from the point of view of applications,
might be that this very weak remnant of the covalent bond
suffices; that apparently \emph{very little} is needed to convert a
system exhibiting bulk-scale phase separation (e.g.\ an oil-water
mixture) into a system which separates at a smaller scale.

\subsection{Other approaches to the upscaling limit}
In the physico-chemical literature many authors have established
connections between the Helfrich Hamiltonian~\pref{def:EHelfrich} on one
hand and various smaller-scale models of lipid bilayers on the
other~\cite{SzleiferKramerBenShaulGelbartSafran90,OversteegenBlokhuis00,
  LaradjiMouritsen00}. Besides being an interesting scientific issue,
such a connection should also give insight in the dependence of the
coefficients~$k$ and~$\overline k$ on molecular parameters; indeed the
derivation given by Helfrich and others is phenomenological, based on
scaling and dimensionality arguments, and therefore the coefficients are
unknown.


An approach that comes close to ours is that
of~\cite{ChaconSomozaTarazona98}, in which the  authors connect the
Helfrich description with a model in which lipids are represented by a
density and a vectorial quantity that may be interpreted as an alignment
order parameter. As in the examples above, however, the final
coefficients are again found by a numerical method
(minimization of a grand potential energy). Our result here appears to
be unique in giving a fully analytic formula for the bending modulus,
which is not so surprising since the model contains no unscalable
parameters other than $\e$. 

Another way in which our contribution differs from existing work is the
possibility to assess the energy penalties associated with non-optimal
thickness and with curve `ends'.  
In the current paper we only show that these penalities are larger than
$O(\eps^2)$, but determining the actual scale of penalization is an
interesting open question. 

\subsection{Generalization to $d_p(u,v)$ for $p>1$}

The derivation of the energies $\F_\eps$ presented in
Appendix \ref{app:derivation} suggests substituting  the
\Wasserstein distance by the 2-Wasserstein distance. Some of the 
results of this paper carry over to this situation, and in fact to the
$d_p$-metric for all $1<p<\infty$. This follows from the remark that
\begin{align*}
d_1(u,v) &= \min \Big( \int_{\Rn\times\Rn} |x-y| \,d\gamma(x,y)\Big) \\
  &\leq \min \Big( \int_{\Rn\times\Rn} |x-y|^p \,d\gamma(x,y)\Big)^{1/p}
           \Big( \int_{\Rn\times\Rn} d\gamma(x,y)\Big)^{(p-1)/p}\\
   &= d_p(u,v),
\end{align*}
where we take total mass $M$ equal to one for simplicity. Those statements of Theorem~\ref{the:main} that only depend on the upper 
bound~\pref{eq:liminf-bounded} therefore carry over without change for functionals such as 
\begin{align*}
\F^p_\e(u,v) &= \e\int\mod{\nabla u} + \frac1{\e}d_p(u,v) \\
\tilde \F^p_\e(u,v) &= \e\int\mod{\nabla u} + \frac1{p\e^p}d_p(u,v)^p
\end{align*}
To be concrete, sequences $(u_\e,v_\e)$ along which the associated functionals
$\G^p_\e$ or $\tilde\G^p_\e$ remain bounded will converge along subsequences to a system of closed $W^{2,2}$-curves. 

The Gamma-convergence result of this paper, however, can not be proved in this way: there is a gap between this lower bound and the upper bound that follows from a generalization of Section~\ref{sec:limsup}; this gap results from the strict convexity of the function $z\mapsto z^p$ for $p>1$. Nonetheless, we believe a similar result to be true for all $1<p<\infty$.

\subsection{Generalizations}
There are several other interesting directions in which further research could
continue, among them
\begin{itemize}
\item generalizations to higher dimension and higher codimension,
\item characterisation of local minimizers of the energies $\F_\eps$, 
\item formulation of consistent evolution equations on the mesoscale,
  \emph{i.e.} for the densities $(u_\e,v_\e)$. 
\end{itemize}
In higher dimensions, and for limit structures of dimension larger than
one, a concept similar to the `systems of curves' will probably not be
appropriate, and we expect that a 
varifold approach will be better suited. Some results of the present
paper are 
easily generalized but there are also fundamental differences.


To investigate how close the proposed mesoscale approximation is
to the Helfrich energy or to the elastica functional it is interesting to
compare (local) minimizers of the energy $\F_\eps$  
to the variety of shapes found as local minimizers for the Helfrich
energy. 

A natural idea for extending the mesoscale description of lipid bilayers to
a time-dependent evolution is to set up a gradient flow for the energies
$\F_\eps$, such as in~\cite{Fraaije93} or~\cite{BlomPeletier04}. The
latter corresponds to a gradient flow based on the 2-Wasserstein 
distance.

\newpage
\begin{appendix}
\section{Derivation of $\F_\eps$}
\label{app:derivation}

The functional $\F_\eps$ that is the basis of this paper arises in a
simple model for a water-lipid system. In this model a lipid is
represented by a two-bead chain: a head bead and a tail bead connected
by a spring. The water molecules are represented by  a third type of
bead. 

The state space at the microscopic level is given by the positions
$X^i_t$,  $X^i_h$, and $X^j_w$ of the lipid tail, lipid head, and water
beads; the lipids are numbered by $i=1,\ldots,\nl$ and the water beads
by $j=1,\ldots,\nw$. 
Assuming that the beads are confined to a space $\Omega\subset\R^d$, the
full microscopic state space for the system is then 
\[
\X = \Omega^{2\nl+\nw}.
\]
Elements $X=(X_t^1,\ldots,X_t^\nl,X_h^1,\ldots,X_h^\nl,
X_w^1,\ldots,X_w^\nw)\in \X$ are called microstates.

We describe the system in terms of probabilities on $\X$, i.e. the
(probabilistic) state $\psi$ is a probability measure on $\X$:
\[
\psi\in \E, \qquad \text{where} \qquad
\E = \left\{\,\psi : \X \to \R^+, \quad \int_\X \psi = 1 \,\right\}.
\]
We assume that neither the microstates themselves nor the 
measure $\psi$ can be observed at the continuum level; 
the observables are three derived quantities, the volume fractions
of tails, of heads, and of water.

For a given probability measure $\psi\in \E$, the water volume fraction
$r_w(\psi):\Omega\to\R^+$ is defined by
\[
r_w(\psi)(x) = v \sum_{j=1}^\nw \int_{\X} \psi\, \delta(X_w^j-x) \, dX
\qquad \text{for all }x\in\Omega.
\]
Here $v$ is the volume fraction of a single bead. The tail and head
volume fractions $r_t(\psi)$ and $r_t(\psi)$ are defined similarly.

To specify the behaviour of the system we introduce two free energy
functionals. The `ideal' free energy $\Fid:\E\to\R$ models the effects
of entropy and the interactions between beads of the same molecule;
the `non-ideal' free energy $\Fnid:\E\to\R$ represents the interactions
between the beads of different molecules. The total free energy is the 
sum of the two,
\[
F(\psi) = \Fid(\psi) + \Fnid(\psi).
\]
For $\Fid$ we assume zero temperature and postulate
\[
\Fid(\psi) = \int_\X \psi \Hid,
\]
where the function $\Hid:\X\to\R$ is the internal energy of a microstate,
and the superscript `id' again refers to a restriction to interaction
within a single lipid molecule. While remarking that many different
choices are found in the literature, here we choose simply to implement
a spring by penalizing head-tail distance, 
\[
\Hid(X) = \frac k2\sum_{i=1}^\nl \mod{X_t^i-X_h^i}^p.
\]
A natural choice is $p=2$, which gives the energy of a linear
spring. Even more realistic seems a general distance term that becomes
infinite if $\mod{X_t^i-X_h^i}$ exceeds a certain maximal value.

For the non-ideal free energy $\Fnid$ we make an important
simplifying assumption: $\Fnid$ can be written as a function
of only the observables, 
\[
\Fnid(\psi)= \Fnid(r_w(\psi),r_h(\psi),r_t(\psi)).
\]
Typical terms in the non-ideal energy 
$\Fnid$ are a convolution integral, in which
proximity of hydrophilic (heads and water) beads and hydrophobic
tail beads is penalized: 
\[
\mu\int_\Omega \int_\Omega 
    \bigl(r_w(\psi)(x)+r_h(\psi)(x)\bigr)\,r_t(\psi)(y)\, \kappa(x-y) \, dxdy,
\]
and a compressibility term that penalizes 
deviation from unit total volume:
\[
\frac K2 \int_\Omega \bigl(r_w(\psi)(x)+r_t(\psi)(x) + r_h(\psi)(x)-1\bigr)^2\, dx.
\]
In these expressions we take two limits:
\begin{itemize}
\item In the limit $K\to\infty$, we find that the observables satisfy an
  incompressibility condition 
\begin{equation}
\label{eq:incompressibility}
r_t(\psi)+r_h(\psi)+ r_w(\psi) \equiv 1.
\end{equation}
With this condition the convolution integral above can be written as
\begin{equation}
\label{int:conv}
\mu\int_\Omega \int_\Omega 
    \bigl(1-r_t(\psi)(x)\bigr)\,r_t(\psi)(y)\, \kappa(x-y) \, dxdy.
\end{equation}
\item Replacing the fixed function $\kappa$ by a rescaled version
$\kappa_\delta$, defined by
\[
\kappa_\delta(x) := \delta^{-d}\kappa(x/\delta),
\]
the integral~\pref{int:conv} Gamma-converges in the following way,
\[
\frac 2\delta \int_\Omega \int_\Omega 
    \bigl(1-r_t(\psi)(x)\bigr)\,r_t(\psi)(y)\, \kappa_\delta(x-y) \, dxdy
    \stackrel{\delta\to0}\longrightarrow 
    K_{1,d} \left(\int \kappa\right) F^{\mathrm{int}}\bigl(r_t(\psi)\bigr)
\]
where the limit interface functional $F^{\mathrm{int}}$ is given by
\begin{equation}
F^{\mathrm{int}}(u) = \begin{cases}
\int_\Omega \mod{\nabla u}\, dx & \text{if $u\in \BV\bigl(\Omega;\{0,1\}\bigr)$}
\\
\infty & \text{otherwise}.
\end{cases}
\label{result:Davila}
\end{equation}
Here $K_{1,1} = 1$, $K_{1,2} = 2/\pi$, and $K_{1,3} = 1/2$
\cite{AlbertiBellettiniCassandroPresutti96,Davila02}. 
\end{itemize} 
    
Putting the ideal and non-ideal energies together, and applying the
simplifications above, we find (up to rescaling) 
\[
F(\psi) := \Fid(\psi) + \Fnid(\psi)
 = \int_\X \sum_{i=1}^\nl\mod{X_t^i-X_h^i}^p\psi(X)\, dX + 
   F^{\mathrm{int}}\bigl(r_t(\psi)\bigr).
\]
We now assume that we minimize $F$ under all variations of $\psi$ that conserve the observables $r_{t,h,w}(\psi)$. By a convexity argument, using the fact that  the functions $r_{t,h,w}$ do not distinguish between different lipids or water molecules, it follows that all lipids are identically distributed, as are all water molecules:
\[
\psi(X) = \prod_{i=1}^\nl \psi_\ell(X^i_t,X_h^i)
  \prod_{j=1}^\nw \psi_w(X^j_w),
\]
and that the total energy becomes a function of the observables alone,
by minimization over all other degrees of freedom (again we disregard
rescaling) 
\begin{align*}
&F(\rho_t,\rho_h,\rho_w) := F^{\mathrm{int}}(\rho_t) +{}\\
&\qquad  {}+\inf_{\psi_\ell}\Bigl\{\int_{\Omega\times\Omega}
  \mod{X_t-X_h}^p\psi_\ell(X_t,X_h)\, dX_tdX_h: 
    r_t(\psi_\ell) = \rho_t, \ r_h(\psi_\ell) = \rho_h\Bigr\}  
\end{align*}
with the remaining constraint $\rho_t+\rho_h+\rho_w \equiv 1.$
Remark that the final infimum is exactly the Wasserstein distance of
degree $p$, $d_p(\rho_t,\rho_h)^p$. 

\medskip
We now recover the functional $\F_\eps$ in~\eqref{def:Fe} for $\eps=1$ 
by defining  $u:=\rho_t$, $v:=\rho_h$ and by choosing $p=1$. 
Combining the pure-phase condition $u(x) = \rho_t(x)\in\{0,1\}$ for all $x$~\pref{result:Davila} with the incompressbility condition~\pref{eq:incompressibility} gives $v(x)=\rho_h(x)\in[0,1]$ and $uv=0$. It is easy to see that there is nothing to be gained in letting $v$ take values in the interior $(0,1)$, and we can therefore consider $\F_1$ to be given on the set of admissible functions
\begin{multline*}
\K_1 := \biggl\{ (u,v)\in 
   \BV\bigl(\R^2;\{0,1\}\bigr)\times L^1\bigl(\R^2;\{0,1\}\bigr): \\
  \left.\int u = \int v =M , \ uv = 0\text{ a.e.}\right\}.
\end{multline*}
The remaining dependence on $\eps$ is a consequence of simple rescaling.

\section{Ring solutions}
\label{app:ring_solutions}
We consider $r_1<r_2<R<r_3<r_4$ with
\begin{align}
  r_2^2 \,&=\, \tfrac12(R^2 + r_1^2), \qquad
  r_3^2 \,=\, \tfrac12(R^2 + r_4^2),\label{eq:a-r2r3}\\
  2R \,&=\, r_1+r_4\label{eq:a-R}
\end{align}
and $u,v$ given by
\begin{gather*}
  u\,=\, \Chi_{B_{r_3}(0)\setminus B_{r_2}(0)},\qquad v\,=\,
  \Chi_{B_{r_2}(0)\setminus B_{r_1}(0)} +\Chi_{B_{r_4}(0)\setminus
  B_{r_3}}.
\end{gather*}
In this setting the unique monotone optimal transport map $S$ from $u$
to $v$ is radially symmetric, $S(x)=S(|x|)$ and determined
by the requirement that
\begin{alignat}2
  \int_{r_2}^s 2\pi r\,dr \,&=\, \int_{r_1}^{S(s)}2\pi
  r\,dr&\qquad&\text{ for } 
  r_2<s<R,\label{eq:a-cond-S-1}\\ 
  \int_{s}^{r_3} 2\pi r\,dr \,&=\, \int_{S(s)}^{r_4}2\pi r\,dr
  &\qquad&\text{ for } 
  R<s<r_3.\label{eq:a-cond-S-2}
\end{alignat}
Then
\begin{align}
  d_1(u,v)\,&=\, \int |x-S(x)|u(x)\,dx\notag\\
  &=\, \int_{r_2}^R 2\pi r(r-S(r))\,dr + \int_R^{r_3} 2\pi
  r(S(r)-r)\,dr.\label{eq:a-d1-1}
\end{align}
To compute the first integral we introduce new coordinates
\begin{gather}
  m(r)\,=\,\pi r^2-\pi r_2^2\quad\text{ with inverse }\quad
  r(m)\,=\, \big(r_2^2+\frac{1}{\pi}m\big)^{1/2}.\label{eq:a-def-m1}
\end{gather}
In this coordinates \eqref{eq:a-cond-S-1} implies
\begin{gather}
  m(r)\,=\, m(S(r)) -m(r_1),\\
  S(r(m))\,=\, r\big(m+m(r_1)\big)\,=\,r\big(m-m(R)\big),
\end{gather}
where in the last equality we have used that
\begin{gather*}
  m(r_1) + m(R)\,=\,\pi\big(r_1^2-2r_2^2+R^2\big)\,=\, 0
\end{gather*}
by \eqref{eq:a-r2r3}.
The first integral in \eqref{eq:a-d1-1} therefore becomes
\begin{align}
  \int_{r_2}^R 2\pi r(r-S(r))\,dr \,&=\, \int_0^{m(R)}
  \big(r(m)-r(m-m(R))\big)\,dm\notag\\
  &=\, \int_0^{m(R)} r(m)-r(-m)\,dm\notag\\
  &=\, \frac{2\pi}{3}\Big[\Big(r_2^2 +\frac{1}{\pi}m\Big)^{3/2}
  +\Big(r_2^2 -\frac{1}{\pi}m\Big)^{3/2}\Big]_0^{m(R)}\notag\\
  &=\, \frac{2\pi}{3}\Big[\Big(r_2^2 +\frac{1}{\pi}m(R)\Big)^{3/2}
  +\Big(r_2^2 -\frac{1}{\pi}m(R)\Big)^{3/2}-2r_2^3\Big].\label{eq:a-d1-2} 
\end{align}
If we now introduce $t=(r_4-r_1)/2$ then
\begin{gather*}
  R-r_1\,=\,r_4-R\,=\, t,\qquad m(R)\,=\, \frac{\pi}{2}(2Rt-t^2),
\end{gather*}
and \eqref{eq:a-d1-2} yields
\begin{align}
  \int_{r_2}^R 2\pi r(r-S(r))\,dr  \,&=
  \frac{2\pi}{3}
  \Big[R^3+(R-t)^3 -2R^3\Big(1-\frac{t}{R}+\frac{t^2}{2R^2}\Big)^{\frac{3}{2}}\Big].
  \label{eq:a-d1-3} 
\end{align}
If we do the analogous calculations for the second integral in
\eqref{eq:a-d1-1} we obtain that
\begin{align}
  \int_R^{r_3} 2\pi r(S(r)-r)\,dr \,&=
  \frac{2\pi}{3}
  \Big[R^3+(R+t)^3 -2R^3\Big(1+\frac{t}{R}+\frac{t^2}{2R^2}\Big)^{\frac{3}{2}}\Big].
  \label{eq:a-d1-4} 
\end{align}
which is the expression on the right-hand side of \eqref{eq:a-d1-3} with
$t$ substituted by $-t$.\\
We consider $R\to\infty$ with $t$ of order one and obtain by a Taylor expansion
in $1/R$
\begin{align*}
  & d_1(u,v) +\int |\nabla u|\\
  =\,&   \frac{2\pi}{3}
  \Big[2R^3+(R+t)^3 +(R-t)^3
  -2R^3\Big(1+\frac{t}{R}+\frac{t^2}{2R^2}\Big)^{\frac{3}{2}}
  -2R^3\Big(1-\frac{t}{R}+\frac{t^2}{2R^2}\Big)^{\frac{3}{2}}\Big]\\
  & +2\pi R\Big[\Big(1+\frac{t}{R}+\frac{t^2}{2R^2}\Big)^{\frac{1}{2}}
  +\Big(1-\frac{t}{R}+\frac{t^2}{2R^2}\Big)^{\frac{1}{2}}\Big]\\
  =\,& \pi Rt^2 + \frac{\pi}{16}\frac{t^4}{R}+ 4\pi R
  +\frac{\pi}{2}\frac{t^2}{R} + O(R^{-3}) \\
  =\,& 2\pi Rt \Big[\Big(\frac{t}{2} +\frac{2}{t}\Big) +
  \frac{1}{4R^2} + \frac{(2-t)}{32R^2}\big(t^2+2t-4\big) + O(R^{-2})O(|t-2|) \Big]\\
  =\,& 2\pi Rt\Big[ 2 +\frac{1}{4}(t-2)^2 +\frac{1}{4}R^{-2}+ O(R^{-2})O(|t-2|)\Bigr],
\end{align*}
which is the desired asymptotic development.
\end{appendix}

\bibliography{helf-lib}
\bibliographystyle{plain}
\end{document}